\documentclass{aa}
\usepackage{graphicx,natbib}
\usepackage[varg]{txfonts}
\usepackage{afterpage}
\usepackage{color}
\usepackage{upgreek}
\usepackage{soul}
\usepackage{hyperref}

\newcommand{\Prot}{\ensuremath{P_{\mathrm{rot}}}}
\newcommand{\Bm}{\ensuremath{\langle B\rangle}}
\newcommand{\Bz}{\ensuremath{\langle B_z\rangle}}
\newcommand{\Bq}{\ensuremath{\langle B_\mathrm{q}\rangle}}
\newcommand{\vsi}{\ensuremath{v\,\sin i}}
\newcommand{\kms}{km\,s$^{-1}$}
\defcitealias{2020A&A...639A..31M}{Paper~I}
\defcitealias{2022A&A...660A..70M}{Paper~II}
\defcitealias{2009A&A...498..961R}{the Renson catalogue}

\bibpunct[ ]{(}{)}{;}{a}{}{,}

\begin{document}

\title{Long-period Ap stars discovered with TESS data: \\Cycles~3 and 4}

\author{G.~Mathys\inst{1}
 \and D.~L.~Holdsworth\inst{2,3}
 \and D.~W.~Kurtz\inst{4,2}}
 
\institute{European Southern Observatory,
  Alonso de Cordova 3107, Vitacura, Santiago,
  Chile\\\email{gmathys@eso.org} 
\and
Jeremiah Horrocks Institute, University of Central Lancashire, Preston
PR1 2HE, UK 
\and
South African Astronomical Observatory, P.O. Box 9, Observatory 7935,
Cape Town, South Africa 
\and
Centre for Space Physics, North-West University, Mahikeng 2735, South
Africa} 

\date{Received $\ldots$ / Accepted $\ldots$}

\titlerunning{Long period Ap stars}

\abstract{One of the most challenging aspects of the Ap stars is the
  extreme differentiation of their rotation periods, which span more
  than five orders of magnitude. The physical origin of this
  differentiation remains poorly understood. The consideration of the
  most slowly rotating Ap stars represents a promising approach to
  gain insight into the processes responsible for the rotational
  braking to which the Ap stars are subject. However, historically,
  the study of these stars focused primarily on the most strongly
  magnetic among them. This bias introduced an ambiguity in the
  conclusions that could be drawn, as it did not allow the distinction
  between the rotational and magnetic effects, nor the investigation
  of possible correlations between rotational and magnetic
  properties. We previously showed that the identification
  of super-slowly rotating Ap (ssrAp) star candidates (defined as Ap
  stars that have rotation periods $\Prot>50$\,d) through
  systematic exploitation of the available TESS photometric
  observations of Ap stars is an effective approach to build a sample
  devoid of magnetic bias. This approach rests on the presence of
  brightness spots on the surface of Ap stars that are not distributed
  symmetrically about their rotation axes and show long-term
  stability, hence are responsible for photometric
  variations with the stellar rotation period. In our previous
  analyses of TESS Cycle~1 and Cycle~2 data, we interpreted the Ap
  stars showing no such variability over the 27-d duration of a TESS
  sector as being ssrAp star candidates. Here, we apply the same
  approach to TESS Cycle~3 and Cycle~4 observations of Ap stars. We
  show, however, that two issues that had not been fully appreciated
  until now may lead to spurious identification of ssrAp star
  candidates. On the one hand, a considerable fraction of the Ap stars
  in the existing lists turn out to have erroneous or dubious spectral
  classifications. On the other hand, the TESS data processing may
  remove part of the variability signal, especially for stars with
  moderately long periods 
  ($20\,\mathrm{d}\lesssim\Prot\lesssim50$\,d). After critical
  evaluation of these effects, we report the identification of  25 new
  ssrAp star candidates and of 8 stars with moderately long
  periods. Combining this list with the lists of ssrAp stars from
  Cycles~1 and 2 and with the list of ssrAp stars that were previously
  known but whose lack of variability was not detected in our study,
  we confirmed at a higher significance level the conclusions drawn in
  our earlier work. These include the lower rate of occurrence of
  super-slow rotation among weakly magnetic Ap stars than among
  strongly magnetic ones, the probable existence of a gap between
  $\sim$2 and $\sim$3\,kG in the distribution of the magnetic field
  strengths of the ssrAp stars, and the much higher rate of occurrence
  of rapid oscillations in ssrAp stars than in the whole population of
  Ap stars. The next step to gain further understanding of the ssrAp
  stars will be to obtain high-resolution spectra of those for which
  such observations have not been made yet, to constrain their
  rotation velocities and their magnetic fields. 
   }


\maketitle

\section{Introduction}
\label{sec:intro}
More than a century ago, \citet{1913AN....173....1L} reported that a number of spectral lines
in the spectrum of $\alpha^2$~CVn (= HD~112413) show intensity
variations. \citet{1913AN....195..159B} then found that these variations
are periodic and derived an approximate value of their period,
$P=5\fd50$. In the spectrum of $\alpha^2$~CVn, there are spectral
lines that appear more intense or weaker than those of the majority of
the stars of spectral type A. Such anomalies were observed in a
fraction of the A stars, which were accordingly assigned the spectral
type Ap in the Harvard classification (``p'' stands for
``peculiar''). It was subsequently found that at least a fraction of
the Ap stars display periodic line intensity variations such as those
of $\alpha^2$~CVn. Such variability was recognised as typical of Ap
stars, and $\alpha^2$~CVn became the prototype of the class. Even now
Ap stars are frequently referred to as $\alpha^2$~CVn 
variables.

The discovery by \citet{1947ApJ...105..105B} of a magnetic
field in the Ap star HD~118022 (= 78~Vir) soon led to the realisation
that such fields were ubiquitously present in Ap stars \citep[see][and
references therein]{1958ApJS....3..141B}. The star HD~125248 (=
CS~Vir) became the first Ap star in which the periodic variability of
the magnetic field was established \citep{1951ApJ...114....1B}. 
Whether this variability, which is now known to be
characteristic of Ap stars in general, was intrinsic (due to
``magnetic oscillations'') or of geometric origin was unclear at
first. Arguments supporting the latter interpretation were presented
by \citet{1956PASP...68...92D}, who developed the oblique rotator
model, according to which the observed magnetic and spectral
variations result from the changing aspect of the visible hemisphere
of a rotating star whose surface presents inhomogeneities that are not
symmetric about the rotation axis.

Taking advantage of the advent of
photoelectric photometry, \citet{1968ApJ...154..945S} carried out the
first systematic study of the brightness variations of a sample of Ap
stars. He concluded that the periods of these variations were not in
contradiction with the magnetic variations within the framework of the
oblique rotator theory. Further evidence in support of the oblique
rotator model was presented by \citet{1970stro.coll..254P}, who took
advantage of the increasing number of stars with known magnetic,
spectral and/or photometric variation periods to confirm at a much
higher level of statistical significance the period 
vs. line width relation first reported by \citet{1956PASP...68...92D}.
This relation is
illustrated in Fig.~1 of \citet{1971ApJ...164..309P}.

However, \citet{1970stro.coll..254P} acknowledged that 6 stars were
known at the time to show variations with periods longer than 100\,d
that appeared secure. He reckoned that this number could not be
reconciled with a rotational origin of the variability, if one assumed a
Maxwellian distribution of the rotational velocities of the Ap stars,
similar to their distribution for the normal upper main sequence
stars. Therefore, the possibility that long period variations in Ap
stars were not due to rotation but to some other, unknown mechanism,
remained debated for a long time. It still was, to some extent,
an open question, as recently as 30 years ago
\citep{1993ASPC...44..547H}.

\begin{table*}
  \caption{Known ssrAp stars.}
  \label{tab:known_ssrAp}
\small
  \begin{tabular*}{\textwidth}[]{@{}@{\extracolsep{\fill}}rllrlrc}
\hline\hline\\[-4pt]
    \multicolumn{1}{c}{HD/HDE}&Other ID&Spectral type&$\Prot$
                                                   (d)&Reference&$B_0$
                                                                  (kG)&roAp\\[4pt]
    \hline\\[-4pt]
    \multicolumn{7}{c}{Stars with accurately determined periods}\\[4pt]
    \hline\\[-4pt]
50169&BD~$-1$~1414&A3p SrCrEu&10600&\protect{\citet{2019A&A...624A..32M}}&5.1&\\
9996&HR~465&B9p CrEuSi&7937&\protect{\citet{2014AstBu..69..315M}}&4.7&\\
965&BD~$-$0~21&A8p SrEuCr&6030&\protect{\citet{2019A&A...629A..39M}}&4.3&\\
166473&CoD~$-$37~12303&A5p SrEuCr&3836&\protect{\cite{2020A&A...636A...6M}}&7.1&roAp\\
94660&HR~4263&A0p EuCrSi&2800&\protect{\citet{2017A&A...601A..14M}}&6.2&\\
187474&HR~7552&A0p EuCrSi&2345&\protect{\citet{1991A&AS...89..121M}}&5.4&\\
\tablefootmark{a}&CPD$-$62~2717&B8\tablefootmark{b}&1765&\protect{\citet{2023MNRAS.522.5931C}}&9.9&\\
    59435&BD~$-$8~1937&A4p SrCrEu&1360&\protect{\citet{1999A&A...347..164W}}&3.0&\\
18078&BD~$+$55~726&A0p SrCr&1358&\protect{\citet{2016A&A...586A..85M}}&3.4&\\
93507&CPD~$-$67~1494&A0p SiCr&556&\protect{\citet{1997A&AS..123..353M}}&7.2&\\
2453&BD~$+$31~59&A1p SrEuCr&518.2&\protect{\citet{2017PASP..129j4203P}}&3.7&\\
51684&CoD~$-$40~2796&F0p SrEuCr&371&\protect{\citet{2017A&A...601A..14M}}&6.0&\\
61468&CoD~$-$27~4341&A3p EuCr&322&\protect{\citet{2017A&A...601A..14M}}&6.8&\\
110274&CoD~$-$58~4688&A0p EuCr&265.3&\protect{\citet{2008MNRAS.389..441F}}&4.0&\\
69544&CD~$-$50~3225&B8p Si&236.5&\protect{\citet{2015A&A...581A.138B}}\\
188041&HR~7575&A6p SrCrEu&223.78&\protect{\citet{2017A&A...601A..14M}}&3.7&\\
221568&BD~$+$57~2758&A1p SrCrEu&159.10&\protect{\citet{2017PASP..129j4203P}}\\
66821&CPD~$-$54~1483&A0p Si&154.9&\protect{\citet{2015A&A...581A.138B}}\\
116458&HR~5049&A0p EuCr&148.39&\protect{\citet{2017A&A...601A..14M}}&4.7&\\
126515&BD~$+$1~2927&A2p CrSrEu&129.95&\protect{\citet{1997A&AS..123..353M}}&12.7&\\
340577&BD~$+$25~4289&A3p SrCrEu&116.7&\protect{\citet{2016AJ....152..104H}}\\
263361&BD~$+$19~1468&B9p Si&88.9&\protect{\citet{2015A&A...581A.138B}}\\
149766&CD~$-$51~10384&B8p Si&87.3&\protect{\citet{2015A&A...581A.138B}}\\
\tablefootmark{c}&BD~$-$7~1884&SrCrEu&75.97&\protect{\citet{2016AJ....152..104H}}\\
158919&CPD~$-$65~3465&B9p Si&75.72&\protect{\citet{2015A&A...581A.138B}}\\
102797&CPD~$-$75~757&B9p Si&75.02&\protect{\citet{2002MNRAS.331...45K}}\\
    8441&BD~$+$42~293&A2p Sr&69.51&\protect{\citet{2017PASP..129j4203P}}\\
5797&BD~$+$59~163&A0p CrEuSr&68.05&\protect{\citet{2018PASP..130d4202D}}\\
123335&HR~5292&Bp He wk SrTi&55.215&\protect{\citet{Hensberge:2007oz}}\\
103844&CPD~$-$75~767&A0p Si&52.95&\protect{\citet{2015A&A...581A.138B}}\\
200311&BD~$+$43~3786&B9p SiCrHg&52.0084&\protect{\citet{1997MNRAS.292..748W}}&8.4&\\
256476&TYC~736-1842-1&A2p Si&51.23&\protect{\citet{2016AJ....152..104H}}\\
184471&BD~$+32$~3471&A9p SrCrEu&50.8&\protect{\citet{2012AN....333...41K}}\\
    87528&CPD~$-$62~1415&B8/9\,III\tablefootmark{d}&50.35&\protect{\citet{2015A&A...581A.138B}}\\[4pt]
    \hline\\[-4pt]
    \multicolumn{7}{c}{Stars with well established period lower limits}\\[4pt]
    \hline\\[-4pt]
201601&$\gamma$~Equ&A9p
                     SrEu&$\gtrsim35,500$&\protect{\citet{2016MNRAS.455.2567B}}&3.9&roAp\\
    213258&BD$+$35~4815&B9p&$\gtrsim18,000$&\protect{\citet{2023A&A...670A..72M}}&3.8&roAp\\
    116114&BD$-$17~3829&F0p
                         SrCrEu&$>17,700$&\protect{\citet{2022MNRAS.514.3485G}}&6.0&roAp\\
    101065&CD$-$46~7232&F3p
                         Ho&$>15,700$\rlap{\tablefootmark{e}}&\protect{\citet{2018MNRAS.477.3791H}}&&roAp\\
    55719&HR~2727&A3p
                   SrCrEu&$>14,000$&\protect{\citet{2022MNRAS.514.3485G}}&6.3&\\
    177765&CD$-$26~13816&A5p
                          SrEuCr&$>13,500$&\protect{\citet{2022MNRAS.514.3485G}}&3.4&roAp\\
    137949&33~Lib&F0p
                   SrEuCr&$\gg10,000$&\protect{\citet{2022MNRAS.514.3485G}}&4.7&roAp\\
    165474&BD$+$12~3382B&A7p SrCrEu&$>9,900$&\protect{\citet{2022MNRAS.514.3485G}}&6.6&\\
    176232&10~Aql&A6p Sr&$\gg4,400$&\protect{\citet{2019MNRAS.483.3127S}}&&roAp\\
    29578&CD$-$54~902&A4p SrEuCr&$\gtrsim4,000$\rlap{\tablefootmark{f}}&\protect{\citet{2022MNRAS.514.3485G}}&3.1&\\[4pt]
    \hline\\[-4pt]
  \end{tabular*}
  \tablefoottext{a}{TYC~8979-339-1.}
  \tablefoottext{b}{Spectral type from \citet{1976A&AS...23..283L}.}
    \tablefoottext{c}{TYC~5395-1139-1.}
    \tablefoottext{d}{Spectral type from \citet{1975mcts.book.....H}.}
    \tablefoottext{e}{We adopted the time base covered by the $\Bz$
      measurements as lower limit, not the heavily extrapolated value
      of 188\,yr proposed by \citet{2018MNRAS.477.3791H}.}
    \tablefoottext{f}{See text.}
  \end{table*}

New light was shed on this issue by the discovery of a large
population of Ap stars with resolved magnetically split lines, as a
result of an extensive, systematic search project carried out by
Mathys and collaborators \citep[and references
therein]{1997A&AS..123..353M,2017A&A...601A..14M}. The resolution of
the spectral lines of these stars  into their magnetic components
results from the combination of their strong magnetic fields (with
intensities of several kG) and of their low projected equatorial
velocities \vsi\ (of a few \kms). Their identification was based on these
properties. For most of them, constraints on their variation periods
were derived only in a second step, from magnetic measurements
obtained at series of different epochs. On a statistical basis, one
must expect the majority of the stars with small  values of \vsi\ to
be genuine slow rotators (as opposed to being seen at a low inclination
angle $i$ of the rotation axis to the line of sight). On the other
hand, of the 50 Ap stars with resolved magnetically split lines in the
sample of \citet{2017A&A...601A..14M} for which reliable values of the
variation period $P$, or lower limits thereof, were available, 31 (or
62\%) had $P>30$\,d. Thus, stars selected on the basis of the
sharpness of their spectral lines, hence on their probable slow
rotation, tend in their majority to have long variation periods. This
is the reverse argument of that of \citet{1970stro.coll..254P},
according to which stars selected for their long variation periods
have low projected equatorial velocity. It strengthens the evidence
that all Ap stars, regardless of the length of their periods, owe
their variability to the change of aspect of their visible hemisphere
over a rotation cycle, as interpreted in the framework of the oblique
rotator model. Within this context, the magnetic, spectral and
photometric variations occur with the rotation period, and the surface
inhomogeneities remain stable over very long timescales, generally
covering many rotation periods. 

An unexpectedly large number of Ap stars showing resolved magnetically
split lines in the visible were discovered.\footnote{More were found
  showing resolved magnetically split lines in the infrared
  \citep{2019ApJ...873L...5C}, but they are of lesser relevance in the
  context of this study. Indeed, as the Zeeman effect depends
  quadratically on the wavelength while the wavelength dependence of
  the Doppler effect is only linear, for a given magnetic field
  strength, magnetically split lines can be resolved in faster
  rotating stars in the infrared than in the visible.} This, in turn,
led to the realisation of the existence of a much larger population of
very slowly rotating Ap stars than was previously estimated. It also
became apparent that Ap stars with rotation periods of the order of 300\,yr
must definitely exist, and that some Ap stars may even have much
longer periods \citep{2017A&A...601A..14M}. On the other hand,
extensive studies of large samples of Ap stars based on ground- and
space-based photometric surveys have confirmed that the vast majority
of Ap stars have rotation periods beween 1 and 10 days
\citep{2012MNRAS.420..757W,2015A&A...581A.138B,2015AN....336..981B,2016AJ....152..104H,2017MNRAS.468.2745N,2018A&A...619A..98H,2020MNRAS.493.3293B,2023A&A...676A..55L}. With
the shortest periods known of the order of 0.5\,d and the
longest ones set to be of the order of 300\,yr at least, the rotation rates of
Ap stars range over more than five orders of magnitude. Understanding
how such a large differentiation is achieved in stars whose other
physical properties appear similar represents a major challenge. In
this context, achieving a better characterisation of the extreme slow
rotation tail of the period distribution must be expected to provide
some of the most valuable constraints to guide theoretical
developments. This is the main motivation underlying studies such
as the one presented here.

\citet{2020pase.conf...35M} introduced the nomenclature super-slowly
rotating Ap (ssrAp) stars to refer to the group of the longest period
Ap stars. By definition, the members of this group have rotation
periods $\Prot>50$\,d. While this definition involves a certain degree
of arbitrariness, it may be noted that for an Ap star with
$\Prot=50$\,d observed equator-on, the thermal broadening of the
spectral lines of Fe (which are often used to diagnose various
physical properties, such as the magnetic field) is of the same order
as their rotational broadening. The latter is also close to, or
slightly below, the 
resolution limit ($R=\Delta\lambda/\lambda\sim10^5)$ of most
high-resolution spectrographs currently used for  Ap star studies.

The main limitation to the knowledge of the distribution of the
rotation periods of the most slowly rotating Ap stars is the time
interval over which relevant observations of each of them has been
obtained. In most cases, data sampling a full rotation cycle have not
been acquired yet. Actually, accurate determination of the value of a
rotation period requires the consideration of measurements covering
substantially more than one rotation
cycle. \citet{2020pase.conf...35M} compiled a list of the ssrAp stars
for which accurate periods had been derived from the analysis of data
suitably distributed over a long enough time span. Furthermore, in a
number of cases, meaningful lower limits of the periods of
ssrAp stars can be set from the consideration of observations sampling
only a fraction of a rotation period. Table~\ref{tab:known_ssrAp}
presents a list of the ssrAp stars that are presently known either on
the basis of accurately determined period values or on reliably
established period lower limits. The former part represents an updated
version of Table~1 of \citet{2020pase.conf...35M}, to which one recent
period determination was added.

While we tried to be as complete as possible in compiling this list,
we may have overlooked the odd ssrAp star with a published
period value or lower limit. However, we have also deliberately
omitted from the list several stars for which long periods or limits
thereof have been proposed in the literature, whenever critical
evaluation suggests that the published values are not reliable
enough. In particular, we did not include HD~95699 (= CD$-$41~6291) whose
published period ($P=57\fd176$, \citealt{2002MNRAS.331...45K}) may be
of instrumental origin; HD~110066 (= HR~4816) for which the period
values proposed by \citet{1981A&AS...44..265A}, $P=13.5$\,yr or
$P=27$\,yr, were not confirmed by more recent observations
\citep{2017PASP..129j4203P}; nor HD~150562 (= CoD$-$48~11127), for which
we could not confirm, on the basis of our own
frequency analysis of the magnetic data of \citet{2022MNRAS.514.3485G},
the value $\Prot=2100$\,d of the rotation period
derived by these authors. We do not regard either the
values $\Prot=4000$\,d or $\Prot=9370$\,d that they proposed
for HD~29578 as fully convincing, but we agree that the former
represents a reliable lower limit of the rotation period. On the other
hand, in those cases for which the observations 
obtained until now do not adequately sample a full variation cycle,
the list contains only stars that show definite variability and for
which a period shorter than 50\,d can be unambiguously ruled
out. Accordingly, stars such as HD~92499 (= CoD$-$42~6407), HD~117290 (=
CoD$-$48~8252) or HD~213637 (= BD$-$20~6447), for which tentative lower
limits of the period are given in Tables~1 and 2 of
\citet{2017A&A...601A..14M}, are not included in
Table~\ref{tab:known_ssrAp}. Neither is HD~75445 (=CD$-$38~4907),
whose variability remains to be established \citep{2022MNRAS.514.3485G}.

The table contains 34 stars whose rotation period has been accurately
determined, and 10 stars for which the observations obtained until now
lend themselves to setting a meaningful lower limit of the
period. Column~1 gives the HD number of the stars (when available) and
Col.~2 another 
identification. The spectral types listed in Col.~3 are from
\citet{2009A&A...498..961R}, except when otherwise mentioned in a
footnote. Columns 4 and 5 contain the value of the 
period, when it has been accurately determined, or its lower limit,
and the corresponding literature source. For those stars whose
spectral lines are resolved into their magnetically split components,
the phase-averaged value $B_0$ of the mean magnetic field modulus
$\Bm$ (the
line intensity weighted average over the visible stellar hemishphere
of the modulus of the magnetic vector) appears in Col.~6. For stars
for which the available $\Bm$ measurements sample well the full
rotation cycle, $B_0$ is the average of $\Bm$ over the rotation
period; otherwise, it is an average over the observed phases. Most entries
in Col.~6 are from Tables~13 and 14 of
\citet{2017A&A...601A..14M}; for a few stars, they are complemented by
more recent data from
\citet{2019A&A...624A..32M,2019A&A...629A..39M,2020A&A...636A...6M,2023A&A...670A..72M},
\citet{2022MNRAS.514.3485G} and \citet{2023MNRAS.522.5931C}. The
rapidly oscillating Ap (roAp) stars are identified in Col.~7. 

The longest period whose value could be accurately determined
until now is that of HD~50169, $\Prot=10,600$\,d. Among the stars of
Table~\ref{tab:known_ssrAp} for which only a lower limit of the period
could be derived, only one, HD~29578, may possibly have a
significantly shorter period. This illustrates one of the difficulties
of characterising the slow rotation tail of the period distribution:
Ap stars in general have not yet been systematically studied over a
long enough time interval to cover a full variation cycle of the most
slowly rotating ones. For a more detailed discussion of this
fundamental limitation, see \cite{2019A&A...624A..32M}. 

Among the 44 stars of Table~\ref{tab:known_ssrAp}, 26 are known to
show resolved magnetically split lines. These stars have magnetic
fields stronger than the bulk of the Ap stars. Their predominance
among the known ssrAp stars certainly in part results from a selection
bias. On the one hand, the most strongly magnetic stars are
particularly interesting to study to understand better how the
magnetic field affects the other physical properties and
processes. Thus, they have been the subject of more systematic
searches and studies than stars with weaker fields. On the other hand,
the consideration of magnetic variations generally represents the
method of choice to constrain the longest rotation periods, because
the ratio of their amplitudes to the measurement uncertainties 
is often much larger than for photometric variations and because long-term
instrumental effects can be more easily handled for magnetic measurements
than for photometric measurements. These arguments have been developed
in greater detail by \citet{2020pase.conf...35M}.

Nevertheless, one should wonder if the rate of occurrence of
super-slow rotation is different in weakly magnetic Ap stars than in
strongly magnetic ones. Little information is available in the
literature about the magnetic fields of the 18 stars of
Table~\ref{tab:known_ssrAp} that are not known to show resolved
magnetically split lines. Most of them do not appear to have been
oberved at a spectral resolution sufficient to derive meaningful
constraints about their magnetic fields. The exceptions include
HD~8441, for which \citet{2012AstL...38..721T} report
that there is no significant magnetic intensification of the spectral
line intensities; HD~184471, where the magnetic
splitting of the line Fe~{\sc ii}~$\lambda$\,6149.2\,\AA\ may have
been marginally resolved \citep{2017A&A...601A..14M}; HD~101065
(Przybylski's star), for which a value of 1.6\,kG of the mean quadratic
magnetic field was derived by \citet{2018MNRAS.477.3791H}; and
HD~176232, for which several differential line broadening studies,
yield field strengths (similar to the mean quadratic magnetic field)
averaging 1.3\,kG
\citep{2000A&A...357..981R,2002MNRAS.337L...1K,2003A&A...409.1055L}. For
a definition of the mean quadratic magnetic field, see
\citet{2006A&A...453..699M}; the value of this field moment is
generally close to that of the mean magnetic field modulus (see
Sect.~\ref{sec:conc} for details). The mean magnetic field modulus of
these four stars must be weaker than
2\,kG. \citet{2011AstL...37...20S} also noted that the magnetic field
of HD~5797 does not exceed 3\,kG.

In summary, of the 31 stars of Table~\ref{tab:known_ssrAp} for
which constraints on the magnetic field are available, 26 have
$B_0\ge3.0$\,kG and 5 have $B_0<3.0$\,kG. To which extent this
difference is the result of a selection bias that definitely exists or
is the reflection of a genuine excess of super-slow rotators among
strongly magnetic Ap stars is an important question that needs to be
addressed. To this effect, usage of a method that allows ssrAp stars
to be identified independently of the strength of 
their magnetic field is essential. We have proposed an approach that
fulfils 
this requirement. It is based on the argument that Ap stars that do
not show photometric variations when observed with TESS over a 27-d
sector are very probably ssrAp stars, except for a few that have $20
\le P_{\rm rot} \le 50$\,d (still very slowly rotating, but not ssrAp,
by definition) and even fewer with an unlikely pole-on geometrical 
configuration. In previous papers, we applied it to
TESS Cycle~1 \citep[\!, hereafter Paper~I]{2020A&A...639A..31M} and
Cycle~2 \citep[\!, hereafter Paper~II]{2022A&A...660A..70M}
data. Here, we present its application to analyse TESS Cycles~3 and 4
observations.

Section~\ref{sec:ssrAp} is devoted to the description of the process
of selection of the ssrAp star candidates. Two sources of spurious
identifications are discussed in detail: contamination of the original
Ap star lists by misclassifications and removal of intermediate period
($20\,\mathrm{d}\lesssim P\lesssim50$\,d) variations from the TESS
signal as part of its processing. In Sects.~\ref{sec:roap} and
\ref{sec:dsct}, we report the presence of roAp stars and of a
$\delta$~Sct star among the ssrAp star candidates. The properties of
the ssrAp star candidates are discussed in Sect.~\ref{sec:disc}. In
Sect.~\ref{sec:conc}, we draw conclusions about the statistical trends
revealed by this study and their implications for our understanding of
the rotational and magnetic properties of Ap stars. 

\section{Super slowly rotating Ap (ssrAp) star candidates}
\label{sec:ssrAp}
\subsection{Search strategy}
\label{sec:strategy}
This new step of our project for systematic search of ssrAp star
candidates is based on the consideration of the observations obtained
during Cycles~3 and 4 of the TESS mission. To select the targets to be
analysed, we built a list of the stars whose spectral type contains ``CR'',
``EU'', ``SR'', ``SI'' or ``P'' in the catalogue of \citet[][,
hereafter the 
Renson catalogue]{2009A&A...498..961R}. This list contains
3110 stars that have a TESS Input Catalog (TIC) identification
number. Among the latter, 484 have been 
observed in Cycle~3 (with 748 corresponding light curves, because for
some of the stars, photometric data were obtained in several sectors)
and 594 have 
been observed in Cycle~4 (with 1320 corresponding light curves). We
restricted our search to light curves that have 2\,min cadence data
processed  with the TESS SPOC (Science Processing Operations Center; 
\citealt{2016SPIE.9913E..3EJ}) pipeline. This is consistent with the
previous papers. We extracted the PDCSAP flux data (for more details,
see Sect.~\ref{sec:sap}), rejecting datapoints with bad-data flags. 

Within the  framework of the oblique rotator model, the Ap stars show
photometric variability as a result of the non-uniformity of their
surface brightness, whose distribution is to first order symmetric
about the magnetic axis. The identification of potential ssrAp star
candidates from TESS data is based on the consideration that Ap stars 
that do not show photometric variations over one TESS 27-d sector have
a very high probability of having long rotation periods. This assumes
that the inclination angle $i$ of the rotation axis to the line of
sight and the obliquity angle $\beta$ between the magnetic and
rotation axes are not close to zero. We have shown in detail in
\citetalias{2020A&A...639A..31M} and \citetalias{2022A&A...660A..70M}
that the rate of occurrence of very low $i$ or $\beta$ is expected to
be very low. This discussion will not be repeated here.

Following this approach, we carried out an automated pre-selection of
potential ssrAp stars by calculating on a sector-by-sector basis the
Fourier transforms of the 2068 above-mentioned light curves. Those
stars showing peaks in the range of $0 - 10$\,d$^{-1}$ with a S/N
ratio exceeding 10 were rejected. This left 96 potential ssrAp
candidates in Cycle~3 and 151 in Cycle~4. 

The areas covered by TESS in Cycles~1 and 3 show considerable
overlap. As a result, 42 of the 60 ssrAp star candidates listed in Table~1
of \citetalias{2020A&A...639A..31M} are among the 96 potential ssrAp
star candidates identified in Cycle~3. From this point on, they will be
left out of the Cycle~3 list to avoid unnecessary duplication. On the
other hand, we checked that for most of the 18 ssrAp star candidates absent
from the Cycle~3 list, no 2-min data were obtained in Cycle~3, while
the omission of the others can be attributed to low frequency noise
causing spurious peaks with $\rm{S/N}>10$ in the $0 - 10$\,d$^{-1}$
frequency range.

Similarly, although the sky coverage of TESS Cycle~4 differs
considerably from that of Cycle~2, 28 of the 151 stars returned by the
automatic search procedure for Cycle~4 were already listed in Table~A.1
of \citetalias{2022A&A...660A..70M}, and 1 was present in Table~1 of
\citetalias{2020A&A...639A..31M}. These stars were accordingly 
removed from the Cycle~4 list. After completion of these steps, there
were 54 stars left in the Cycle~3 list and 122 stars left in the Cycle
4 list. Two stars, TIC~152803574 and TIC~372617495, appear in both
lists (see also Appendix~\ref{sec:misclass} and
Sect.~\ref{sec:visual}). 

\subsection{Classification issues}
\label{sec:class}
The initial automatic identification of the stars from
\citetalias{2009A&A...498..961R} for 
which TESS 2-min observations were obtained in Cycles~3 and 4 needs to
be refined for a number of reasons. In particular, the resulting lists
include stars 
flagged with either ``/'' or ``?'' in the first column of the
catalogue. \citet{2009A&A...498..961R} assigned the ``/'' flag to stars
that were once improperly considered as chemically peculiar. The Cycle~3
and 4 lists contained, respectively, 2 and 3 such stars. Making use of
the classification and literature information available through
SIMBAD, we confirmed that none of these 5 stars is indeed an Ap star.

The flag ``?'' in \citetalias{2009A&A...498..961R} identifies stars whose peculiar
nature is doubtful. Consistent with our approach of
\citetalias{2022A&A...660A..70M}, to ensure that the final list of
ssrAp star candidates is restricted as much as possible to bona fide
Ap stars, we excluded the stars flagged ``?'' by
\citet{2009A&A...498..961R} unless evidence of their
peculiar nature was subsequently brought forward, according to the
information available through SIMBAD. Such evidence was found only for
one of the 13 stars flagged ``?'' from the Cycle~3 list, the roAp star
TIC~349945078 \citep{2019MNRAS.487.2117B}. No convincing evidence that
any of the 31 flag ``?'' stars of the Cycle~4 list is a bona fide Ap
star was found. 

After exclusion of the stars flagged ``/'' or ``?'' (except for
TIC~349945078), 40 entries were left in the Cycle~3 list, and 88 in
the Cycle~4 list. Some of these are affected by another ambiguity that
is apparent in \citetalias{2009A&A...498..961R}. Namely, the spectral type
reported for some stars both contains a hyphen (``--''), which
distinguishes Am stars, and lists an overabundant chemical element
(e.g., Sr),  which is the attribute of an Ap star. Of the 6 stars of
this type from the Cycle~4 list, 5 are definitely Am stars, and one,
TIC~2934856, is an Ap star \citep{2023ApJ...943..147S}. The 5 Cycle~3
stars with similar classification information are all Am or normal A
stars.

Moreover, for some stars, the notes to \citetalias{2009A&A...498..961R} indicate
that the validity of the classification appearing in the main
table is questionable -- most often, there exists an ambiguity between
Ap and Am classifications. We reviewed these cases individually. Only
TIC~248354858 appears to be an Am star; it was excluded from the
Cycle~3 list.

After the exclusion of the Am and normal A stars discussed in the two
paragraphs above, the Cycles~3 and 4 candidate lists were reduced to 34
and 83 stars, respectively. As the next step, information available in
the literature (as listed 
in SIMBAD) for the remaining stars was reviewed. Particular attention
was paid to spectral classification and abundance analyses, period and
\vsi\ determinations, and magnetic field determinations.

Among the remaining 34 stars of the Cycle~3 list, we found 4 that have
been definitely confirmed to have peculiarities that are different
from those of magnetic Ap stars: one Am star, one HgMn star, one
F~str~$\uplambda\,4077$ star, and one Ba star. Furthermore, there is one
normal late B star that was classified as such in a study specifically
devoted to the identification of Ap stars. Also, in his extensive,
systematic study of the stellar magnetic fields,
\citet{1958ApJS....3..141B} reported a lack of lines, of which there
are only traces, in the spectrum of HD~45827. A high-resolution
spectrum ($R\simeq70,000$)  from our collection, recorded with the AURELIE spectrograph
at the Observatoire de Haute-Provence (OHP), does not show any lines
either. Accordingly, HD~45827 appears very unlikely to be an Ap
star. Finally, for one star, the only references listed in SIMBAD are
the two editions of the Renson catalogue
\citep{1991A&AS...89..429R,2009A&A...498..961R} and no MK
spectral type is given. The source of the Ap classification reported
in the Renson catalogue is unknown. This star cannot be considered a
bona fide Ap star.

The arguments presented in the previous paragraph lead us to exclude
the 7 stars to which they refer from the list of the Cycle~3 ssrAp
candidates, on account that either they are definitely not Ap stars, or
their Ap nature is not convincingly established. Thus, the number of
entries in the list is reduced to 27.

A more severe issue was identified when carrying out a similar
literature review for the Cycle~4 list of candidate ssrAp stars. For
51 of these 83 stars, the only source that we could identify for the
Ap classification reported by \citet{2009A&A...498..961R} is the
catalogue of spectral classifications of 
stars in an area of 140 square degrees in the Galactic anticentre
direction that was compiled by the Abastumani Astrophysical
Observatory \citep{Abastumani}. The 208 Ap stars discovered as part of
the study performed to build this catalogue were listed by
\citet{1990AJ.....99..379K}.

The automatic search procedure that we applied found that 51 of these
208 stars, or 25\%, do not show photometric variations at low frequencies
typical of Ap star rotation. Taken at face value, this suggests that
the rate of occurrence of super slow rotation among Ap stars in the
direction of the Galactic anticentre is much higher than that rate among Ap
stars in general, which does not considerably exceed 10\%\
\citep{2017A&A...601A..14M,2020pase.conf...35M,2017MNRAS.468.2745N,2018MNRAS.475.5144S,2023A&A...676A..55L}. That
the rate of occurrence of super slow rotation is much higher
than the average value in a limited region of the sky such as a patch
around the Galactic anticentre appears implausible. It is much more
likely that a considerable fraction (more than half) of the stars
classified as Ap in the Abastumani catalogue are actually normal A (or
Am)
stars. This is all the more probable since all of those stars for
which an MK spectral type was available in SIMBAD (mostly from the
Michigan Spectral Survey) were classified as normal A stars.
Thus, while some of the stars from the list of
\citet{1990AJ.....99..379K} must almost certainly be genuine Ap stars,
none of them can be regarded as a bona fide Ap star unless there
exists some other convincing evidence of its peculiarity. Such
evidence was only found for 2 of the 51 Abastumani stars that we
identified as candidate ssrAp stars through the above-described
automatic analysis of the TESS data. For these two stars, short
photometric periods possibly attributable to rotation were reported in
the literature, so that they may not be super slowly rotating
anyway. This will be further discussed below. 
The other 49 stars for which the only identified Ap classification
source is the Abastumani catalogue were removed from the Cycle~4 list
of ssrAp star candidates, leaving only 34 entries in this list. 

Based on the information available in the literature, 9 more of these
34 stars were not considered as bona fide magnetic Ap stars. Five have
unambiguous, different classifications: 2 are F~str~$\uplambda\,4077$
stars, 1 is a HgMn star, 1 is an Am star and 1 is a normal F0
star. One more is an ambiguous
case: depending on the source of the classification, it may either be
an Ap Si star or an Am star. In 2 cases, we could not identify the source of the
peculiar classification; only one has an MK spectral type reported in
SIMBAD, as a normal A star. For the ninth one, the abundances derived
from the analysis of spectra recorded at a spectral reasolution
$R\simeq34,000$ are mostly normal. These
9 stars were also excluded from the Cycle~4 ssrAp candidate list. One
of them, the F~str~$\lambda\,4077$ star TIC~152803574, was also
observed in Cycle~3, and removed from the Cycle~3 list of the ssrAp
star candidates on account of misclassification.

In summary, a total of 64 stars that were assigned an Ap spectral type
in \citetalias{2009A&A...498..961R} without any indication of classification
uncertainty were removed from our selection of ssrAp star candidates
because, after critical evaluation, we do not regard them as bona fide
Ap stars. These stars are listed in Tables~\ref{tab:abastumani} and
\ref{tab:misclass}. 

\begin{figure}
  \centering  
  \includegraphics[width=1.0\linewidth,angle=0]{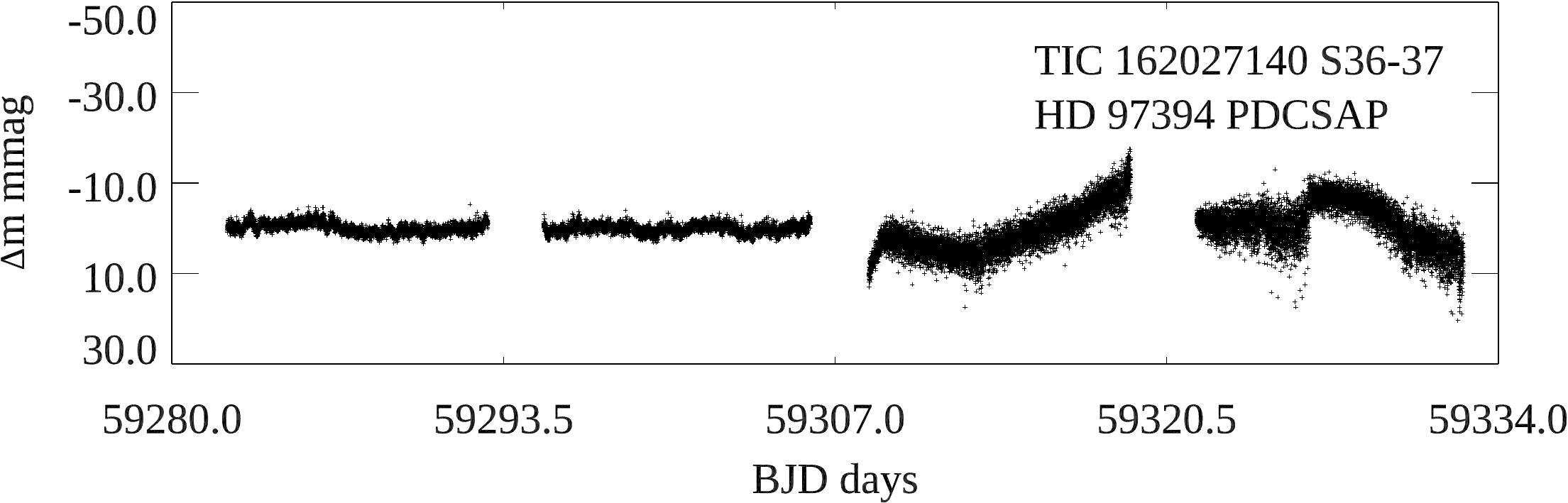}
  \includegraphics[width=1.0\linewidth,angle=0]{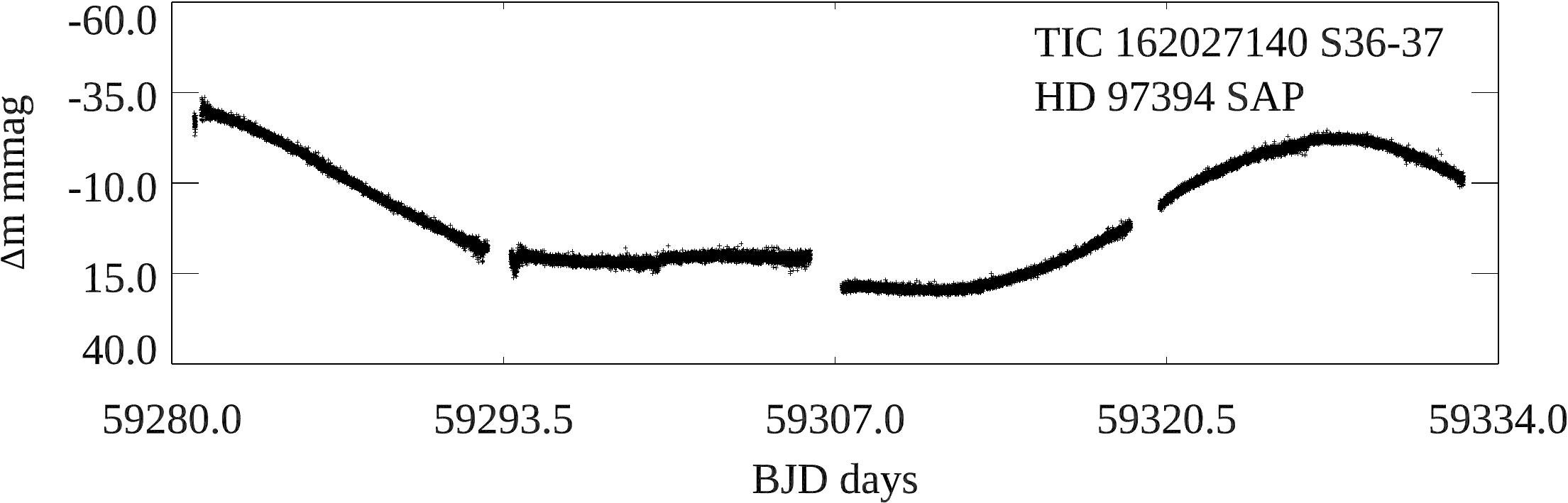}
 \includegraphics[width=1.0\linewidth,angle=0]{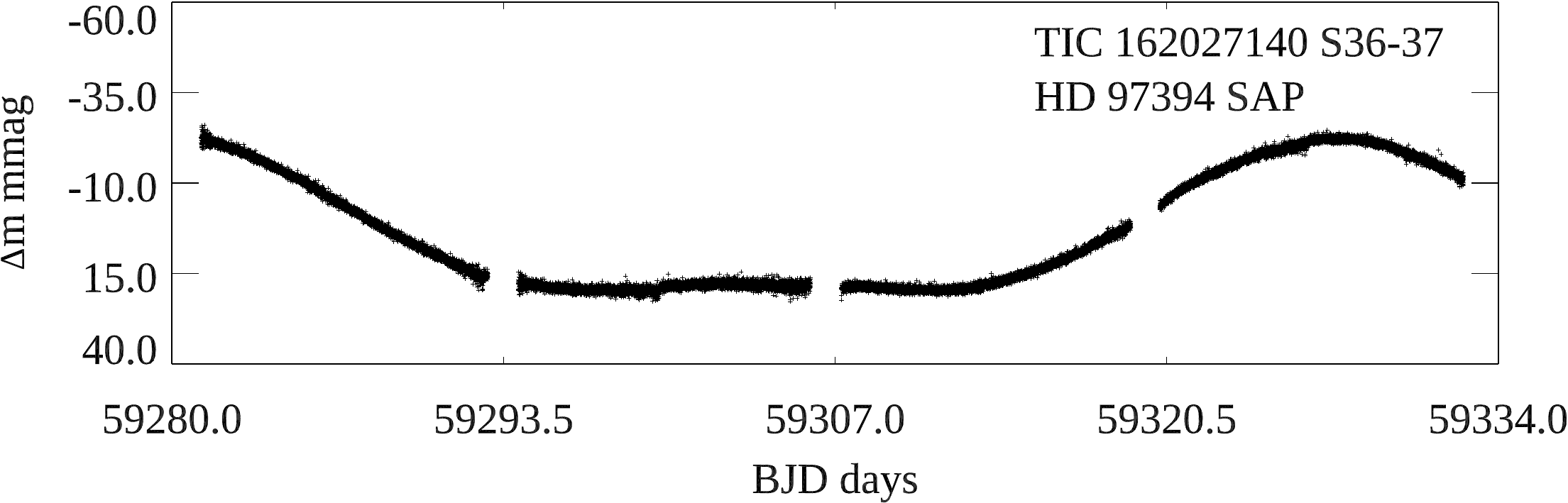}
  \caption{TIC\,162027140 (HD\,97394). Top panel: The light curve of the S36-37 PDCSAP data where the removal of long term trends by the CBV has generated low frequency problems. Middle panel: the SAP data where the $47.38$-d rotational variation can be seen. There is also an instrumental zero-point shift between  S36 and S37. Bottom panel: The zero-point shift has been adjusted. 
    }
  \label{fig:162027140_lc}
\end{figure}

\subsection{Visual inspection of the amplitude spectra}
\label{sec:visual}
For the remaining 27 stars of the Cycle~3 list and 25 stars of the
Cycle~4 list, we proceeded to the visual inspection of the amplitude
spectra to confirm the lack of low-frequency variability that led to
their selection by the automatic search procedure. Five of the
Cycle~3 stars and 13 of the Cycle~4 stars proved to show rotational
variability with frequencies $f>0.049$\,d$^{-1}$. The latter include
the two above-mentioned Ap stars from the Abastumani catalogue
for which we regarded the peculiar nature as confirmed since short
periods had been reported in the literature. However, only for one of
them does the value of the published period coincide with the one
derived from analysis of the TESS data.

The exclusion of the stars for which super slow rotation was ruled
out through visual inspection of the amplitude spectra left 22 and 12
ssrAp candidates in the Cycle~3 and Cycle~4 lists,
respectively. However, the star TIC~372617495 (=\,HD~48953) appears in
both lists. Accordingly, we are left with 33 stars that, after critical
evaluation, we regard as bona fide Ap stars, for which the
above-described analysis of the TESS data does not show any low
frequency variability. In \citetalias{2020A&A...639A..31M} and
\citetalias{2022A&A...660A..70M}, we concluded that Ap stars selected
in that way must have rotation periods considerably longer than one
TESS 27-d sector, except for the infrequent occurrence of very small
inclination angle $i$ or obliquity angle $\beta$. Thus, it appeared
that almost all of these stars must be ssrAp stars, according to the
definition of \citet{2020pase.conf...35M}, that is, having
$\Prot>50$\,d. In the next section, we shall show that, in fact,
variations with moderately long periods
($20\,\mathrm{d}\lesssim\Prot\lesssim50$\,d) may have escaped detection
up to this point. 

\begin{figure*}
  \centering  
\includegraphics[width=0.48\linewidth,angle=0]{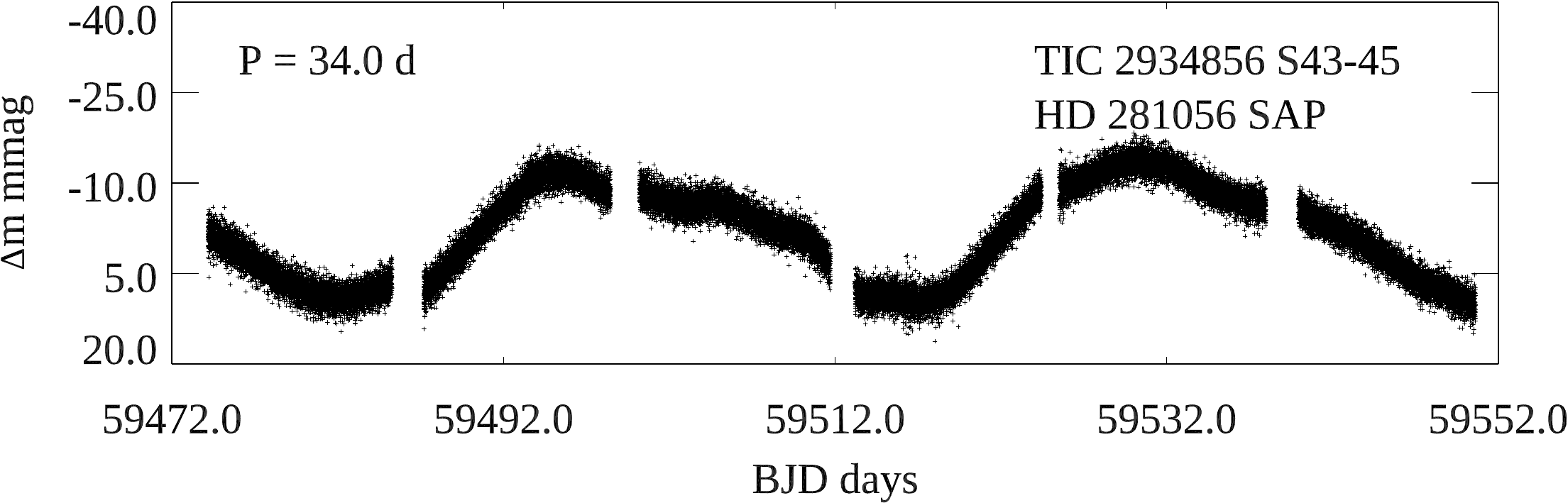}  
\includegraphics[width=0.48\linewidth,angle=0]{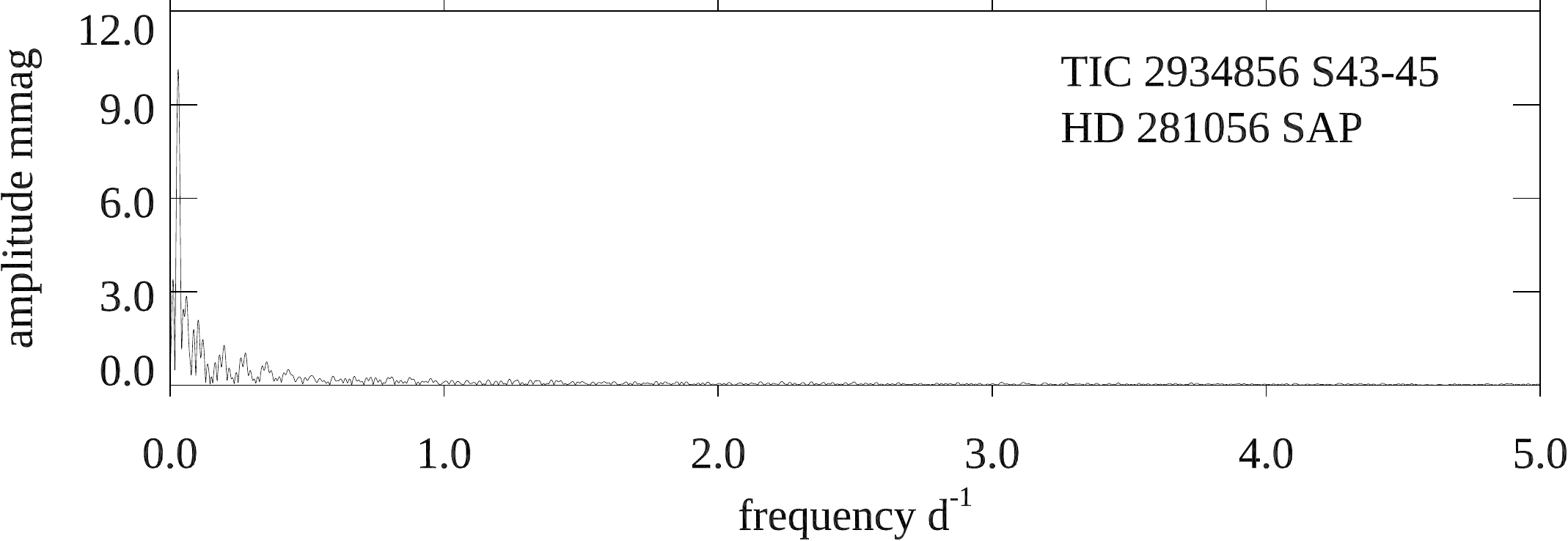}
\includegraphics[width=0.48\linewidth,angle=0]{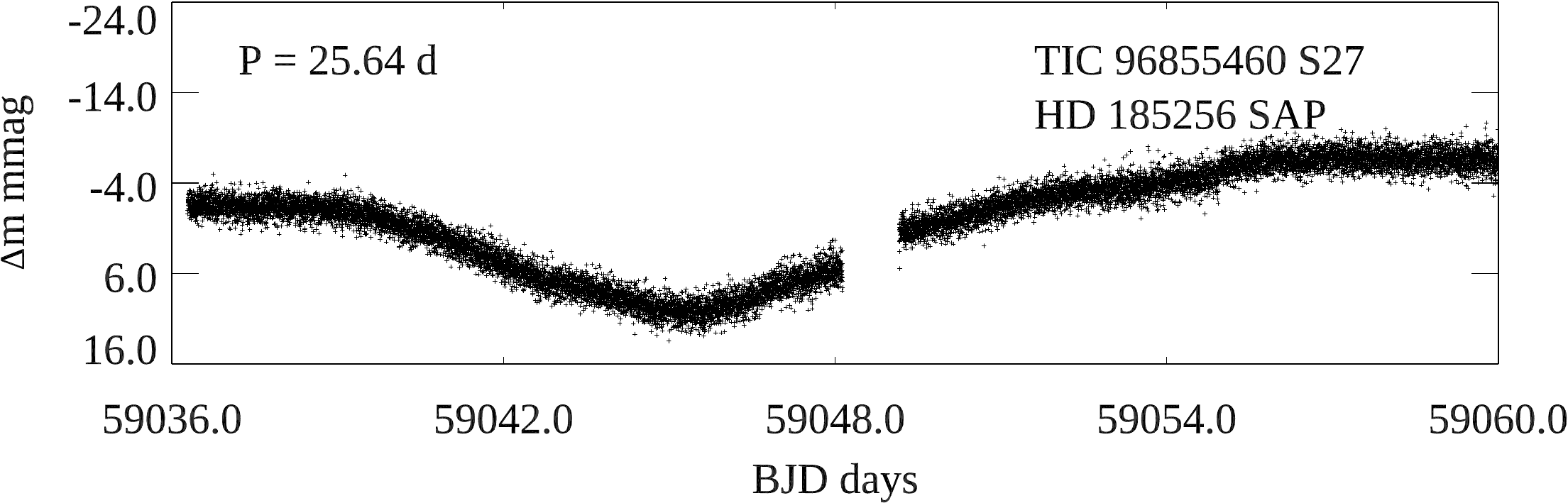}
\includegraphics[width=0.48\linewidth,angle=0]{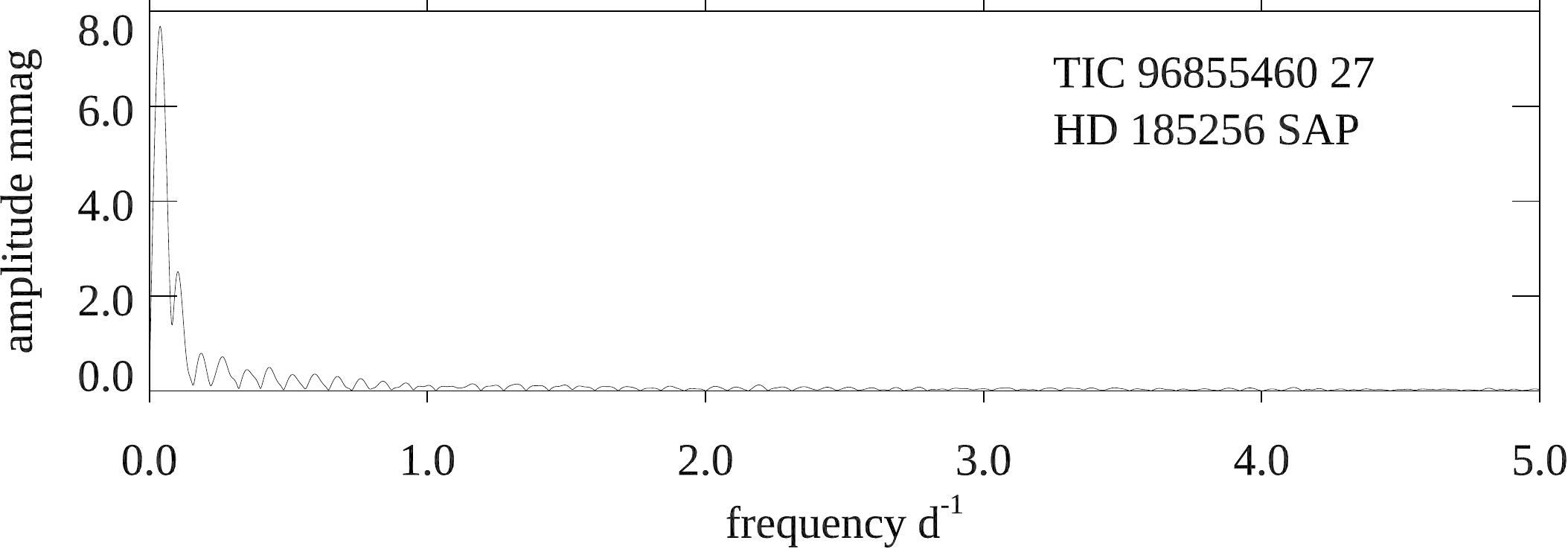}
\includegraphics[width=0.48\linewidth,angle=0]{162027140_S36-37_SAP_adjusted_lc.pdf} 
\includegraphics[width=0.48\linewidth,angle=0]{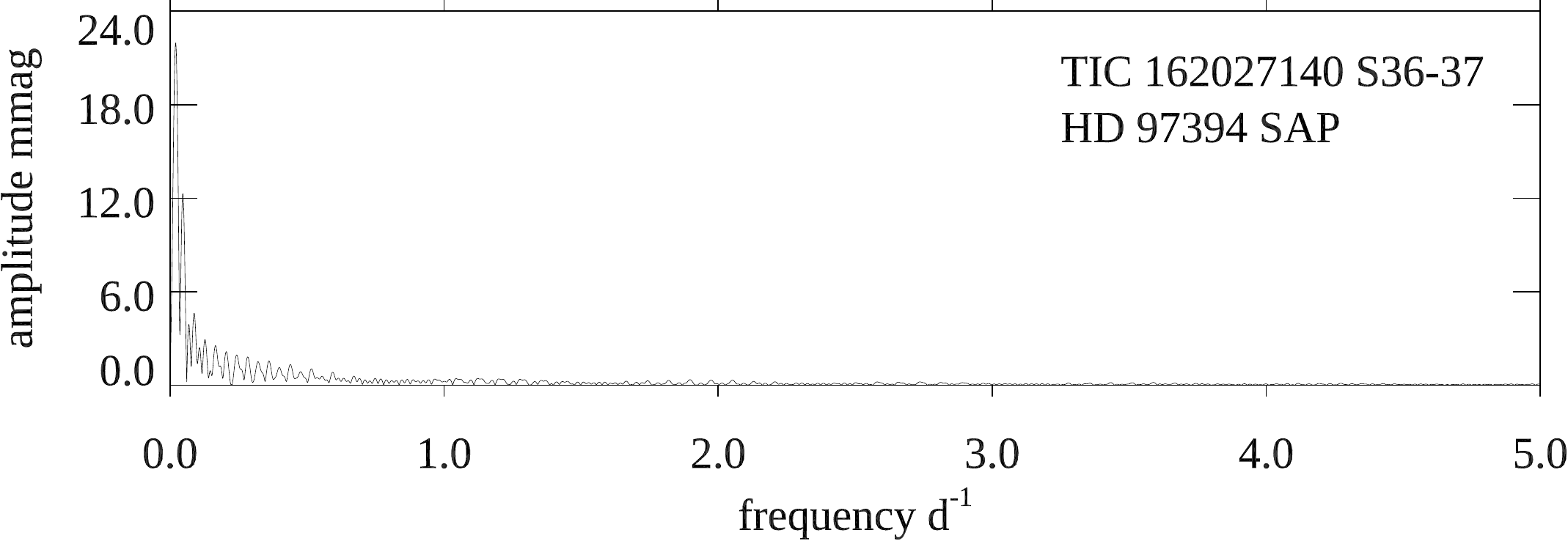}
\includegraphics[width=0.48\linewidth,angle=0]{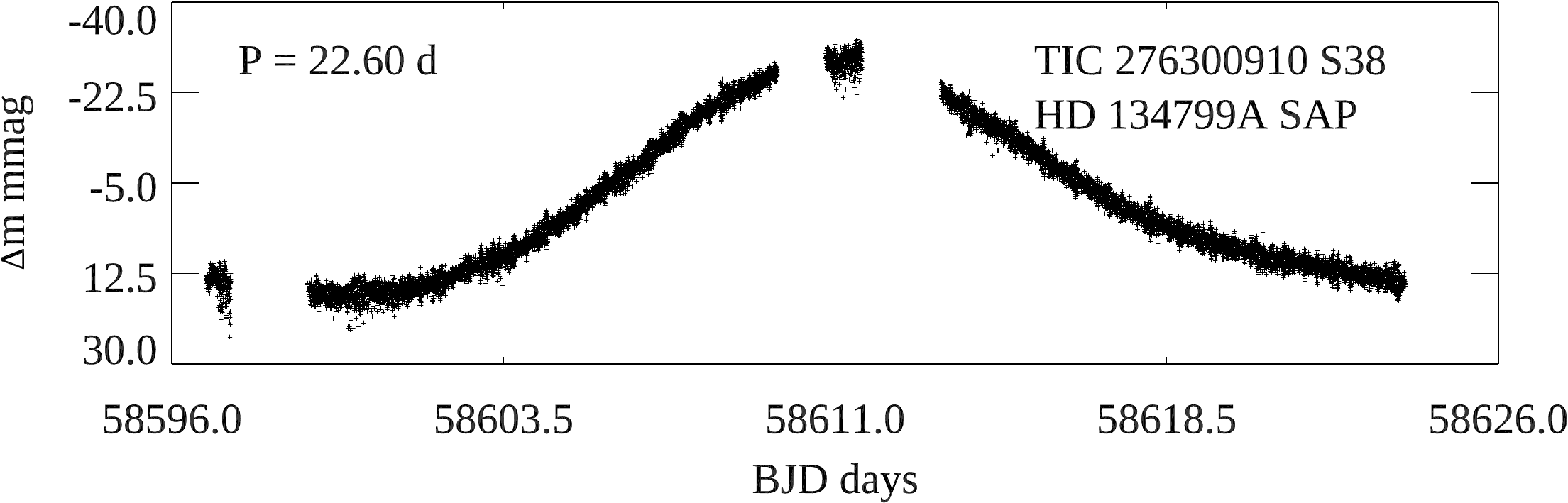}
\includegraphics[width=0.48\linewidth,angle=0]{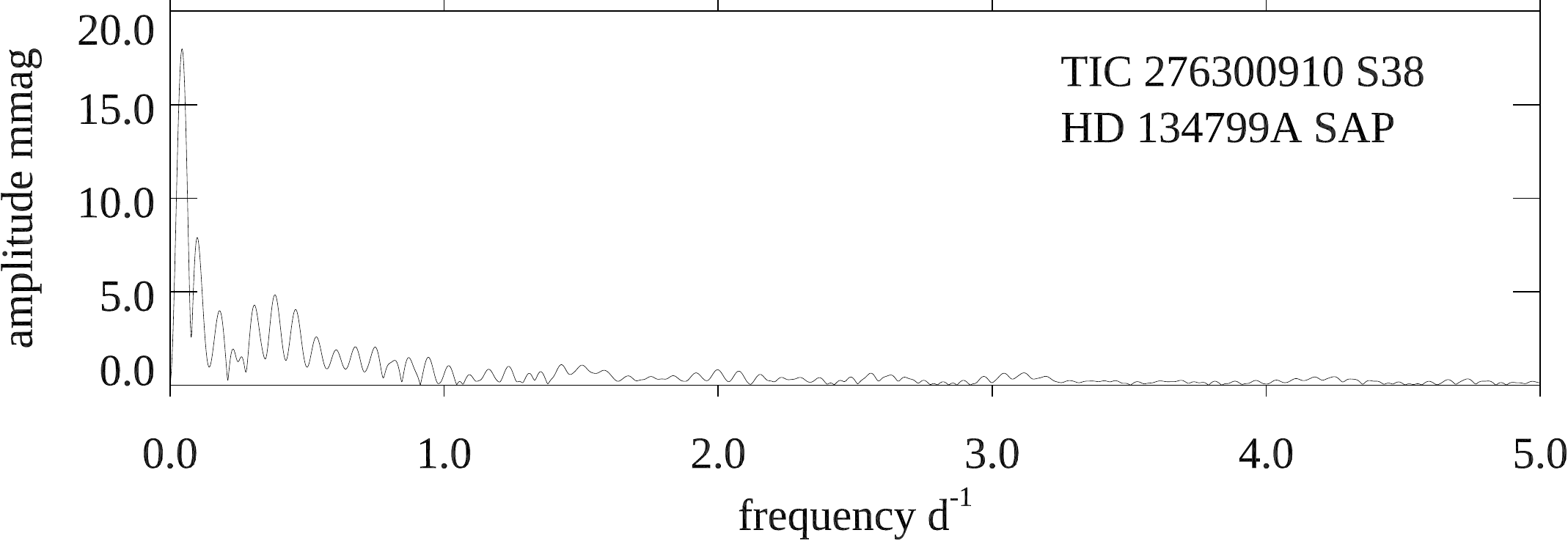}
\includegraphics[width=0.48\linewidth,angle=0]{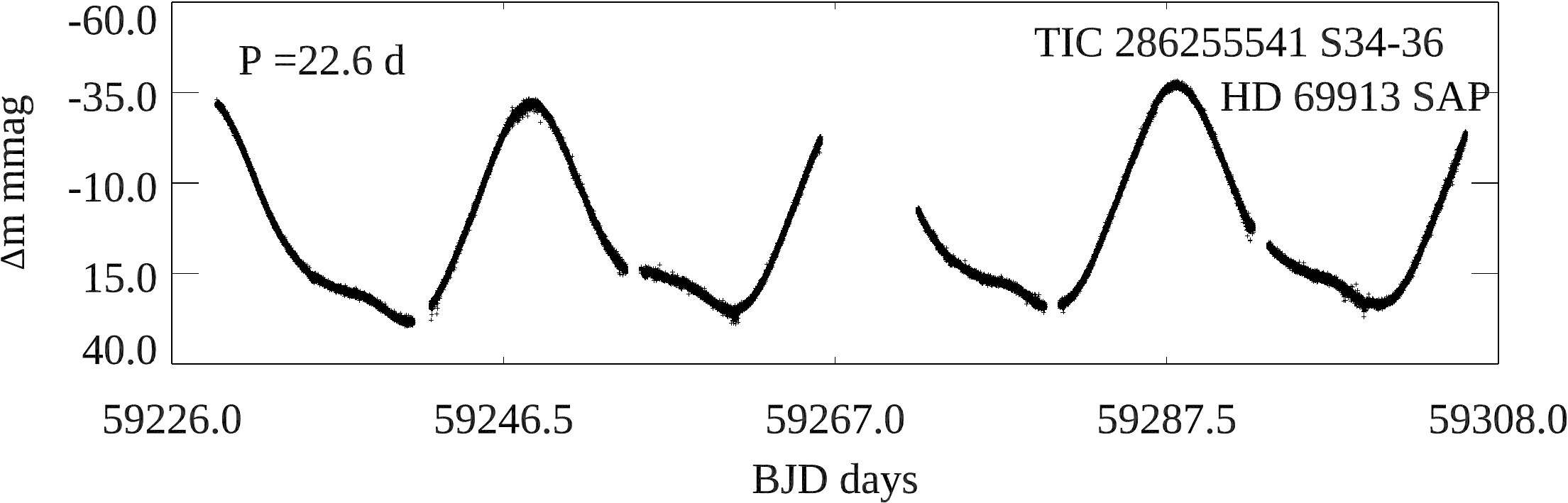}
\includegraphics[width=0.48\linewidth,angle=0]{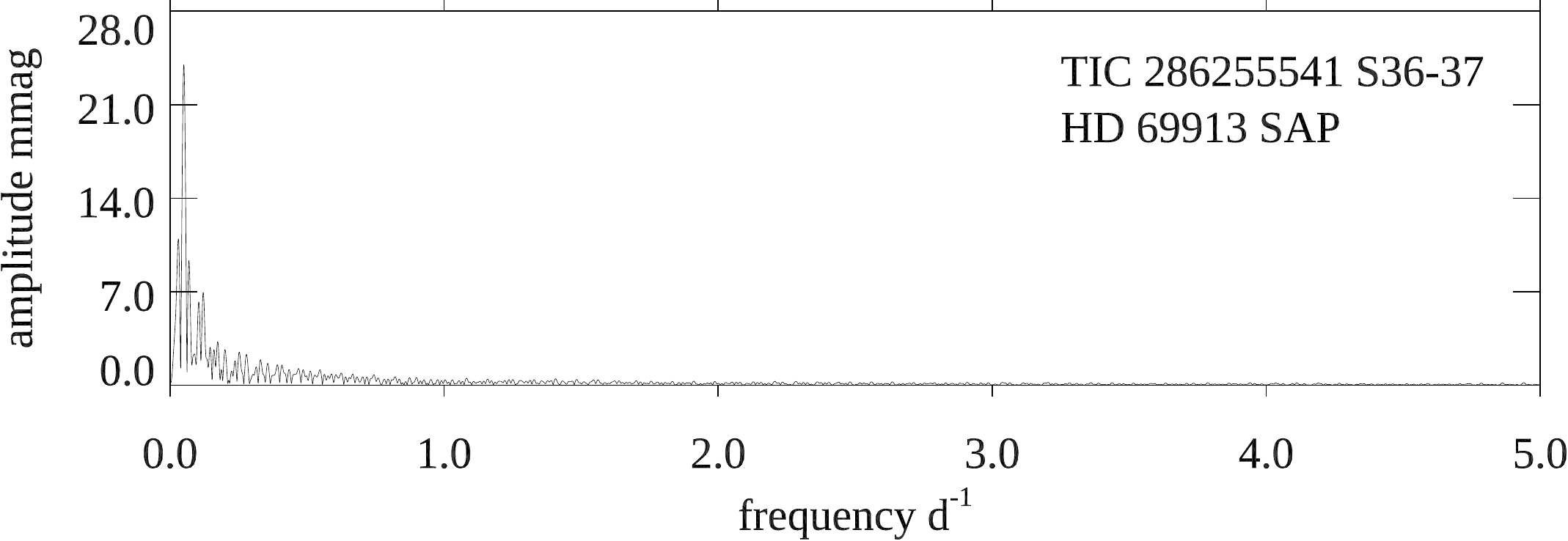}
\caption{ The SAP light curves and low-frequency amplitude spectra for Ap stars with measurable periods $P_{\rm rot} \gtrsim 20$\,d. Some periods in the range $20 - 30$\,d for those stars with only one sector are uncertain because of the non-sinusoidal shape of $\upalpha^2$\,CVn light curves.  }
  \label{fig:rotation_lc}
\end{figure*}

\begin{figure*}
  \centering  
\includegraphics[width=0.48\linewidth,angle=0]{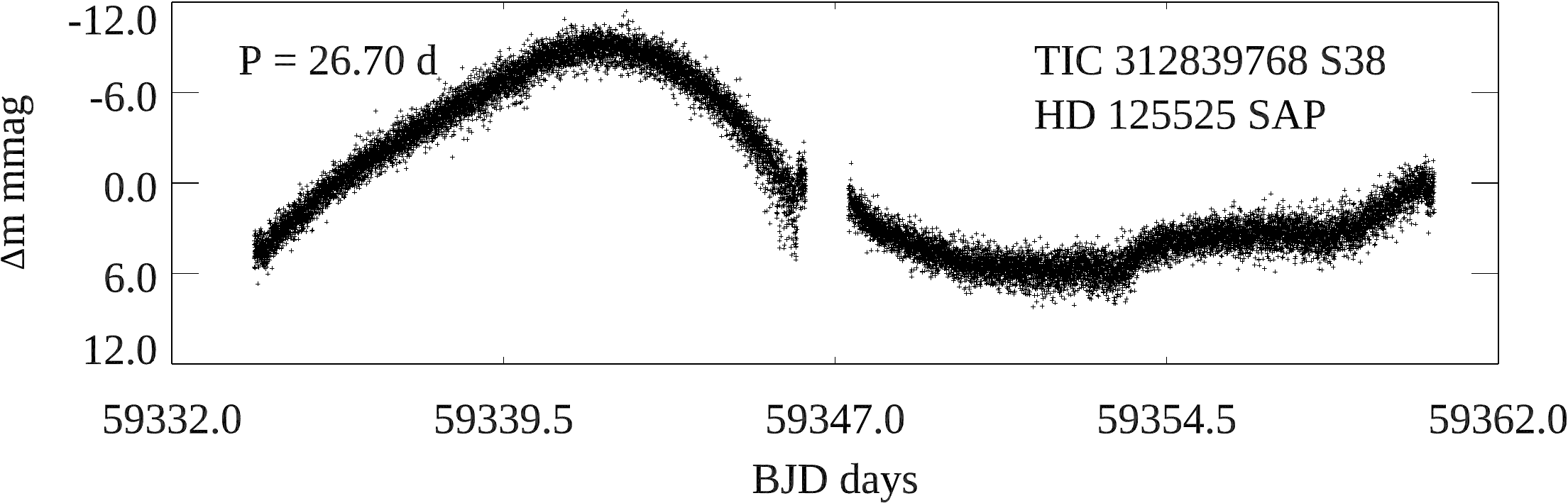}
\includegraphics[width=0.48\linewidth,angle=0]{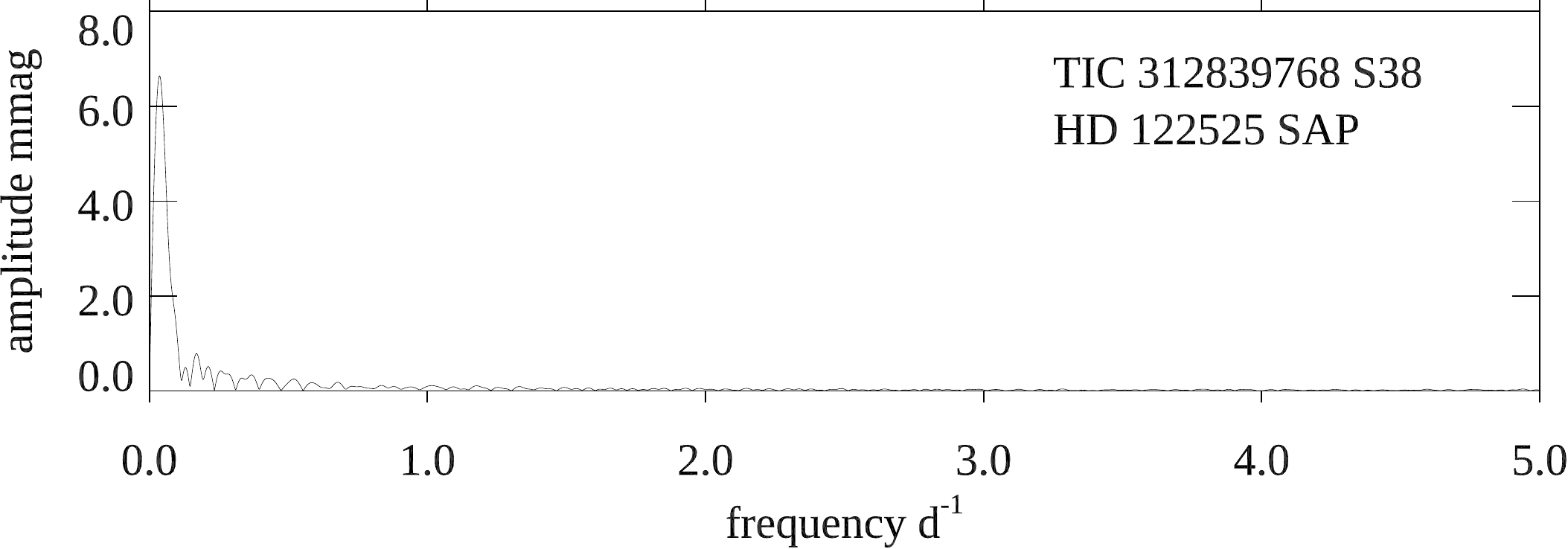}
\includegraphics[width=0.48\linewidth,angle=0]{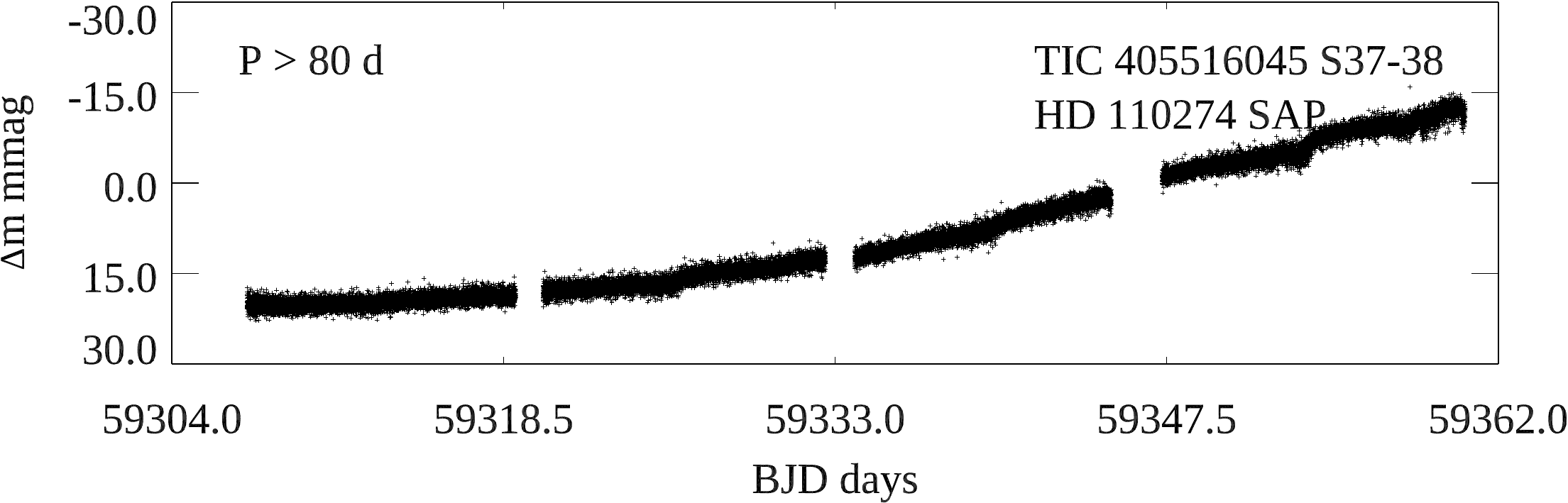}
\includegraphics[width=0.48\linewidth,angle=0]{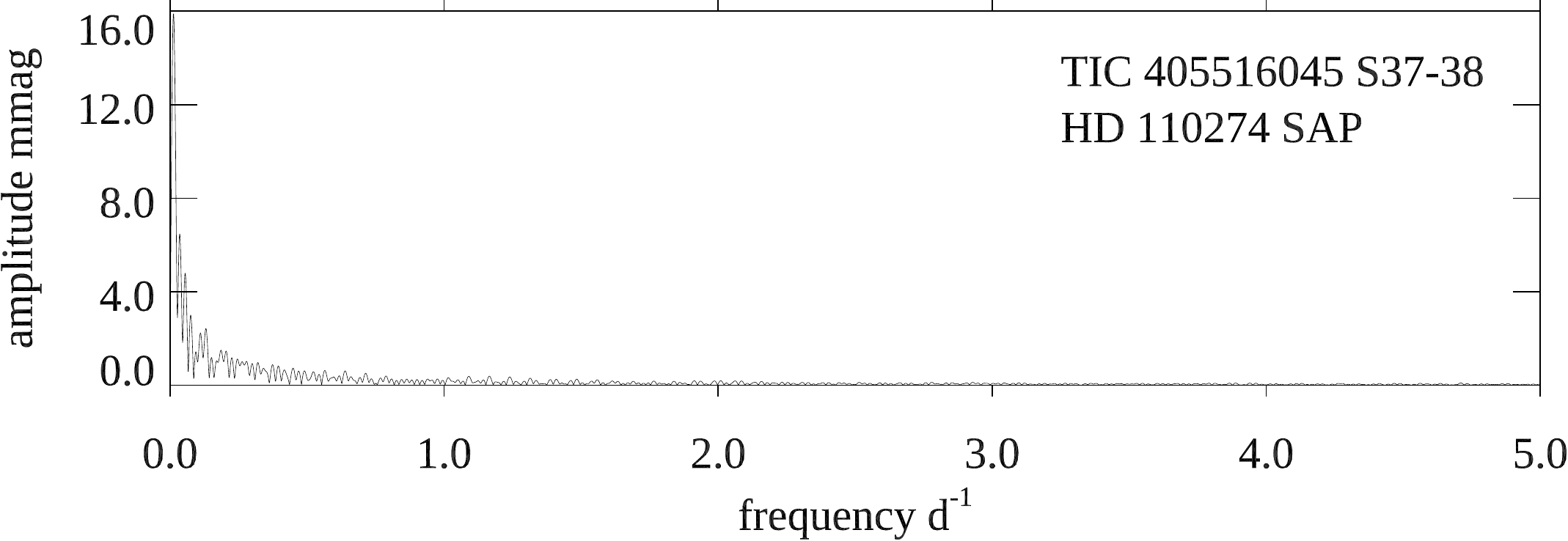}
\includegraphics[width=0.48\linewidth,angle=0]{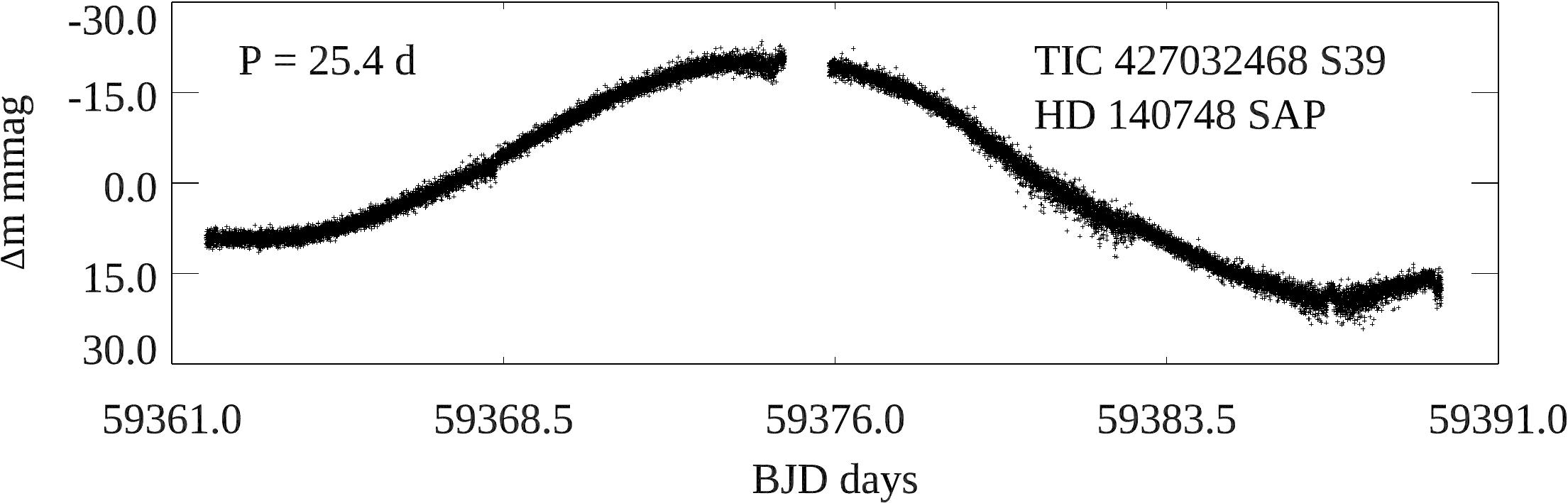}
\includegraphics[width=0.48\linewidth,angle=0]{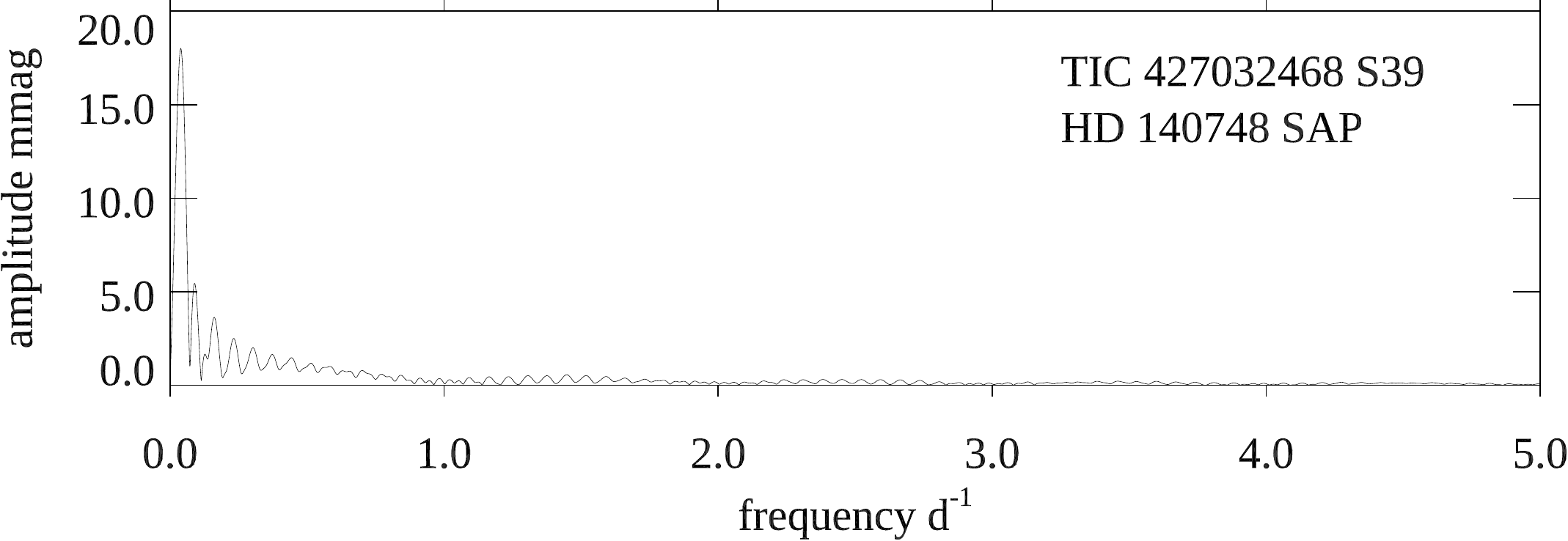}
\includegraphics[width=0.48\linewidth,angle=0]{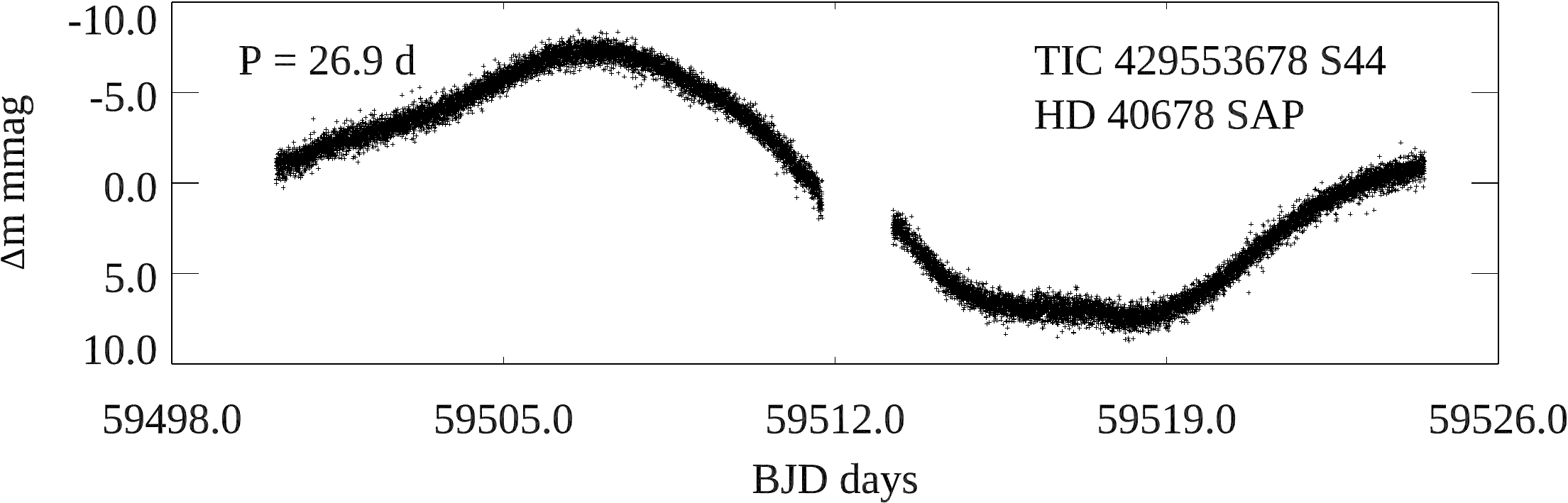}
\includegraphics[width=0.48\linewidth,angle=0]{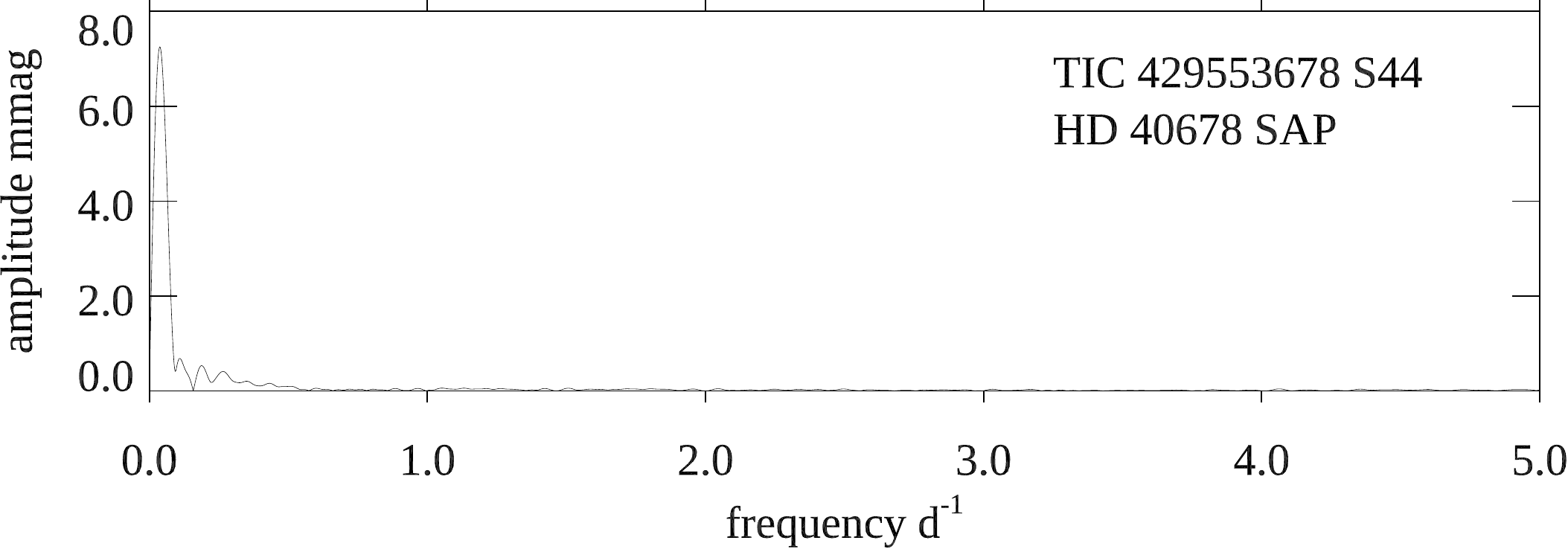}
\caption{The SAP light curves and low-frequency amplitude spectra for Ap stars with measurable periods $P_{\rm rot} \gtrsim 20$\,d -- continued.  }
  \label{fig:rotation_lc2}
\end{figure*}

\subsection{Long period stars with measurable rotation periods $20 \lesssim P_{\mathrm{rot}} \lesssim 50$\,d}
\label{sec:sap}

TESS data are delivered in two forms, the simple aperture photometry
(SAP) and pre-search data conditioning SAP (PDCSAP). The SAP suffers
from instrumental drifts and zero point shifts that are modelled with
ensembles of stars on the same CCD to produce the co-trending basis
vectors (CBV) that are then used to remove these instrumental effects
sector-by-sector. The primary purpose is to optimise the data to
search for exoplanet transit signals. One consequence of the
application of the CBV is that long term astrophysical trends can also
be removed in the PDCSAP data. \citet{2023arXiv230705664C} discuss
this in detail in a study of rotation signals in TESS data for cool
stars. 

For stars with rotation periods of the order of, or greater than, the TESS sector length (27\,d), some have amplitudes greater than the instrumental variations at low frequencies so that we can determine the rotational period despite the instrumental noise. Because of the many kinds of instrumental variations that occur, caused by, e.g., telescope operations, changing background, changing contamination, and data download gaps, we are cautious in determining periods from the SAP data.  Our purpose in this work is to find ssrAp stars with rotation periods longer than 50\,d for spectroscopic follow-up to study the magnetic fields. In this section we only show cases of Ap stars that have rotation periods that can be determined from the SAP data. There are 9 such stars, 8 of which have periods in the range $20 \lesssim P_{\rm rot} < 50$\,d, and one of which is a bona fide ssrAp star with $P_{\rm rot} \gtrsim 80$\,d. 

 Our technique for searching for ssrAp stars, as described in the
previous sections,  is to find Ap stars that show no low-frequency
variations in the PDCSAP data. Stars with $P_{\rm rot} \lesssim 20$\,d
do not pass this test, as the signal from the rotation is generally
seen. However, there is a grey area for $20 \lesssim P_{\rm rot}
\lesssim 50$\,d where the CBV remove much or all of the rotation
signal in the PDCSAP data, but the signal can still be seen in the SAP
data. These stars are not ssrAp stars,  but they are still of
interest, hence we note them here.

Indeed, their rotation periods are longer than those of the vast
majority (about 90\% -- see for instance Fig.~1 of
\citealt{2017MNRAS.468.2745N} and Fig.~4 of
\citealt{2023A&A...676A..55L}) of the Ap stars. They have lost a
considerable amount of angular momentum and their consideration as an
intermediate case between the most numerous stars with rotation
periods of a few days and the extreme ssrAp stars can be expected to
provide additional insight into the braking mechanisms that have been
(and possibly are still) at play. Furthermore, their projected
equatorial velocity is at most of the order of 7\,\kms, so that their
most magnetically sensitive spectral lines, such as Fe~{\sc
  ii}~$\uplambda$\,6149.2\,\AA, can be partly resolved into their
magnetically split components for magnetic fields with a mean modulus
of the order of 3\,kG. Such fields are weak enough to be of relevance
to constrain the possible connection between magnetic field strength
and the rate of occurrence of (super) slow rotation in Ap stars (see
\citetalias{2022A&A...660A..70M} and Sect.~\ref{sec:conc}). 

\subsubsection{A comparison of the SAP and PDCSAP data for an example star: TIC\,162027140 (HD\,97394)} 
\label{sec:compar}
We illustrate here the impact of the CBV on long-term rotational
variations in the stars that show variability in the SAP data but not
in the PDCSAP data with the example of TIC\,162027140
(HD\,97394). This star shows remarkable overabundances of some rare
earth ions up to a million times solar and has resolved magnetically
split lines, seen on UVES spectra \citep{2011MNRAS.415.2233E}, that
clearly indicate a long rotation period. This star is an ssrAp star candidate.  

 Yet the PDCSAP data are problematic. This can be seen in Fig.\,\ref{fig:162027140_lc} where a $47\fd38$ rotational variation is apparent in the SAP data, but is removed by the pipeline application of the CBV in the PDCSAP data. (In the SAP data we adjusted the zero point of S36 to match that of S37.) The star has a double wave with the secondary maximum being almost flat. If we assume that there are two major spots near the magnetic poles, with the magnetic axis inclined to the rotation axis (as is typical of $\upalpha^2$\,CVn Ap stars), then one pole crosses near to the line of sight, and the other skims the observed limb. We have not modelled this, but speculatively it indicates a geometry with the rotational inclination, and magnetic obliquity $i, \beta \sim 45^\circ$. Future long-term magnetic measurements will clarify this.

\subsubsection{Determination of $P_{\rm rot}$ from the SAP data for slowly rotating Ap stars }
\label{sec:Psap}
We show here the SAP light curves and low-frequency amplitude spectra
for stars where a measurement, or estimate, of $P_{\rm rot} \gtrsim 20$\,d is possible. For the stars with periods in  the range $20 -30$\,d and one 27-d sector of data, only one cycle, or less, has been observed making the period determined somewhat uncertain. Also, $\upalpha^2$\,CVn stars often show a double wave variation as a consequence of spots at both magnetic poles, hence the periods derived from one cycle of data may be only half the true period. For stars that have more than one sector of data, in some cases a zero point correction has been applied to particular sectors. Obvious outliers have been trimmed, along with some short sections of data on the edges of gaps where there are data reduction problems. 

We note that \citet{2023A&A...676A..55L} have recently looked at rotation periods for stars classified as magnetic Ap stars with LAMOST spectra. These authors used the Full-Frame Images (FFI) up to Sector 35. For S1-26 the integration time was 30\,min, then for S27-35 it was 10\,min. Hence \citeauthor{2023A&A...676A..55L} do not use the 120-s cadence we used, so are not sensitive to any possible roAp pulsations. There is overlap between their study and ours of only one star, TIC\,2934856 (HD\,281056), where we agree that $P_{\rm rot} = 34$\,d.  

Figures~\ref{fig:rotation_lc} and \ref{fig:rotation_lc2} show the SAP light curves for 9 stars in our sample where the rotation period was determined, or estimated.

\subsubsection{Notes for Figures~\ref{fig:rotation_lc} and
  \ref{fig:rotation_lc2}} 
\label{sec:fignotes}
TIC\,2934856 (HD\,281056): \citet{2023A&A...676A..55L} found $P_{\rm
  rot} = 34\fd35$; we find $P_{\rm rot} = 34\fd0$ in this work. These
are in agreement within the uncertainties.   

TIC\,96855460 (HD\,185256) is a known roAp star; see
Sect.\,\ref{sec:roap}. With the value that we derive for the
rotation period, $\Prot=25\fd64$, this is not an ssrAp star. 

TIC\,162027140 (HD\,97394) has $P_{\rm rot} = 47\fd38$, close to our
lower limit of 50\,d for ssrAp stars.

TIC\,276300910 (HD\,134799A) is not resolved from its visual companion
HD\,134799B in the TESS data. The SAP data show a rotation
signal with $P_{\rm rot} = 22\fd6$ in both the S11 and S38 data, separately, and $P_{\rm rot} = 23\fd0$ for the combined sectors. Because there is only one cycle per sector, and there is a large gap between S11 and S38, this 
value is not inconsistent with that reported by
\citet{2017MNRAS.468.2745N}, $\Prot=25\fd373$.
The period is clearly moderately long, but TIC\,276300910 is not an
ssrAp star according to our definition. 

TIC\,286255541 (HD\,69913) has data from S8-9, S34-36, and S61-63. All
phase well with the determined period $P_{\rm rot} = 19\fd98$. This
is in good agreement with the value $\Prot=19\fd975$ derived by
\citet{2015A&A...581A.138B} from the analysis of ASAS-3 photometric
data. 

TIC\,312839768 (HD\,125525) has a derived period $P_{\rm rot} =
26\fd70$ equal to the the length of the S38 data, also $26\fd70$,
and the amplitude is relatively low. The similarity of this value with
the one that \citet{2016AJ....152..104H} determined using data from
the ASAS-3 archive suggests that the variations seen in the SAP light
curves is of stellar origin rather than due to an instrumental
effect.

TIC\,405516045 (HD\,110274) shows a long-term drift. This star, whose
spectrum shows resolved magnetically split lines
\citep{2008MNRAS.389..441F}, was already a known 
ssrAp star. Its rotation period, $\Prot=265\fd3$, was determined by
\citet{2008MNRAS.389..441F} from the analysis of ASAS-3 photometric
data. While this period is much longer than the time span covered by
the available TESS observations, the variation trend is clearly seen
in the SAP data, which suggests $\Prot\gtrsim80$\,d. 

TIC\,427032468 (HD\,140748) potentially has a double-wave light curve,
in which case $P_{\rm rot} = 50\fd8$ and this is a ssrAp
star. However, this is uncertain, as the period  
value,  $P_{\rm rot} = 25\fd4$, differs considerably from, but is not inconsistent with that published by
\citet{2017MNRAS.468.2745N}, $\Prot=36\fd520$. Because the TESS data do not cover full cycles for these periods, the data can be reasonably phased with all three periods, hence do not distinguish amongst them. 

TIC\,429553678 (HD\,40678): the value of the period that we estimate,
$\Prot=26\fd9$, and the value reported by
\citet{2016AJ....152..104H}, $\Prot=22\fd029$,  based on the analysis of
photometric  data from the ASAS-3 archive, are not incompatible. 

For all of the stars in this subsection, it is difficult to determine precise rotation periods from one cycle, or less, of light curves that are generally not sinusoidal. The values we provide are guides to future studies, particularly of the magnetic fields. 

\begin{table*}
  \scriptsize
  \caption{List of the ssrAp star candidates found by our technique in
    the TESS Cycles~3 and 4 data.}
  \begin{tabular*}{\textwidth}[]{@{}@{\extracolsep{\fill}}rrlrrrcclrlclll}
\hline\hline\\[-4pt]
  &  &Spectral&   &  &    &   &   &   &  &
  &\multicolumn{1}{c}{Lines\tablefootmark{b}}   &  & &TESS  \\
  \multicolumn{1}{c}{TIC}  &  \multicolumn{1}{c}{HD} &type&  \multicolumn{1}{c}{$V$}  & \multicolumn{1}{c}{$T_{\rm  eff}\tablefootmark{a}$}   & \multicolumn{1}{c}{$\log g\tablefootmark{a}$}   &  \multicolumn{1}{c}{roAp}  &  \multicolumn{1}{c}{$\langle B_z\rangle_{\mathrm{rms}}$/$B_0$/$Q_0$} &  Refs  &  \multicolumn{1}{c}{$P_{\rm rot}$}  &  Refs  & \multicolumn{1}{c}{$\vsi$}  & Refs\tablefootmark{c} & Notes &sectors  \\
  &  &  &  \multicolumn{1}{c}{(mag)}  & \multicolumn{1}{c}{(K)}   & \multicolumn{1}{c}{(cm\,s$^{-2}$)}   &   &  \multicolumn{1}{c}{(kG)}  &   &  \multicolumn{1}{c}{(d)}  &    & \multicolumn{1}{c}{(\kms)} &  &  \\[4pt]
  \hline\\[-4pt]
32259138&138777&A3p SrEu&9.73&7350&3.90&&2.1/-/-&1&&&25&1&&51\\
36576010&216018&A7p SrEuCr&7.62&7750&4.05&roAp&1.4/5.6/6.8&2, 3&34.044\tablefootmark{d}&2&r&4&&42\\
49159482&179246&B9p EuCr&9.82&8320&&&&&&&&&&54\\
88202438&192686&A0p Si&8.88&11670&3.75&&&&&&&&&54\\
126975139&126297&A5p CrEuSr&9.49&7640&3.70&&-/-/1.0&5&&&4&5&&38\\
169382402&32996&A0p Si&6.04&10420&3.95&&&&&&15&6&&32\\
189996908&75445&A3p SrEu&7.12&7610&4.00&roAp&0.1/3.0/4.3&3&$> 5000$?&2&r&4&&35\\
209708422&122379&A0p Si&10.45&13120&&&&&&&&&&38\\
253260234&113149&A0p CrEu&10.14&10310&4.15&&&&&&&&&37-38\\
262956098&3988&A0pCrEuSr&8.35&8470&3.60&&-/2.65/-&3&&&r&7&SB2 (7)&39\\
276354649&134874&B9p Si&7.67&12070&&&&&&&30&8&&38\\
282468249&197077&A2p SiSr&9.38&8177&4.15&&&&&&&&&55\\
295698744&119933&A2p SrCrEu&9.27&6640&3.90&&&&&&&&&50\\
299000970&176232&A6p Sr&5.89&7520&3.85&roAp&0.4/-/1.3&9, 10--12&$>12$\,yr&9&2&11&&53-54\\
334327860&111675&A0p EuSr&9.74&8620&3.80&&&&&&&&&38\\
354619745&201601&A9p SrEu&4.68&7490&4.05&roAp&1.0/3.9/5.2&3&$>97$\,yr&13&r&14&&55\\
369969602&128472&A2p CrSrEu&9.87&7260&&&&&&&&&&38\\
372617495&48953&Fp SrEu&6.80&6910&3.70&&&&2.8939\tablefootmark{e}&15&&&&44-45\\
380607580&119794&A2p CrEuSr&9.00&9350&3.95&&&&&&&&&38\\
405516045&110274&A0p EuCr&9.47&7770&3.85&&-/4.0/-&3&265.3&16&r&16&&37-38\\
405557056&155102&A2p Si&6.36&9100&3.80&&&&&&38&17&&51-53\\
410451752&66318&A0p EuCrSr&9.56&10130&4.10&&6.6/14.5/-&3,18,19&&&r&18&&34-37\\
425796196&138146&A0p EuCr&10.07&7980&3.95&&&&&&&&&39\\
438694338&117227&A0p CrSr&9.12&7800&4.35&&&&&&b&F&&38\\
442695956&114568&A0p Si&10.18&12390&&&&&&&&&&38\\[4pt]			
    \hline\\[-4pt]
\end{tabular*}
\tablefoottext{a}{Values retrieved from the TIC where available}
\tablefoottext{b}{r = resolved; b = broad.}
\tablefoottext{c}{F: the line width information is based on visual
  inspection of a FEROS spectrum of our collection.}
\tablefoottext{d}{This value of the period seems spurious. The star
  more likely has a rotation period vastly exceeding 6\,yr, as first
  suggested by \citet{2017A&A...601A..14M}. See text for details.}
\tablefoottext{e}{No low-frequency is seen in the TESS data. See text
  for details.}
\tablebib{(1)~\citet{2017AstBu..72..391R};
  (2)~\citet{2022MNRAS.514.3485G}; (3)~\citet{2017A&A...601A..14M};
  (4)~\citet{1997A&AS..123..353M}; (5)~\citet{2017AstL...43..252R};
  (6)~\citet{1995ApJS...99..135A};
  (7)~\citet{2012MNRAS.420.2727E};  (8)~\citet{1996A&AS..118..231L};
  (9)~\citet{2019MNRAS.483.3127S};
  (10)~\citet{2000A&A...357..981R}; (11)~\citet{2002MNRAS.337L...1K};
  (12)~\citet{2003A&A...409.1055L}; (13)~\citet{2016MNRAS.455.2567B};
  (14)~\citet{1979AN....300..213S};
  (15)~\citet{2012MNRAS.420..757W}; (16)~\citet{2008MNRAS.389..441F};
  (17)~\citet{2012A&A...537A.120Z}; (18)~\cite{2003A&A...403..645B};
  (19)~\cite{2006A&A...450..777B}.}
  \label{tab:ssrAp}
\end{table*}

\begin{table*}
  \scriptsize
  \caption{List of the stars with moderately long rotation periods
    ($20\,\mathrm{d}\lesssim\Prot\lesssim50$\,d) found in 
    the TESS Cycles~3 and 4 data.}
  \begin{tabular*}{\textwidth}[]{@{}@{\extracolsep{\fill}}rrlrrrcclrrlclll}
\hline\hline\\[-4pt]
  &  &Spectral&   &  &    &   &   &   &  & &
  &\multicolumn{1}{c}{Lines\tablefootmark{b}}   &  & &TESS  \\
  \multicolumn{1}{c}{TIC}  &  \multicolumn{1}{c}{HD} &type&
                                                            \multicolumn{1}{c}{$V$}
                  & \multicolumn{1}{c}{$T_{\rm  eff}\tablefootmark{a}$}   & \multicolumn{1}{c}{$\log g\tablefootmark{a}$}   &  \multicolumn{1}{c}{roAp}  &  \multicolumn{1}{c}{$\langle B_z\rangle_{\mathrm{rms}}$/$B_0$/$Q_0$} &  Refs  &  \multicolumn{1}{c}{$P_{\rm rot}$} &  \multicolumn{1}{c}{$P_{\rm litt}$}  &  Refs  & \multicolumn{1}{c}{$\vsi$}  & Refs\tablefootmark{c} & Notes &sectors  \\
  &  &  &  \multicolumn{1}{c}{(mag)}  & \multicolumn{1}{c}{(K)}   &
                                                                    \multicolumn{1}{c}{(cm\,s$^{-2}$)}   &   &  \multicolumn{1}{c}{(kG)}  &   &  \multicolumn{1}{c}{(d)}  & \multicolumn{1}{c}{(d)}  &    & \multicolumn{1}{c}{(\kms)} &  &  \\[4pt]
  \hline\\[-4pt]
2934856&281056&A5-&10.68&8910&3.45&&&&34.0&34.3&1&&&&43-45\\	
96855460&185256&F0p SrEu&9.96&7100&4.00&roAp&0.7/-/1.4&2, 3&25.64&&&5.5&3&&27\\
162027140&97394&A5p EuCrSr&8.76&8360&3.75&&-/3.1/-&4&47.38&&&r&4&&36-37\\
276300910&134799\rlap{A}&A7p SrCrMg&7.94&7840&4.25&&&&22.6&25.373&1&&&$\delta$~Sct\tablefootmark{d}&38\\	
286255541&69913&B9p Si&8.20&9880&3.50&&&&19.98&19.975&5&&&&34-36\\	
312839768&122525&A0p EuSrCr&8.71&8330&3.55&&-/-/1.7&6&26.70&26.072&7&7&6&&38\\
427032468&140748&A2p EuCr&9.22&8490&3.80&&&&50.8&36.52&1&vs&F&&39\\	
429553678&40678&A0p SiSr&7.38&9520&3.70&&&&26.9&22.029&7&&&&44\\[4pt]
    \hline\\[-4pt]
\end{tabular*}
\tablefoottext{a}{Values retrieved from the TIC where available}
\tablefoottext{b}{r = resolved; vs = very sharp.}
\tablefoottext{c}{F: the line width information is based on visual
  inspection of a FEROS spectrum of our collection.}
\tablefoottext{d}{The $\delta$~Sct pulsation signal probably
  originates from the companion, HD~134799B. See text for details.}
\tablebib{(1)~\citet{2017MNRAS.468.2745N};
  (2)~\citet{2004A&A...415..685H}; (3)~\citet{2013MNRAS.431.2808K};
  (4)~\citet{2011MNRAS.415.2233E}; (5)~\citet{2015A&A...581A.138B};
  (6)~\citet{2017AstL...43..252R}; (7)~\citet{2016AJ....152..104H}.}
 \label{tab:mod_slow}
\end{table*}

\subsection{TESS Cycles~3 and 4 ssrAp candidates}
\label{sec:ssrAp_3-4}
In summary, of the 9 stars whose SAP light curves are discussed in
Sect.~\ref{sec:fignotes}, only TIC\,405516045 is a (definite) ssrAp
star. Although TIC\,162027140 and TIC\,427032468 are borderline cases,
we exclude the remaining 8 stars of this section 
from the final list of ssrAp star candidates
identified as part of our systematic search based on the TESS Cycles~3
and 4 photometric data. Thus, this final list comprises 25 stars, for
which the relevant data are presented in Table~\ref{tab:ssrAp}. The
amplitude spectra for the stars in this table are shown 
in Appendix~\ref{sec:ampspec}.
We also present in Table~\ref{tab:mod_slow} similar data
for the 8 stars 
with moderately long rotation periods from Sect.~\ref{sec:fignotes},
given their relevance in the context of the understanding of
the differentiation of rotation in Ap stars (as argued in Sect.~\ref{sec:sap}). 

\begin{figure*}
  \centering
\includegraphics[width=0.48\linewidth,angle=0]{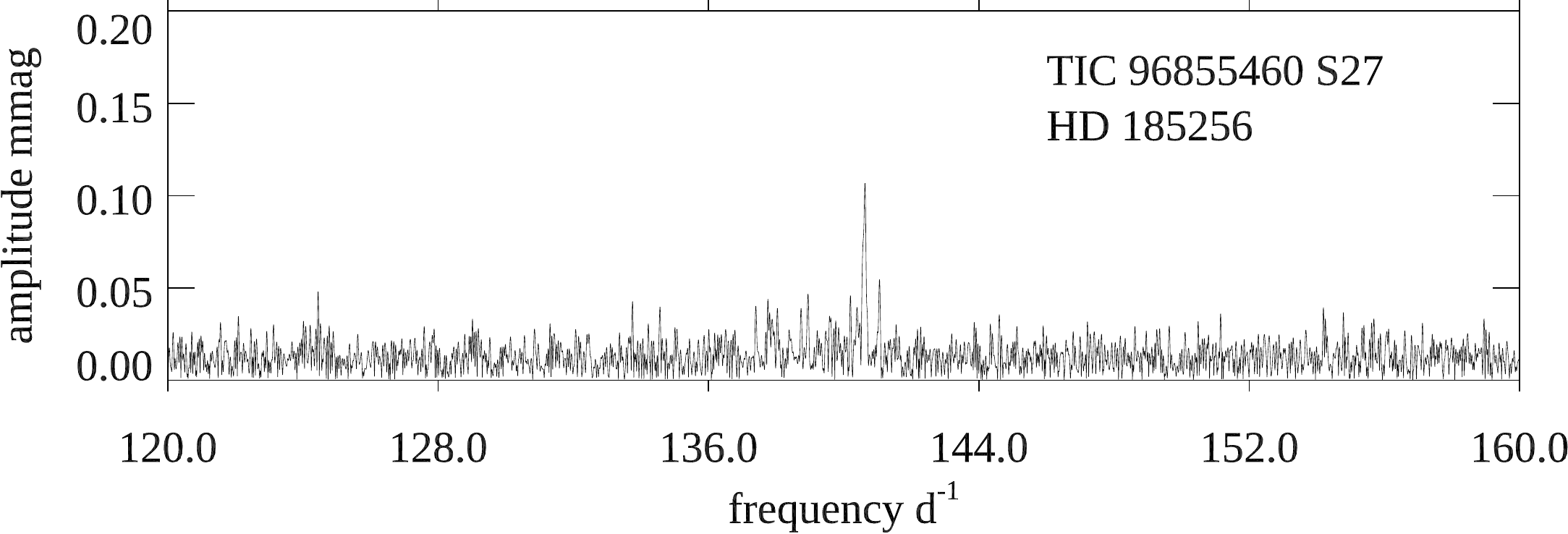}  
\includegraphics[width=0.48\linewidth,angle=0]{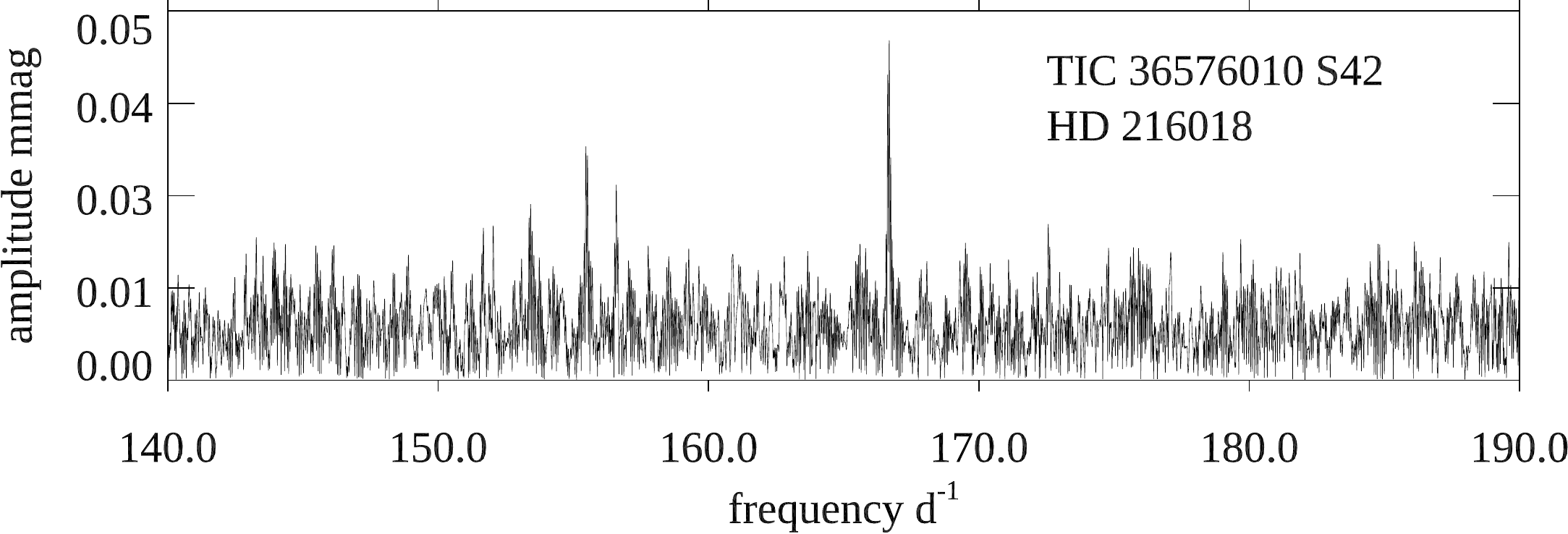}  
\includegraphics[width=0.48\linewidth,angle=0]{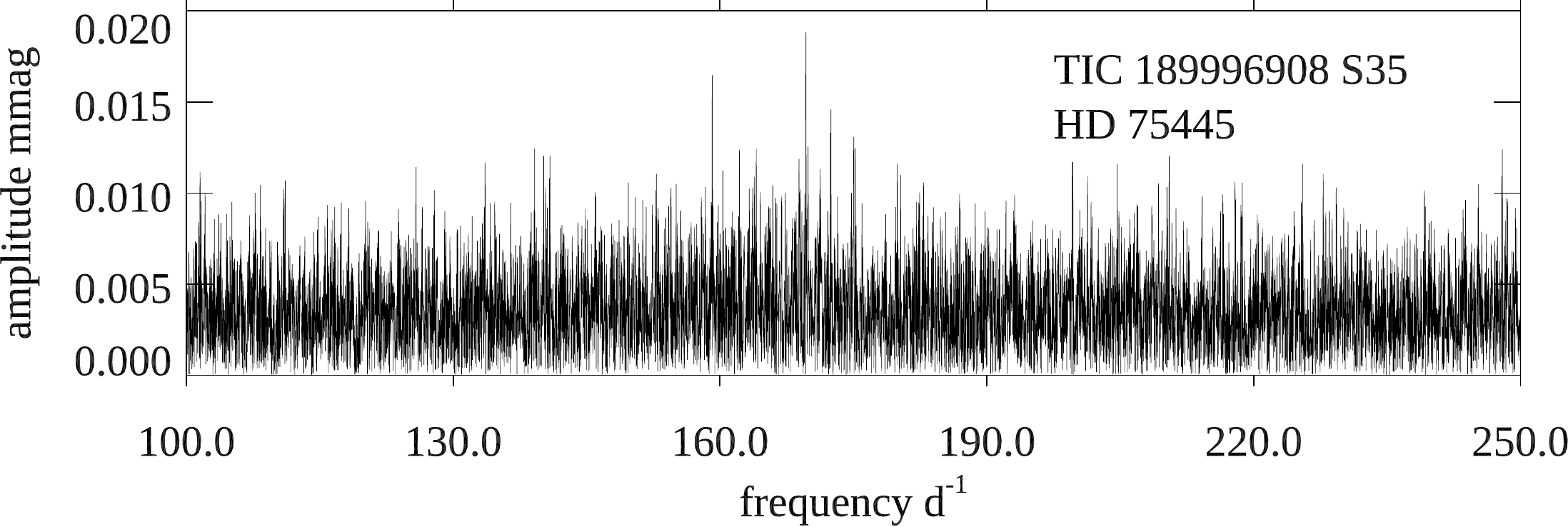}
\includegraphics[width=0.48\linewidth,angle=0]{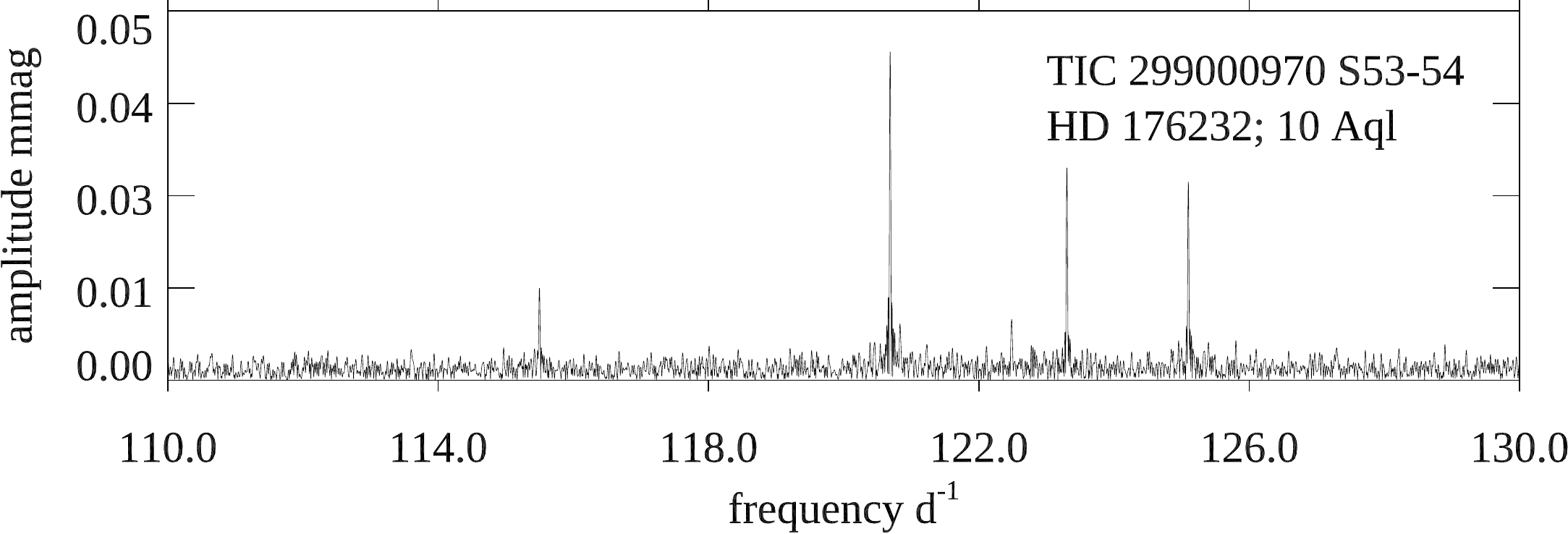}
\includegraphics[width=0.48\linewidth,angle=0]{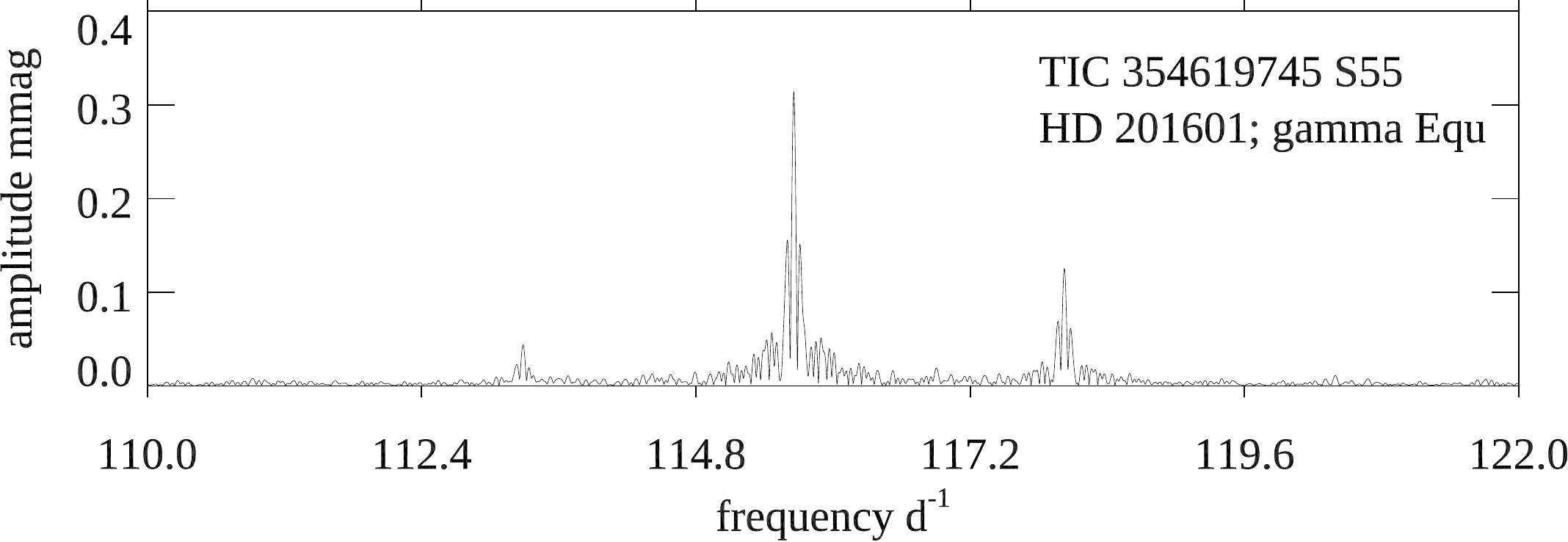}
\includegraphics[width=0.48\linewidth,angle=0]{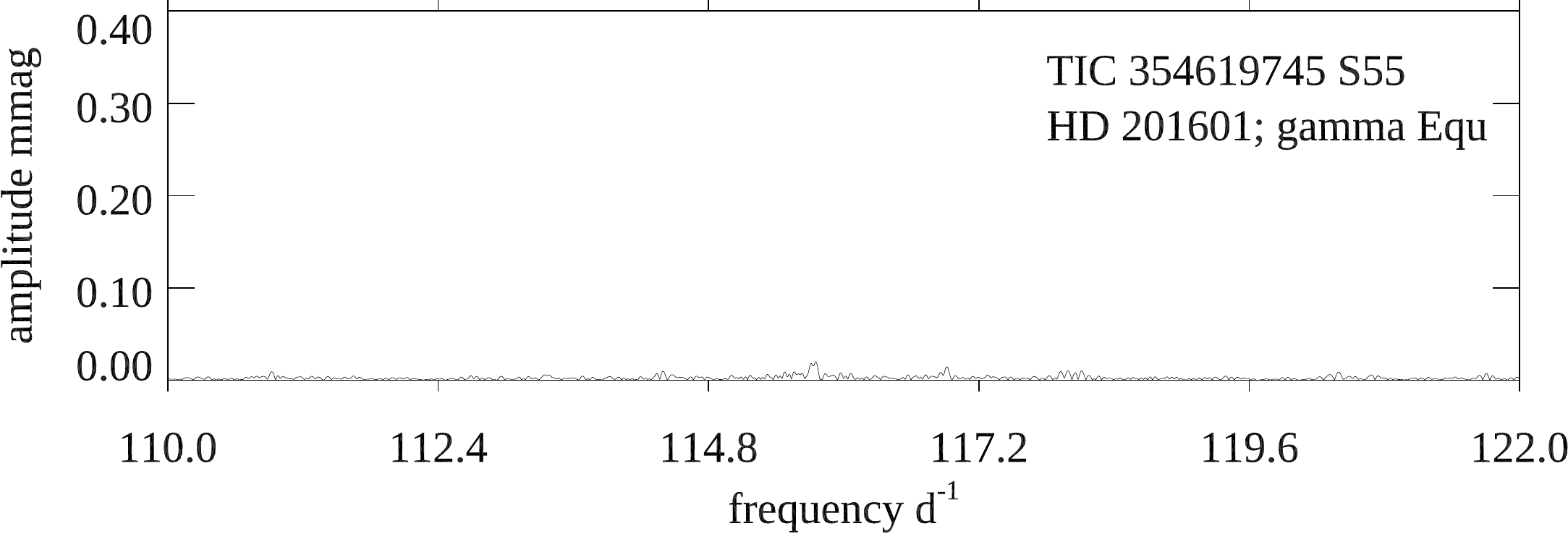}
\includegraphics[width=0.48\linewidth,angle=0]{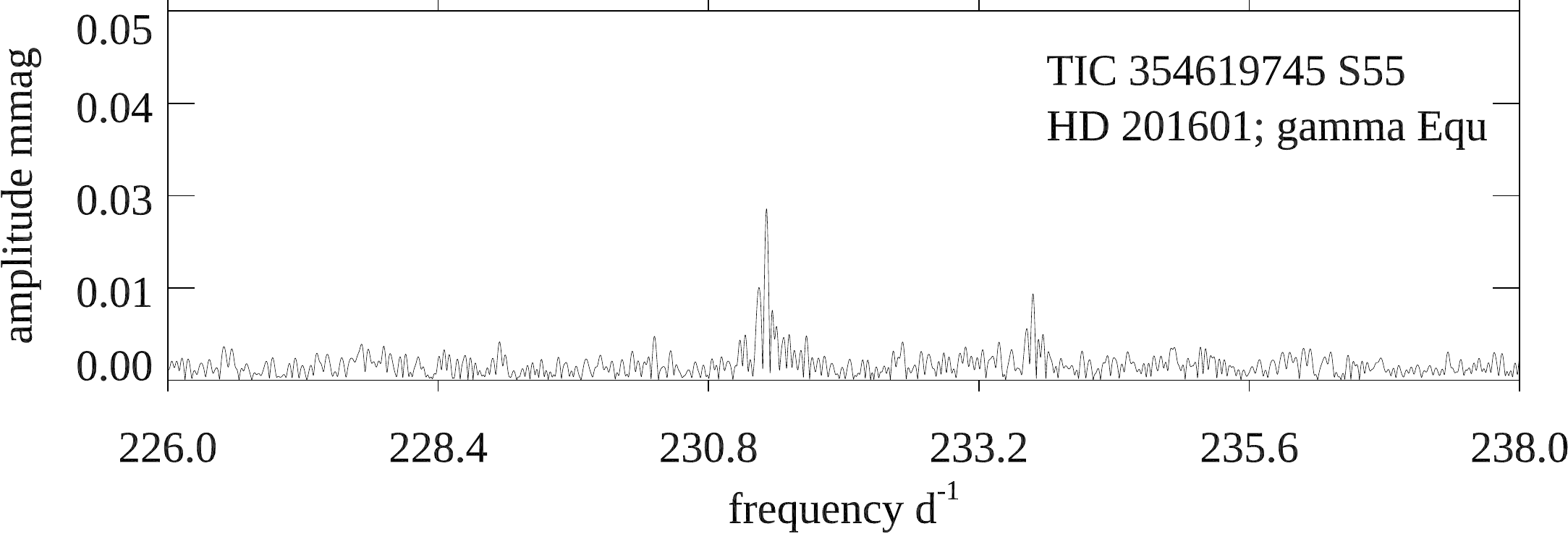}
\caption{Amplitude spectra for 5 roAp stars discussed in the text. The bottom three panels are for $\gamma$\,Equ, showing the three mode frequencies, the amplitude spectrum of the residuals after pre-whitening those three frequencies, and the $2\nu_1$ harmonic and $\nu_1+\nu_2$ combination term.}
  \label{fig:ssrAp2-roap}
\end{figure*}

 The stars in both tables are ordered by increasing TIC number, as
given in Column~1. The HD numbers are listed in Column~2, as alternative
identifications. The spectral types appearing in Column~3 are from
\citetalias{2009A&A...498..961R}; the $V$ magnitudes in Column~4 were
extracted from 
SIMBAD; the effective temperatures $T_\mathrm{eff}$ and surface
gravities $\log g$ in Columns~5 and 6 were taken from the TIC. The roAp
stars (see Sect.~\ref{sec:roap}) are identified in Column~7. Column~8
presents the magnetic data available in the literature, with the
relevant references in Column~9. The values of up to three magnetic
moments are given; all of them correspond to the mean of the
respective moment over a rotation period (if known and adequately
sampled by the existing measurements) or over the observations that
have been obtained (otherwise). The first one is the root-mean square
longitudinal magnetic field, $\Bz_\mathrm{rms}$, as defined by
\citet{1993A&A...269..355B}. In essence, this is the quadratic mean of
the mean longitudinal magnetic field (the line-intensity weighted
average over the stellar disc of the component of the magnetic vector
along the line of sight). The second one is the average value $B_0$
over a rotation cycle of the mean magnetic field modulus \Bm\ (that
is, the same quantity as listed in Column~3 of 
Table~13 of \citealt{2017A&A...601A..14M}). The third field moment,
$Q_0$, is the average value over a rotation period of the mean
quadratic magnetic field \Bq; this is the same quantity as appearing
in Column~10 of Table~13 of \citet{2017A&A...601A..14M}.

Column~10 of Table~\ref{tab:mod_slow} has no counterpart in
Table~\ref{tab:ssrAp}. It contains the values of the moderately long
periods that were determined in the course of this study (see
Sect.~\ref{sec:Psap}). The data presented in Columns~10 to 15 of
Table~\ref{tab:ssrAp} are of the same nature as those given in
Columns~11 to 16 of Table~\ref{tab:mod_slow}. The following
description refers to the latter. The period values from the
literature appear in Column~11.  They come from the references
specified in Column~12. Column~13 contains the \vsi\ values (as
numbers), when they could be found, or indications about the
resolution or width of the spectral lines (as letters). The references
from which this information originates appear in Column~14. For some
stars, a note was added in Column~15. The last column indicates in
which TESS 27-d sectors the data analysed in this study were
obtained.

\section{The roAp stars}
\label{sec:roap}

Four of the 25 ssrAp stars announced in this paper, and one of the stars
with $20\,\mathrm{d}\lesssim\Prot\lesssim50$\,d are also roAp
stars. Thus, there are 4 roAp stars among the 25 ssrAp stars listed in Table~\ref{tab:ssrAp}, 
making up 16\% of the sample. In
\citetalias{2020A&A...639A..31M} and
\citetalias{2022A&A...660A..70M}, we found 22\% and 19\% in the
southern (Cycle~1) and northern (Cycle~2) TESS data,
respectively. These fractions are similar and are significantly
greater than the 5.5\% overall occurrence rate for roAp stars among all
Ap stars studied in TESS data \citep{2021MNRAS.506.1073H,TESS2}. Thus,
there is a positive correlation 
with the occurrence of roAp pulsation and slow rotation, even though
some roAp stars have rotation periods as short as 2\,d.

We discuss the 5 roAp stars in this work individually below. Their
amplitude spectra are shown in Fig.~\ref{fig:ssrAp2-roap}.

\begin{figure*}
  \centering
\includegraphics[width=0.48\linewidth,angle=0]{276300910_S38_SAP.pdf}
\includegraphics[width=0.48\linewidth,angle=0]{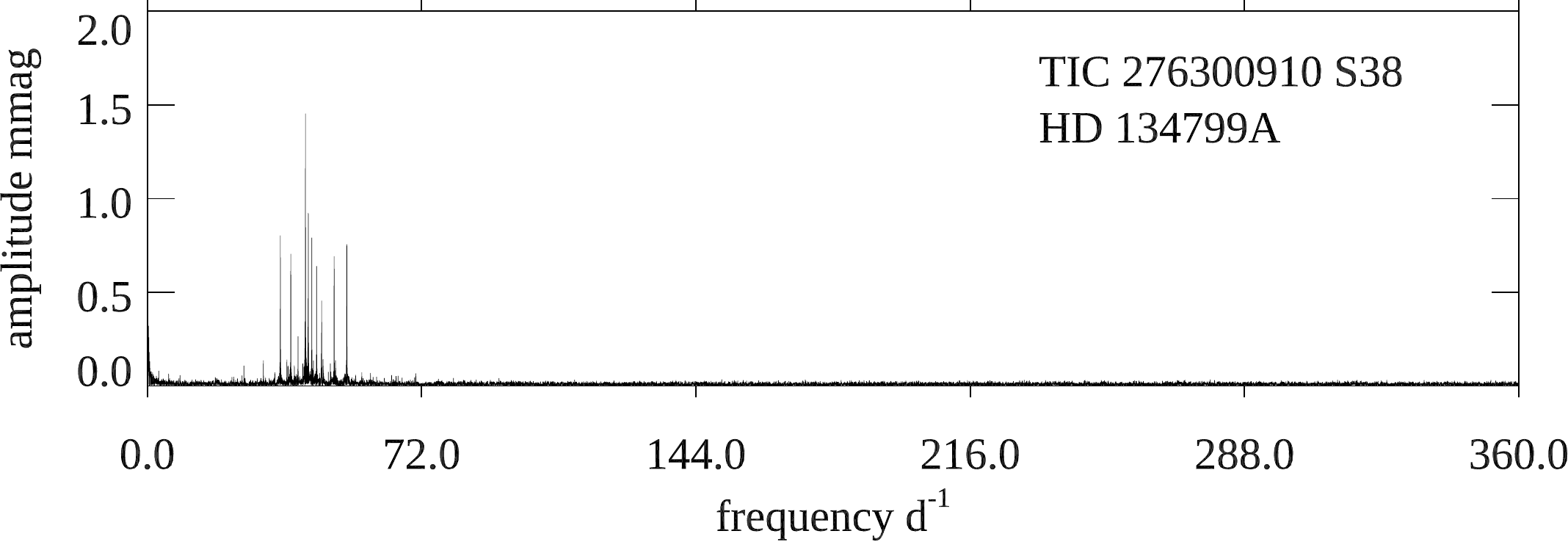}
\caption{Amplitude spectra for TIC\,276300910 (HD\,134799) show clear rotation in the S11 and S38 SAP data with $P_{\rm rot} = 22\fd6$. The low frequency amplitude spectrum shown here is for the SAP data. This star shows rich multi-periodic $\delta$~Sct pulsations, but those  probably arise in TIC\,276300912 (HD\,134799B) which is an A5 star unresolved from TIC\,276300910 in the TESS photometry. 
  }
  \label{fig:dsct}
\end{figure*}

 \subsection{TIC\,36576010 (HD\,216018)}
\label{sec:hd216018}
TIC\,36576010 (HD\,216018) was discovered to be an roAp star by \citet{2022MNRAS.510.5743B}. There are two significant peaks seen in the amplitude spectrum at $155.464$\,d$^{-1}$ and $166.686$\,d$^{-1}$ with amplitudes of $32 \pm 6$\,$\upmu$mag and $45 \pm 6$\,$\upmu$mag, respectively. These differ in frequency by 130\,$\upmu$Hz, which is possibly twice the large separation. With these low amplitudes and low signal-to-noise ratios, pulsations in this star will be difficult to study further.

\subsection{TIC\,96855460 (HD\,185256)}
\label{sec:hd185256}

TIC\,96855460 (HD\,185256) is a known roAp star. \citet{1995IBVS.4209....1K} reported a single pulsation mode with a frequency of $140.8$\,d$^{-1}$. The TESS S27 data show a single peak at $140.6$\,d$^{-1}$. The pulsation was also detected in WASP data with a frequency of 141.18\,d$^{-1}$ \citep{2014MNRAS.439.2078H}. This star is not an ssrAp star, as seen in Fig.\,\ref{fig:rotation_lc}.

\subsection{TIC\,189996908 (HD\,75445)}
\label{hd140748}
TIC\,189996908 (HD\,75445) is a known roAp star with an exceedingly
low pulsation amplitude of $20-40$\,m~s$^{-1}$ in radial velocity
measurements of Nd~{\sc ii} and Nd~{\sc iii} lines
\citep{2009A&A...493L..45K}. Those authors found multiple pulsation
frequencies in the $1.8-2.0$\,mHz range ($155 - 173$\,d$^{-1}$). We
find two $6\,\sigma$ peaks at 159\,d$^{-1}$ and 170\,d$^{-1}$,
consistent with the spectroscopic results. \cite{2009A&A...493L..45K}
also find a marginal rotational variation with a period of $P_{\rm
  rot} = 29\fd5$ in  ASAS data. We do not see this period in the
S8-9, S35 SAP data. From the consideration of mean magnetic field
modulus measurements, \citet{2022MNRAS.514.3485G} suggest that the
rotation period must be much longer than 14\,yr, if the star shows any
variability. 

\subsection{TIC\,299000970 (HD\,176232; 10\,Aql)}
\label{sec:hd176232}  
 TIC\,299000970 (HD\,176232; 10\,Aql) is a known roAp
 star. \citet{1990MNRAS.244..372H} discovered rapid oscillations in
 10\,Aql and identified three pulsation frequencies at 107, 120, and
 124\,d$^{-1}$. \citet{2002MNRAS.337L...1K} found low amplitude
 (30\,m~s$^{-1}$) radial velocity pulsational variations in a
 spectroscopic study of the Nd~{\sc iii}~$\uplambda\,6145$\,\AA\ line at
 frequencies 
 similar to those of \citeauthor{1990MNRAS.244..372H}. The amplitude
 spectrum for 10\,Aql in Fig.\,\ref{fig:ssrAp2-roap} shows multiple
 pulsation frequencies. The four highest amplitude peaks are at
 115.4934, 120.6831, 123.2995, and 125.0940\,d$^{-1}$ with the
 uncertainty in the final digit. These are in the same range as found
 by \citeauthor{1990MNRAS.244..372H}. The amplitudes are mostly stable
 over S53-54 and there is no indication of rotational modulation. The
 separations in frequency for those four peaks are in the $20 -
 50$\,$\upmu$Hz range, which is plausibly representative of the large
 separation and half of that, as expected for dipole and quadrupole
 modes that are typical for roAp stars.

\subsection{ TIC\,354619745 (HD\,201601; $\gamma$\,Equ)}
\label{sec:hd201601} 
 TIC\,354619745 ($= \gamma$\,Equ) is a famous ssrAp star with a rotation period of about 1 century \citep{2016MNRAS.455.2567B}. It is also a known roAp star that has been studied photometrically for 40 years.  
 
\citet{1983MNRAS.202....1K} discovered pulsation in $\gamma$\,Equ from
11 nights of ground-based photometry in 1981. He detected only one
pulsation frequency at $115.7$\,d$^{-1}$. In a multi-site campaign
\citet{1996MNRAS.282..243M} obtained 26 nights of ground-based
photometry from 6 observatories over a time span of 40\,d in
1992. They found 4 pulsation frequencies, $115.69$\,d$^{-1}$,
$118.02$\,d$^{-1}$, $120.70$\,d$^{-1}$, and $123.29$\,d$^{-1}$. They
noted the frequency spacing is about $30$\,$\upmu$Hz, which they
interpreted as half the large separation, implying alternating even
and odd degree modes. That led to an asteroseismic luminosity
measurement in good agreement with the astrometric luminosity.   
 
A comprehensive study was performed on 19\,d of data obtained with the
MOST space mission \citep{2008A&A...480..223G}. Those authors give a
history of previous photometric and radial velocity studies of
pulsation in this star. They found pulsation mode frequencies of
$117.90$\,d$^{-1}$, $117.97$\,d$^{-1}$, $123.30$\,d$^{-1}$,
$120.00$\,d$^{-1}$, and $113.26$\,d$^{-1}$, in order of decreasing
amplitude, along with harmonics of the first and third frequencies. We
note that their first two frequencies are not fully resolved in their
data set, suggesting amplitude modulation.  
 
We find three certain significant mode frequencies in the S55 TESS
data, as seen in Fig.\,\ref{fig:ssrAp2-roap}. They are $\nu_1 =
115.6564$\,d$^{-1}$, $\nu_2 = 118.0240$\,d$^{-1}$, and $\nu_3 =
113.2895$\,d$^{-1}$, also in order of decreasing amplitude. The
uncertainties are in the final digit. We also see the harmonic
$2\nu_1$ and the combination term $\nu_1 + \nu_2$. The three
frequencies are not quite equally spaced with separations of
27.4\,$\upmu$Hz. This is half the large separation, so the modes are
probably alternating $\ell = 2, 1, 2$, assuming the mode with highest
amplitude is a dipole. This is consistent with the result of
\citet{1996MNRAS.282..243M}.  

After pre-whitening the three mode frequencies the amplitude spectrum
of the residuals shows very little
(Fig.\,\ref{fig:ssrAp2-roap}). There is a small indication of mild
amplitude modulation of $\nu_1$. The highest amplitude peak at $\nu_1
= 115.6564$\,d$^{-1}$ is in agreement with those found by
\citet{1983MNRAS.202....1K} and \citet{1996MNRAS.282..243M}, but this
peak is not found at all in the MOST data of
\citet{2008A&A...480..223G}, although $\nu_2$ and $\nu_3$
are. Previous studies have suggested that amplitude modulation of the
modes occurs on time scales as short as 1\,d (see
\citealt{2008A&A...480..223G} for a discussion). This is plausible,
since strong amplitude modulation on a time-scale of only days is
known in other roAp stars, most notably HD\,60435
\citep{1987ApJ...313..782M,2019MNRAS.487.2117B}. 

\section{A $\delta$~Sct star}
\label{sec:dsct}

TIC\,276300910 (HD\,134799) shows clear rotation with $P_{\rm rot} = 22\fd6$. Fig.\,\ref{fig:dsct} shows the amplitude spectra for this star, where clear $\delta$~Sct pulsations can be seen. $\delta$~Sct pulsations are rare in magnetic Ap stars and none is known to pulsate with rich, multiple modes as seen in Fig.\,\ref{fig:dsct}. Those probably arise in TIC\,276300912 (HD\,134799B) which is an A5 star in the $\delta$~Sct instability strip that is unresolved from TIC\,276300910 in the TESS photometry.

\section{Discussion}
\label{sec:disc}
\subsection{The ssrAp star candidates}
\label{sec:disc:ssrAp}

Three of the ssrAp star candidates of Table~\ref{tab:ssrAp} were
already known ssrAp stars, listed in Table~\ref{tab:known_ssrAp}:
HD~110274 ($\Prot=265\fd3$),
HD~176232 ($\Prot>12$\,yr) and HD~201601
($\Prot>97$\,yr). (The references for the periods are
given in the tables.) In the spectra of the former two, the
lines are resolved into their magnetically split components. The lines
of HD~176232 are very sharp but it was still possible to constrain the
magnetic field strength by analysing the differential broadening of
lines of different magnetic sensitivities
\citep{2000A&A...357..981R,2002MNRAS.337L...1K,2003A&A...409.1055L}. 
Furthermore, if HD~75445 (which is also identifed as ssrAp star candidate in
Table~\ref{tab:ssrAp} and also shows resolved magnetically split lines)
is variable at all, its period  must be considerably longer than 14
years \citep{2022MNRAS.514.3485G}.

For a fifth star, HD~216018 (TIC~36576010), which has resolved
magnetically split spectral lines, 
\citet{2017A&A...601A..14M} did not detect any significant variation
of the mean magnetic field modulus \Bm\ nor of the mean longitudinal
magnetic field \Bz. He argued that, unless the lack of variability
results from an infrequent observation geometry (very low value of
the angles $i$ or $\beta$), the rotation period of the star must be
much longer than 6 years. However, combining the magnetic measurements
of \citet{2017A&A...601A..14M} with more recent determinations
of \Bm\ and a single value of \Bz\ published by
\citet{2016AstBu..71..302R}, \citet{2022MNRAS.514.3485G} proposed a
value $\Prot=34\fd044$ for the rotation period of
HD~216018. Our own frequency analysis of
the magnetic data set of \citet{2022MNRAS.514.3485G} fails to provide
convincing evidence in support of the period value that they
advocate. Admittedly, there is a peak close to $f=0.0293$\,d$^{-1}$ in
the amplitude spectrum for \Bm, but there are also other peaks of
similar heights at other frequencies, none of which stands out
remarkably above the noise. We could not distinguish any significant
peak in the amplitude spectrum for \Bz. Actually, for this field
moment, combining the single 
measurement of \citet{2016AstBu..71..302R} with those of
\citet{2017A&A...601A..14M} does not add any meaningful constraint as
the uncertainty about possible systematic differences between
longitudinal magnetic field values obtained with different
instrumental configurations may exceed the amplitude of the $\Bz$
curve presented by \citet{2022MNRAS.514.3485G}. Based on these various
considerations, we do not regard the $\Prot=34\fd044$ value of the
rotation period of HD~216018 as firmly established. Furthermore, two
new measurements performed in 2015 yielded values of \Bz\ close to
1.1\,kG (Romanyuk, private communication). The difference of nearly
300\,G between these values and the one obtainedin 2009 by the same group,
with the same instrument \citep{2016AstBu..71..302R}, is highly
significant. The 2015 \Bz\ values are also well outside the range of
variation shown in Fig.~39 of \citet{2022MNRAS.514.3485G} assuming
$\Prot=34\fd044$. This strongly suggests that the very long period
alternative proposed by \citet{2017A&A...601A..14M} is the
correct one. Accordingly, we kept
HD~216018 in the list of candidate ssrAp stars. 

For two other stars of Table~\ref{tab:ssrAp}, period values
inconsistent with super-slow rotation can be found in the
literature. \citet{2012MNRAS.420..757W} reported photometric
variations of HD~48953 (TIC~372617495) with a period of
$P=2\fd8939$ from the analysis of photometric observations
obtained with the STEREO satellites. However, they note that the
observations are affected by blending from another star and systematic
effects, and that the variability signal is weak, so that the value of
the period should be taken with caution. The TESS data for this star
do not show evidence of significant variability, even at a very low
level. Thus, we regard HD~48953 as a bona fide ssrAp star candidate.

For HD~66318
(TIC~410451752), the  amplitude spectrum computed for the S34-37 data
shows 
a $55\,\sigma$ peak at frequency $f=1.287$\,d$^{-1}$. This is
consistent with the value of the period $P=0\fd77688$ derived by
\citet{2019MNRAS.487.3523C} from analysis of S1 data. However,
\citet{2019MNRAS.487..304D} note that the signal of this star is
highly contaminated, so that there is a significant probability that
the observed variations correspond to another star. This appears all
the more plausible since HD~66318 shows resolved magnetically split
spectral lines \citep{2003A&A...403..645B}, whose very clean component
profiles cannot easily be reconciled with a rotation period that, at
$\Prot=0\fd78$, would be one of the shortest known for any Ap
star. It is much more likely that HD~66318 is an ssrAp star. If the
$f=1.287$\,d$^{-1}$ peak does arise from HD~66318, it is thus likely
to be from a g~mode.  

Besides the five stars already mentioned, a sixth ssrAp candidate from
Table~\ref{tab:ssrAp} shows resolved magnetically split lines: HD~3988
(TIC~262956098). Its rotation period is likely (very) long, as is the
case for the majority of the Ap stars with magnetically resolved lines
\citep{2017A&A...601A..14M}.

Determinations of \vsi\ were found in the
literature for 5 of the stars of Table~\ref{tab:ssrAp}
in which the magnetic resolution of spectral lines has not been
observed until now and no estimate of the rotation period (or of a
lower limit) has been found in the literature. For four of them, the
published values of \vsi\ are consistent with the non-detection of
any rotational broadening at the spectral resolution of
the instrument used for the observation. However, the value
$\vsi=38$\,\kms\ reported by \citet{2012A&A...537A.120Z} indicates
that HD~155102 (TIC~405557056) is not a slow rotator. The broadening
of its metal lines can be easily seen in Fig.~1 of
\citet{2006MNRAS.368..247C} by comparing them with the lines of lower
\vsi\ stars of this study. The lack of variability in the TESS
observations of HD~155102 suggests that its rotation and magnetic axes
must be nearly aligned. However, according to
\citet{2006MNRAS.368..247C}, while the Si peculiarity type is
confirmed by the Si overabundance that they determine, the latter is
mild, and one may wonder if the related brightness contrast of the
spots is too low to produce detectable photometric variations in
integrated light. The same doubt arises about another star listed in
Table~\ref{tab:ssrAp}, HD~117227 (TIC~438694338), which has similarly
broad lines, as seen in a FEROS spectrum ($R\simeq48,000$) of our
collection. In any 
event, while HD~117227 and HD~155102 qualify as ssrAp star candidates
according to the selection process described in Sect.~\ref{sec:ssrAp},
inspection of a FEROS spectrum of the former
and consideration of a meaningful published value of \vsi\ for the
latter show that they are
definitely not ssrAp stars. This stresses the importance of using
spectroscopy to confirm the slow rotation of the candidates identified
with TESS. However, the other examples discussed above show that
spectroscopic confirmations are much more frequent than the opposite.
This validates our search method on a statistical basis.

\subsection{The moderately long period stars}
\label{sec:modlong}
As discussed in Sect.~\ref{sec:sap}, the initial selection of
candidate ssrAp stars based on the analysis of the PDCSAP included a
number of stars that actually undergo significant photometric
variations over the 27-d duration of a TESS sector. These variations   are
removed together with the instrumental effects in the data processing
but can be seen in the SAP data and characterised from them. Most of
these stars have periods in the range
$20\,\mathrm{d}\lesssim\Prot\lesssim50$\,d. We identified 8 such stars
in the Cycle~3 and Cycle~4 samples; they are listed in
Table~\ref{tab:mod_slow}.

As already noted, one of them, HD~97394 (TIC~162027140) has resolved
magnetically split lines \citep{2011MNRAS.415.2233E}. For two more, we
found \vsi\ determinations in the literature: HD~122525 (TIC~312839768;
$\vsi=7$\,\kms; \citealt{2017AstL...43..252R}) and HD~185256
(TIC~96855460; $\vsi=5.5$\,\kms;
\citealt{2013MNRAS.431.2808K}). Furthermore, HD~40678 (TIC~427032468)
shows very sharp, unresolved spectral lines in a FEROS spectrum of our
collection. We could not find any information about the line profiles
for the other four stars of Table~\ref{tab:mod_slow}. However, in the
range of rotation periods that they span, some may possibly show
resolved magnetically split lines. Unfortunately, magnetic field
estimates are only available for 3 stars of Table~\ref{tab:mod_slow},
HD~97394, HD~122525 and HD~185256, whose line profiles are also well
characterised.

\section{Conclusion}
\label{sec:conc}
In \citetalias{2020A&A...639A..31M}, we had identified 60 Ap stars
for which TESS data did not show any variability. However,
FEROS spectra from our collection showed that 6
of them have broad or very broad lines. This left 54 ssrAp star
candidates. Sixty-seven new candidates were added in
\citetalias{2022A&A...660A..70M}. Nevertheless, 3 of them had well
established rotation periods comprised between 26\,d and 49\,d. All
three undergo significant photometric variations over the 27-d
duration of a TESS sector. These variations had not been detected in
our systematic search. The reason why they were overlooked was
unambiguously 
identified in the present study: part of the physical signal coming
from the stars is lost in the data processing carried out to generate
the PDCSAP reduced data on which our analysis is based. The usage of
the latter for our purpose is justified, to deal with instrumental
effects. Nevertheless, the non-processed SAP data can still be
exploited to extract valuable information about the rotation of stars
whose brightness modulation is large enough with respect to the TESS
instrumental effects, especially for those with
$20\,\mathrm{d}\lesssim\Prot\lesssim50$\,d. Eight such stars were
identified in this paper, for which no variability had been detected
in the PDCSAP data.

The systematic analysis of the SAP light curves
that led to the detection of the variations has not been carried out
for the ssrAp star candidates of \citetalias{2020A&A...639A..31M} and
\citetalias{2022A&A...660A..70M}. The latter will include
a number of stars with moderately long periods
($20\,\mathrm{d}\lesssim\Prot\lesssim50$\,d), which are not ssrAp
stars as per the original definition of
\citet{2020pase.conf...35M}. In fact, three stars with periods
in this range are definitely present in Table~A.1 of
\citetalias{2022A&A...660A..70M}. More such periods may be constrained
from consideration of the 
SAP data for the other ssrAp star candidates of
\citetalias{2020A&A...639A..31M} and
\citetalias{2022A&A...660A..70M}. This is an issue outside the scope
of the present work, but we plan to address it in a future
study. Suffice it to say here that 
the inclusion or exclusion of the stars with
$20\,\mathrm{d}\lesssim\Prot\lesssim50$\,d in the samples of ssrAp star
candidates should be expected to have a minor impact on the
statistical conclusions that can be reached by analysing these
samples. Indeed, as already stressed above, Ap stars with
$\Prot\gtrsim20$\,d are anyway part of the long-period tail of the
distribution of the Ap star rotation, so that they are relevant for
the understanding of the physical mechanismes responsible for, and
related to, slow rotation.  Thus, at present, conclusions can be best drawn from
consideration of the combined list of 33 stars of Tables~\ref{tab:ssrAp}
and \ref{tab:mod_slow} with the ssrAp candidate lists of our previous
two papers. 

Another source of potential contamination of the list of ssrAp star
candidates from our TESS Cycle~1 and Cycle~2 studies whose criticality
was emphasised in the present work on Cycles~3 and 4 is the
misclassification of a number of stars. Namely, we rejected 31\%\ of
the stars of Cycle~3 that we had originally selected on the basis of
their classification in \citetalias{2009A&A...498..961R}, for which there was no
classification uncertainty flagged in this catalogue, on account that
they definitely were not Ap stars or they did not appear to be bona
fide Ap stars. For Cycle~4, the rejection fraction was even much
higher, 72\%, but the difference can be in large part attributed to
the unreliability of the Ap classifications of a single study
\citep{Abastumani}, which we established. Leaving this source apart,
the rejection rate for Cycle~4 would have been 37.5\%, similar
to that for Cycle~3. 

The vetting that we have performed involves a certain
amount of subjectivity, and we have erred on the side of
conservatism, to minimise the potential contamination of the final
list of non-variable Ap stars by misclassified stars of other
types. But we do not expect more than a minority of the stars that we
did not regard as bona fide Ap stars to have their peculiarity
eventually confirmed. 

In the construction of the list of ssrAp star candidates of
\citetalias{2022A&A...660A..70M}, we had already tried to include only
bona fide Ap stars. To this effect, we had excluded from the selection
the stars that were not in \citetalias{2009A&A...498..961R} (except for a few
for which compelling evidence of peculiarity had been subsequently
found), and we had carried out a critical evaluation of the
peculiarity of the stars flagged with ``/'' or ``?'' in the first
column of \citetalias{2009A&A...498..961R}. While some additional stars for which
the information from the catalogue does not suggest any classification
ambiguity were also discarded on account of clear indications of
misclassification in more recent publications, this was not part of a
systematic vetting such as the one described in this paper for the
Cycles~3 and 4 stars. Thus, the ssrAp candidate list for Cycle~2 may
still contain a small number of non-Ap stars. By contrast, the search
for ssrAp candidates using the TESS Cycle~1 data included all the
stars with an Ap classification found in the TIC for which 2-min
cadence observations were available. Accordingly, one must expect the
list of ssrAp candidates of \citetalias{2020A&A...639A..31M} to be
significantly contaminated by misclassified normal A stars, Am stars,
etc.

Notwithstanding the above-described differences in the critical
evaluation of the ssrAp star candidate lists of
\citetalias{2020A&A...639A..31M}, \citetalias{2022A&A...660A..70M} and
this paper, these lists can be combined to draw meaningful statistical
conclusions. The merged list contains 152 stars that are either ssrAp
star candidates or Ap stars with moderately long periods
($20\,\mathrm{d}\lesssim\Prot\lesssim50$\,d). The 8 stars for which no
TESS variability was detected but that are known to show rotationally 
broadened spectral lines are not included in this number.

Besides these 152 stars, there are 28 known ssrAp stars in
Table~\ref{tab:known_ssrAp} that were not identified as part of our
search for Ap stars showing no low-frequency photometric variability
over 27 days in the TESS Cycles~1 to 4 data that we analysed. The
reasons why these stars were missed by our search include the
non-availability of 2-min observations for some stars,
contamination of the signal by another, variable star (either a
physical companion or a chance alignment), or instrumental
effects. For statistical purposes, these 28 known ssrAp stars can
valuably be added to the 152 ssrAp star candidates (or Ap stars with
moderately long periods) identified from
analysis of the TESS data. This increases the size of the sample to
180 stars. 

Of these 180 (very) slowly rotating Ap star candidates, 38 are roAp
stars. That is, 
roAp stars account for 21\%\ of the Ap stars whose rotation periods
are in the long-period tail of the distribution. The difference with
the fraction of roAp stars among the entire population of Ap star
studied in TESS Cycles~1 and 2, which is of the order of 5.5\%
\citep{2021MNRAS.506.1073H,TESS2}, is 
highly significant. The possible inclusion in the long-period
candidate lists and in the full Ap star population list of stars
whose classification is mistaken and that 
are not actually Ap stars does not challenge this conclusion (even if
every second star of the full Ap star population were misclassified,
the roAp star fraction in this sample would only reach 11\%). Its
implications for our understanding of the origin of rapid oscillations
and of (very) slow rotation are currently unclear. Their discussion is
beyond the scope of this paper.

\begin{figure}
\resizebox{\hsize}{!}{\includegraphics{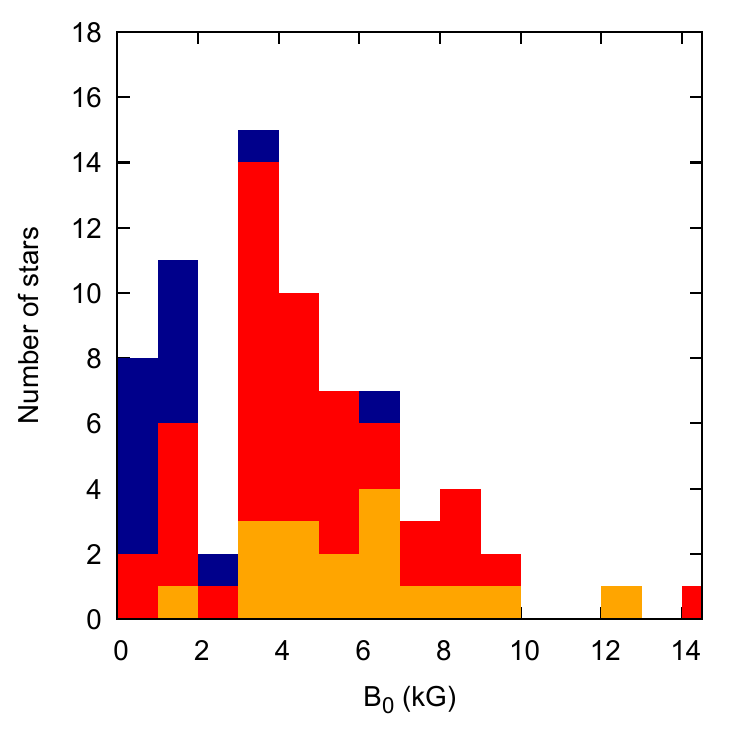}}
\caption{Distribution of the phase-averaged magnetic field strength
  $B_0$ for the long-period Ap stars of Tables~\ref{tab:known_ssrAp},
  \ref{tab:ssrAp} and \ref{tab:mod_slow} of this paper,
  of Table~1 of \citetalias{2020A&A...639A..31M} and of Table~A.1 of
  \citetalias{2022A&A...660A..70M}. The orange and red 
  parts of the histogram corresponds to the stars for which
  measurements of the mean magnetic field modulus or of the mean
  quadratic magnetic field are available; for the remaining stars
  (blue part of the histogram), a
  lower limit of $B_0$ was inferred from the existing mean
  longitudinal magnetic field measurements. Orange distinguishes the
  known ssrAp stars that were not identified as ssrAp star candidates
  on the basis of our TESS-based photometric survey.} 
\label{fig:bavhist}
\end{figure}

For 12 of the 33 stars of Tables~\ref{tab:ssrAp} and
\ref{tab:mod_slow}, at least some magnetic field determinations are
available in the literature. They can be used together with the 42
ssrAp star candidates from \citetalias{2020A&A...639A..31M} and
\citetalias{2022A&A...660A..70M} for which magnetic field measurements
exist to confirm and refine the conclusions
previously reached about the distribution of the magnetic field
strengths in (super) slowly rotating Ap stars. The size of the sample
can be further increased by adding to it those stars that are definitely
known to be ssrAp stars but that were not identifed as ssrAp star
candidates from consideration of the TESS Cycle1 to Cycle~4
data. There are 15 such stars that show resolved magnetically split
lines, for which mean magnetic field modulus measurements have been
obtained. Furthermore, one determination of the mean quadratic
magnetic field of HD~101065 has been performed. Thus, in total there
are 70 known ssrAp stars, ssrAp star candidates or Ap stars with
moderately long period for which magnetic information is available. 
Since non-Ap stars do
not have large-scale organised magnetic fields, the probable presence
of a number of them in Table~1 of \citetalias{2020A&A...639A..31M} and
in Table~A.1 of \citetalias{2022A&A...660A..70M} is irrelevant for the
present discussion.

Figure~\ref{fig:bavhist} shows the distribution of the magnetic field
strength $B_0$ for the stars of Tables~\ref{tab:ssrAp} and
\ref{tab:mod_slow} of the present paper, Table~1 of
\citetalias{2020A&A...639A..31M} and Table~A.1 of
\citetalias{2022A&A...660A..70M}, as well as for the stars of
Table~\ref{tab:known_ssrAp} that have not been identified as ssrAp
star candidates as part of our TESS-based survey. For those stars with
resolved magnetically split lines for which mean
magnetic field modulus values have been derived, $B_0$ is the
average of these values, over a rotation cycle if one at least has
been covered,  or over the available measurements otherwise. As in
\citetalias{2022A&A...660A..70M}, when no mean magnetic field modulus
determinations are available but mean quadratic magnetic field data
have been obtained, $B_0$ was approximated by dividing the
phase-averaged value $Q_0$ of the mean quadratic field by
1.28. Finally, for those stars for which only the mean longitudinal
magnetic field has been measured, the lower limit of $B_0$ was set to
3 times the rms mean longitudinal field $\Bz_\mathrm{rms}$; this is the value
that was used in Fig.~\ref{fig:bavhist}.

For interpretation of this figure, the caveats discussed in
\citetalias{2022A&A...660A..70M} remain relevant. In particular, among
the 12 ssrAp star candidates or moderately long-period Ap 
stars of the present paper for which magnetic field measurements exist, there
is none for which those measurements sample a whole rotation
cycle. Accordingly, the phase-averaged field values listed in
Tables~\ref{tab:ssrAp} and \ref{tab:mod_slow}, which were used to
build Fig.~\ref{fig:bavhist}, sample only those rotation phases at
which observations have been obtained. While the resulting values of
$B_0$ and $Q_0$ should arguably be reasonable first approximations of
the average values over a full rotation cycle of, respectively, the
mean magnetic field modulus and the mean quadratic magnetic field, the
value of $\Bz_\mathrm{rms}$ based on incomplete phase coverage may be
considerably less representative of the actual strength of the mean
longitudinal magnetic field (see \citetalias{2022A&A...660A..70M} for
details). This is why, in Fig.~\ref{fig:bavhist}, the parts of the
histogram based on $\Bz$ measurements are filled in blue, to
distinguish them from those based on \Bm\ or \Bq\ data. These are
filled in red for the ssrAp star candidates, and in orange for those
known ssrAp stars that have not been identified as showing no
photometric variability over one TESS sector. The
distribution of the latter supports the view that the knowledge of the
ssrAp stars prior to our current systematic TESS-based search was
biased towards strongly magnetic Ap stars, but it is otherwise fully
consistent with the distribution derived from consideration of the
ssrAp star candidates from the current project.  

Figure~\ref{fig:bavhist} is an updated version of Fig.~4 of
\citetalias{2022A&A...660A..70M}, which mostly confirms the
conclusions drawn from consideration of the latter. First and
foremost, it confirms that the rate of occurrence of (super) slow
rotation among weakly magnetic Ap stars must be considerably lower
than among strongly magnetic ones. Not only are there fewer stars in
the $0-3$\,kG $B_0$ range than in the $3-6$\,kG $B_0$ range, but
furthermore this trend is opposite to that found by other studies that
consider samples of Ap stars with any rotation periods, as discussed
in detail in \citetalias{2022A&A...660A..70M}.

The intriguing gap in the $B_0$ distribution between 2 and 3\,kG,
already suggested by \citet{2017A&A...601A..14M}, also
appears to be confirmed. Admittedly, compared to Fig.~4 of
\citetalias{2022A&A...660A..70M}, one star, HD~3988, was added in this
histogram bin, on the basis of its mean magnetic field modulus,
$B_0=2.65$\,kG. However, the four spectra from which
this value was derived were acquired over a time interval of only five
days, and they show no significant $\Bm$ variations. This is not
surprising for an ssrAp star candidate, but it implies that the actual
average of the mean magnetic field modulus of HD~3988 may differ
from the value of $B_0$ reported here. Even if it remains between 2
and 3\,kG, the presence of HD~3988 in this histogram bin would not
rule out the existence of a gap in the magnetic field strength
distribution. As already mentioned in
\citetalias{2022A&A...660A..70M}, the presence of this gap suggests
the existence of two distinct populations of ssrAp stars, consisting
of weakly and strongly magnetic stars whose rotational evolution would
follow different paths and involve different physical mechanisms. This
has potentially major implications for the understanding of the origin
of the magnetic Ap stars, which fully justify further investigation
of super-slow rotation in these stars. In particular, it will be
essential to constrain the rotation periods and the magnetic fields of
all ssrAp star candidates so as to establish their distribution on the
basis of a sample as complete and unbiased as possible. 

Perhaps the most remarkable difference between
Fig.~\ref{fig:bavhist} and Fig.~4 of
\citetalias{2022A&A...660A..70M} is the presence in the former of the
star HD~66318, with $B_0=14.5$\,kG, well outside the range covered by
the rest of the sample, in which no star has $B_0>10$\,kG. We expect
very few ssrAp stars with values of the mean magnetic field modulus of the
order of that of HD~66318, since until now no star with $\Prot>150$\,d
is known to have $B_0>7.5$\,kG
\citep{2017A&A...601A..14M}. In particular, the rotation period of the
other star in Fig.~\ref{fig:bavhist} for which $B_0>10$\,G, the well
known ssrAp star HD~126515 ($B_0=12.5$\,kG), is $\Prot=129\fd95$. 
\citet{2003A&A...403..645B} obtained
three measurements of the mean magnetic field modulus of HD~63318, the
second one 49 days after the first one and the third one 11 days after
the second one. That all three measurements yield essentially
the same value of $\Bm$ does not necessarily rule out a rotation
period shorter than 150\,d. Nonetheless, determining the magnetic
field of HD~66318 at 
more epochs in order to constrain its rotation period would represent
a valuable test of the non-occurrence of very strong fields in stars
with periods in excess of $\sim$150\,d.

In summary, the addition of the new ssrAp star candidates identified
from analysis of the TESS Cycles~3 and 4 data to those from Cycles~1
and 2, complemented by the
inclusion of the known ssrAp stars for which the analysed TESS
photometric data do not reveal the lack of variability over 27\,d,
confirms and strengthens the conclusions reached in
\citetalias{2020A&A...639A..31M} and \citetalias{2022A&A...660A..70M}
about the dependence of the rate of occurrence of rapid oscillations
and of magnetic fields of different strengths on the rotational
velocity of Ap stars. To consolidate these conclusions, the following
steps need to be carried out. (1) The list of
ssrAp star candidates from \citetalias{2020A&A...639A..31M} and
\citetalias{2022A&A...660A..70M} must be reviewed for overlooked
variability possibly detectable in the SAP data but removed in the
processed PDCSAP data and for possible misclassifications of non-Ap
stars as peculiar. (2) High resolution spectra of all ssrAp stars
candidates for which such data are not available yet must be
obtained to confirm that they have low projected equatorial velocities
and to determine their magnetic fields. We are now undertaking these
steps, whose results will be reported in future papers. 

\begin{acknowledgements} 
This research has made use of the SIMBAD database, operated at CDS,
Strasbourg, France. This paper includes data collected by the TESS mission,
which are publicly available from the Mikulski Archive for Space
Telescopes (MAST). Funding for the TESS mission is provided
by NASA's Science Mission directorate. Funding for the TESS
Asteroseismic Science Operations Centre is provided by the Danish
National Research Foundation (Grant agreement no.: DNRF106),
ESA PRODEX (PEA 4000119301) and Stellar Astrophysics Centre
(SAC) at Aarhus University. DLH and DWK acknowledge support from the
Funda\c c\~ao para a Ci\^encia e a Tecnologia (FCT) through national
funds (2022.03993.PTDC).
\end{acknowledgements}

\bibliographystyle{aa}
\bibliography{lpn10}

\appendix

\section{Incorrect and dubious Ap classifications}
\label{sec:misclass}
\begin{table}
  \caption{Stars listed as Ap in \citetalias{2009A&A...498..961R}, for which the
    Abastumani catalogue is the only identified source of
    classification as peculiar.}
  \label{tab:abastumani}
  \centering
  \begin{tabular}{rll}
\hline\hline\\[-4pt]
    \multicolumn{1}{c}{TIC}  &Other ID&Spectral type\\[4pt]
    \hline\\[-4pt]
3373254&HD 244248&A5p SiSr\\
3542929&HD 244372&A1p Sr\\
3544794&HD 244352&B9p Si\\
20077256&HD 246148&B9p Si\\
73904395&J05312310+2814129&B9p SiSr\\
74387907&TYC 1858-556-1&B9p SiSr\\
74804951&HD 245423&A3p Si\\
74805804&HD 245353&A1p SiSr\\
75777460&HD 246587&A2p SiCr\\
75856090&HD 246686&A1p Si\\
76031807&HD 246861&B9p Si\\
76098926&HD 246993&B9p Si\\
76213603&TYC 1870-1678-1&A0p Sr\\
77834479&HD 248663&B9p Si\\
78239007&HD 248944&A1p SiSr\\
78789635&HD 40038&A0p Sr\\
78968090&HD 249664&A1p SiSr\\
79042758&TYC 1867-2081-1&B9p SiSr\\
79322287&J06002427+2842163&A0p Si\\
79963265&TYC 1872-1791-1&B9p SiSr\\
80895911&HD 41418&A0p Si\\
80896249&HD 251408&A1p SiCr\\
81436350&HD 252286&A0p Sr\\
115453793&HD 244955&Ap Sr\\
115633147&HD 245191&A0p Si\\
116143102&HD 245725&A0p Si\\
116247916&TYC 2408-446-1&B9p SiSr\\
127712498&TYC 2411-2277-1&B9p Si\\
127837999&TYC 2407-384-1&B9p SiSr\\
127838106&TYC 2407-631-1&A0p SiSr\\
127960970&HD 243523&A0p SiSr\\
239758529&HD 247437&B9p SiSr\\
239802019&HD 247629&A0p Sr\\
239807112&TYC 2405-821-1&B9p SiSr\\
239833172&HD 247794&A0p Sr\\
285856246&HD 243791&A0p Sr\\
308704616&TYC 2409-31-1&B9p SiSr\\
309118102&HD 249382&A0p Si\\
400163968&HD 250149&B9p SiSr\\
429127816&TYC 1855-373-1&A0p SiSr\\[4pt]
    \hline\\[-4pt]
  \end{tabular}
  \end{table}

As mentioned in Sect.~\ref{sec:class}, a critical review of the
information available in the literature led us to conclude that
\citetalias{2009A&A...498..961R} includes a significant number of stars assigned an
Ap spectral type that either are definitely not Ap stars or that
cannot be regarded as bona fide Ap stars, without any indication of
the ambiguity of their classification. For some of these, but not all,
the classification was revised or questioned after publication of the
catalogue. Following this realisation, we undertook a systematic
literature review to confirm as securely as possible that all the
stars that we identified as ssrAp star candidates on account of the
lack of photometric variability in the TESS Cycle~3 and Cycle~4 data
are indeed bona fide Ap stars. As a result, we discarded a number of
stars from our initial selection.

Although our review covered only a small fraction of the Ap stars
featuring in \citetalias{2009A&A...498..961R}, we believe that
the list of those 
among them that are definitely not Ap stars or that we do not consider
as bona fide Ap stars represents a useful piece of information for the
community. There are two groups of such stars. 

 The first one consists of those stars for which the only reference
 that we could identify as the source of the Ap classification is the
 Abastumani catalogue \citep{1990AJ.....99..379K}. We showed in
 Sect.~\ref{sec:class} that more than 50\% of the Ap stars listed in
 this catalogue must be misclassified. Therefore, stars in this group
 cannot be regarded as bona fide Ap stars, unless some other piece of
 evidence can be found that confirms their classification -- for
 instance, the detection of variations that may either be due to
 rotation or to rapid oscillations. No such piece of evidence was
 found for 49 of the 51 stars of this group that were present in our
 initial automated selection of stars that show no photometric
 variability over the duration of a TESS sector.

 These 49 stars are listed in Table~\ref{tab:abastumani}. Columns~1
 and 2 contain the TIC number and another id; the spectral type from
 \citetalias{2009A&A...498..961R}  is given in Col.~3. A number of entries
 in this table must definitely correspond to Ap stars, but on an
 individual basis, their peculiarity needs to be independently
 confirmed. 

The second group of stars with incorrect or dubious classifications as
peculiar contains a few stars for which we were unable to
identify the source of the Ap spectral type as well as stars for which
we found references convincingly indicating a different spectral type
or (in two cases) an ambiguous classification. The stars of this group
are listed in Table~\ref{tab:misclass}, which contains 14
entries. These stars were all identified by the automatic search
procedure as showing no TESS photometric variability over the 27\,d
duration of a sector; for one of them, TIC~152803574, the analysed
TESS observations were obtained both in Cycle~3 and in
Cycle~4. Columns~1 to 3 of the table give the TIC number of the star,
an alternative identification, and the spectral type listed in
\citetalias{2009A&A...498..961R}. The reason why we do not consider the star as a
bona fide Ap star appears in Col.~4 (in a number of cases, under the
form of a different spectral type), with the corresponding reference
(when applicable) in Col.~5.

\begin{table*}[b]
  \caption{Stars listed as Ap in \citetalias{2009A&A...498..961R}, with an incorrect
  or dubious classification as peculiar.}
\label{tab:misclass}
   \begin{tabular*}{\textwidth}[]{@{}@{\extracolsep{\fill}}rllll}
\hline\hline\\[-4pt]
  \multicolumn{1}{c}{TIC}  &Other ID
     &Spectral type&Exclusion reason&Reference\\[4pt]
  \hline\\[-4pt]
     2854318&HD 242991&B9p Si&B9IV HgMn&\protect{\citet{2021A&A...645A..34P}}\\
     7731470&CD$-$31 12076&A5p Sr&Unknown classification source&\\
     43068154&HD 228112&A9p Sr&Normal F0 star&\protect{\citet{1970A&AS....1....1F}}\\
     88091070&HD 192969&A0p SiMg&A0p Si or perhaps Am&\protect{\citet{1970PASP...82..321B}}\\
     127506807&HD 281048&B9p SiSr&Unknown classification source&\\
     152803574&HD 104833&F0p Sr&F str $\lambda\,4077$&\protect{\citet{1994A&A...281..775N}}\\
     156645202&HD 111702&F0p Sr&F str $\lambda\,4077$; composite?&\protect{\citet{1994A&A...281..775N}}\\
     207018530&HD 45827&A7p SrCrEu&No lines\tablefootmark{a}&\protect{\citet{1958ApJS....3..141B}}\\
     231461096&HD 56731&B9p Si&kA6hA7mF0(IIIb)\tablefootmark{b}&\protect{\citet{1989ApJS...70..623G}}\\
     236328832&HD 144999&A7p Sr&Am&\protect{\citet{1988PASP..100.1084B}}\\
     358467049&CPD$-$60 944B&B9p Si&HgMn&\protect{\citet{2014MNRAS.443.1523G}}\\
     385552299&HD 23387&A0p CrSi&Ap character questionable\tablefootmark{c}&\protect{\citet{1998A&A...332..224H}}\\
     405220654&BD$-$14 2015&B9p Sr&B9:Vn&\protect{\citet{1975AJ.....80..131D}}\\
     410033396&HD 202400&F0p Sr&Ba star&\protect{\citet{1994A&A...281..775N}}\\[4pt]
     \hline\\[-4pt]
   \end{tabular*}
   \tablefoottext{a}{See text (Sect.~\ref{sec:class}).}
   \tablefoottext{b}{The star is also studied as an Am star in many
     other references.}
   \tablefoottext{c}{Conclusion derived from an abundance analysis
     based on a spectrum recorded at $R\simeq34,000$.}
 \end{table*}

 \section{The amplitude spectra}
\label{sec:ampspec}

\begin{figure*}[b]
  \centering
  \includegraphics[width=0.48\linewidth,angle=0]{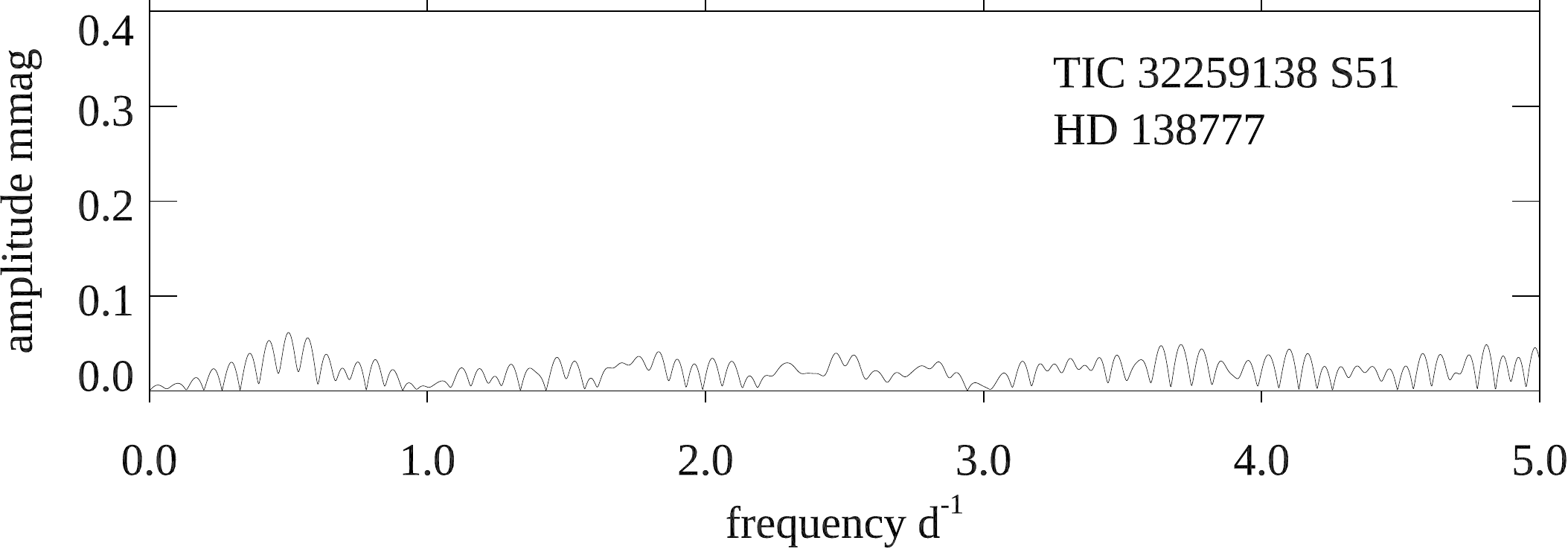}
  \includegraphics[width=0.48\linewidth,angle=0]{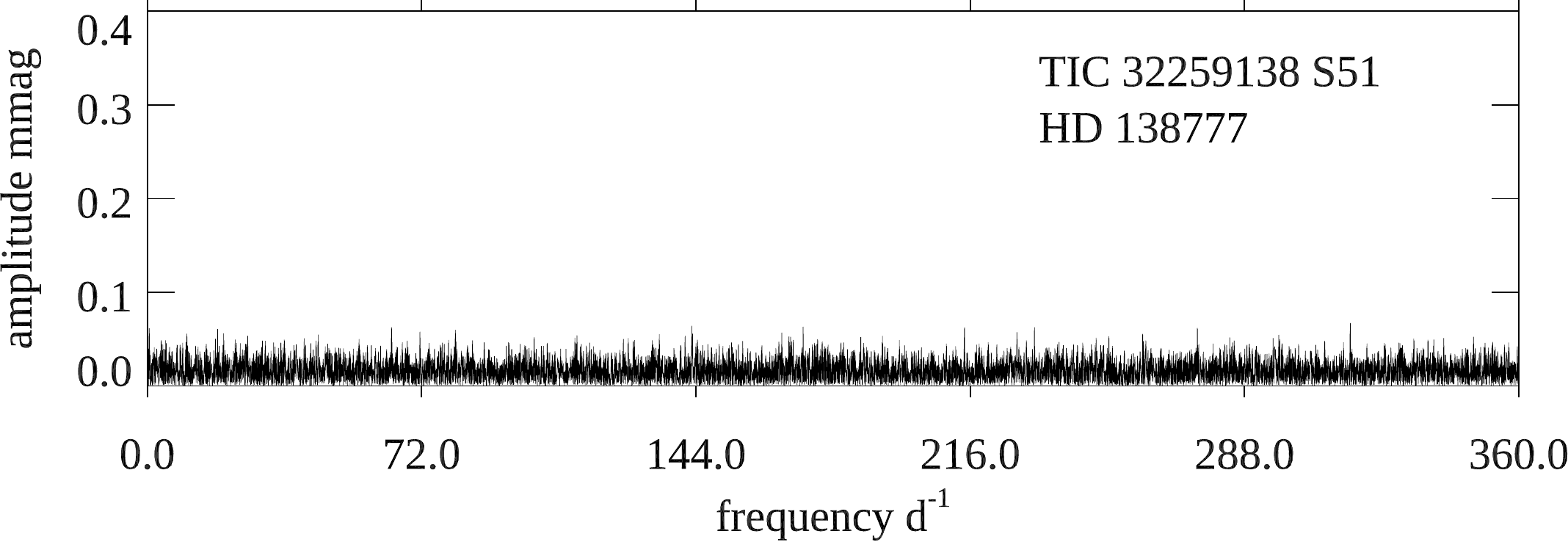}
  \includegraphics[width=0.48\linewidth,angle=0]{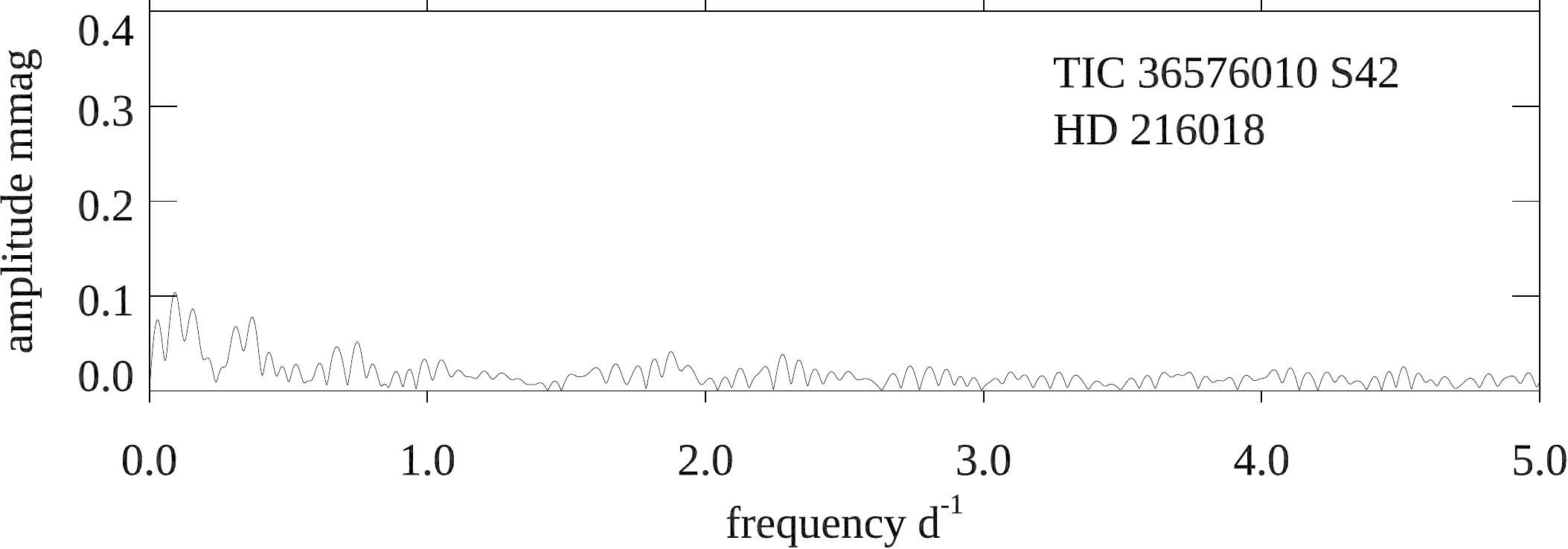}
  \includegraphics[width=0.48\linewidth,angle=0]{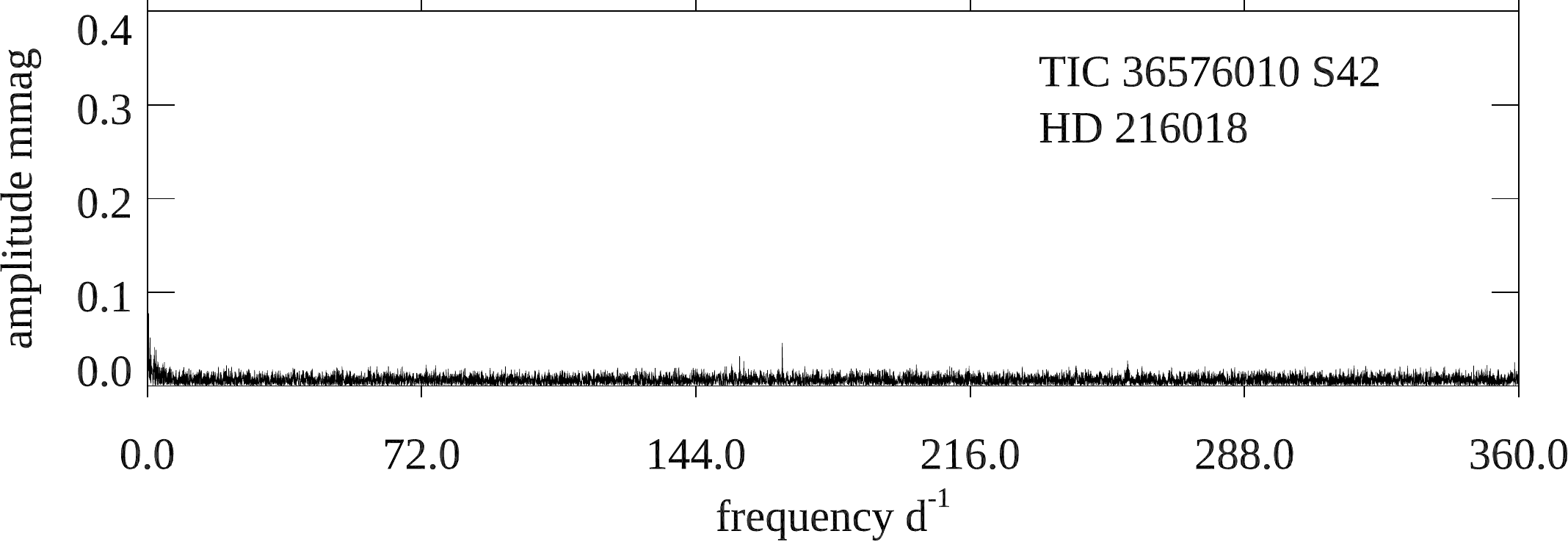}
  \includegraphics[width=0.48\linewidth,angle=0]{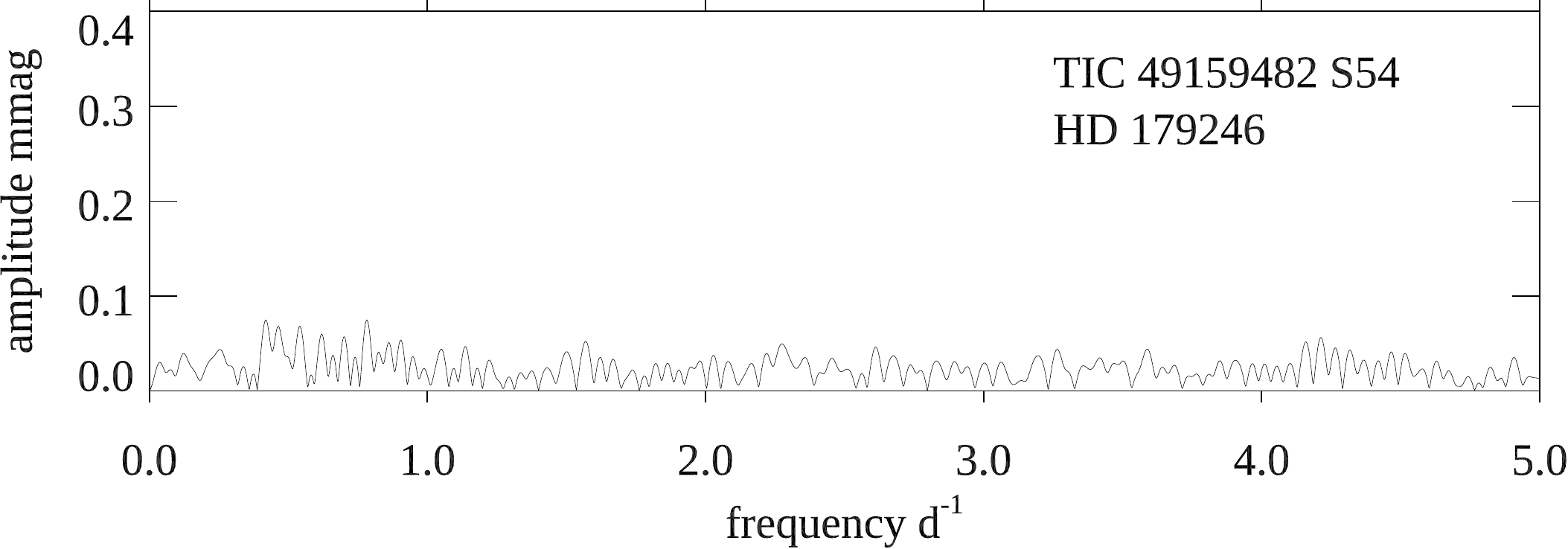}
  \includegraphics[width=0.48\linewidth,angle=0]{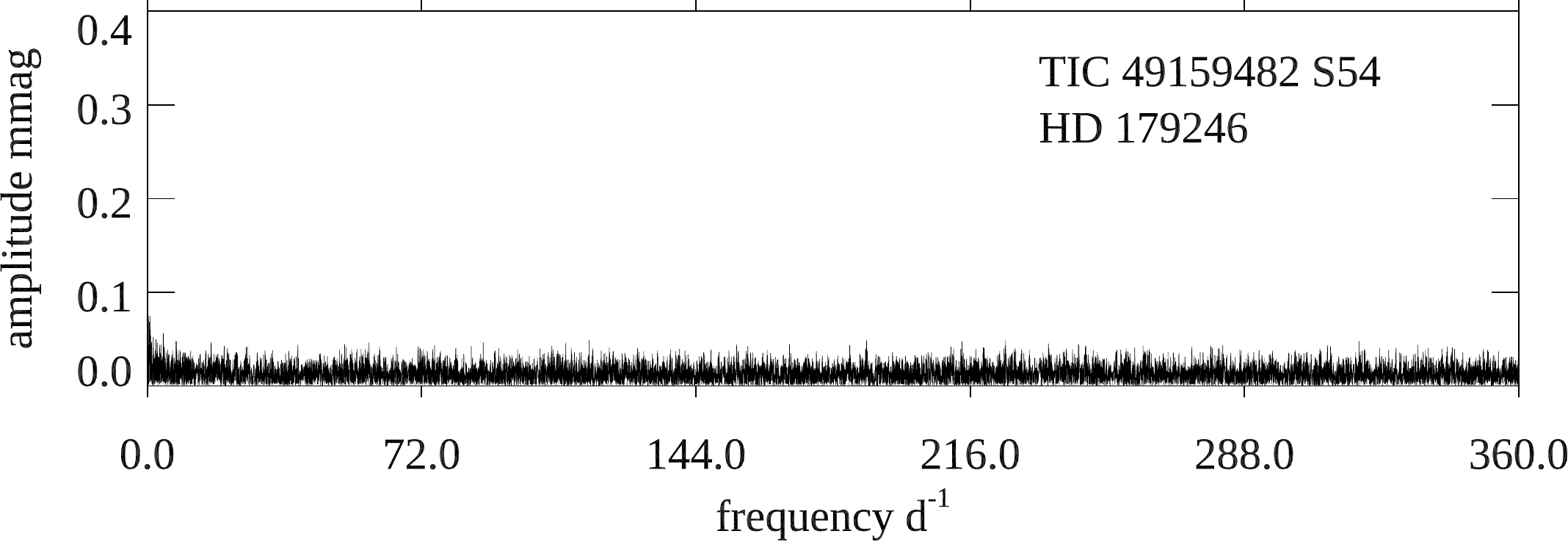}
  \includegraphics[width=0.48\linewidth,angle=0]{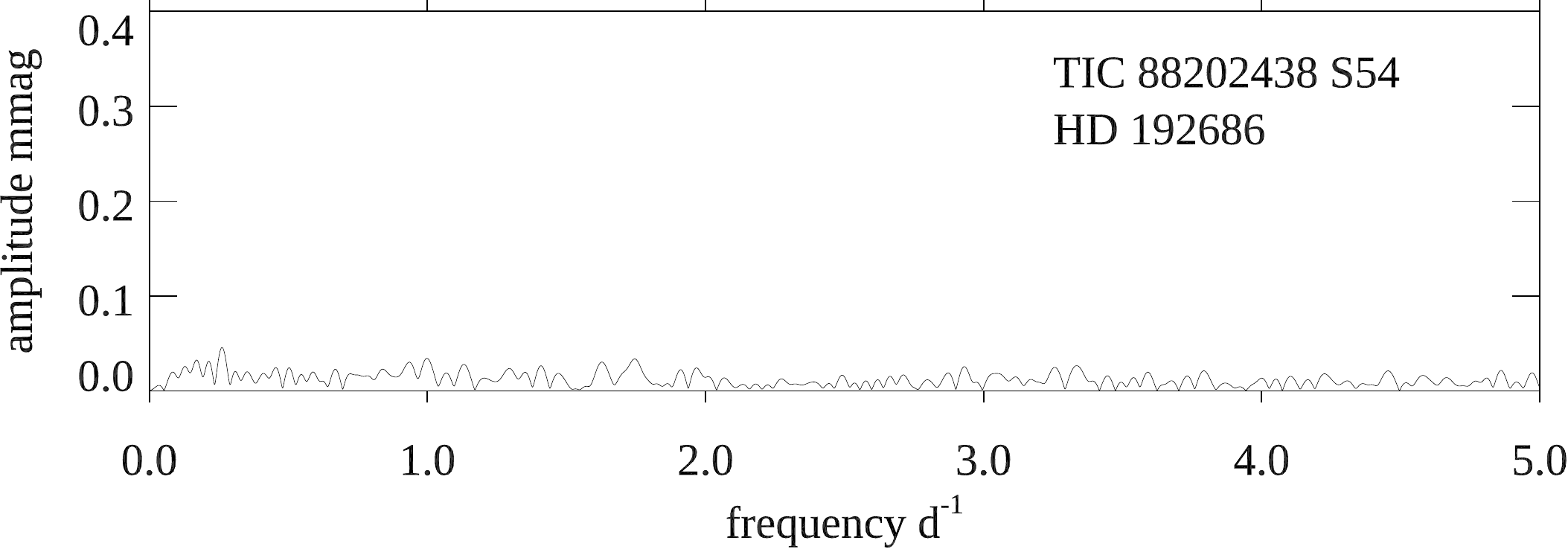}
  \includegraphics[width=0.48\linewidth,angle=0]{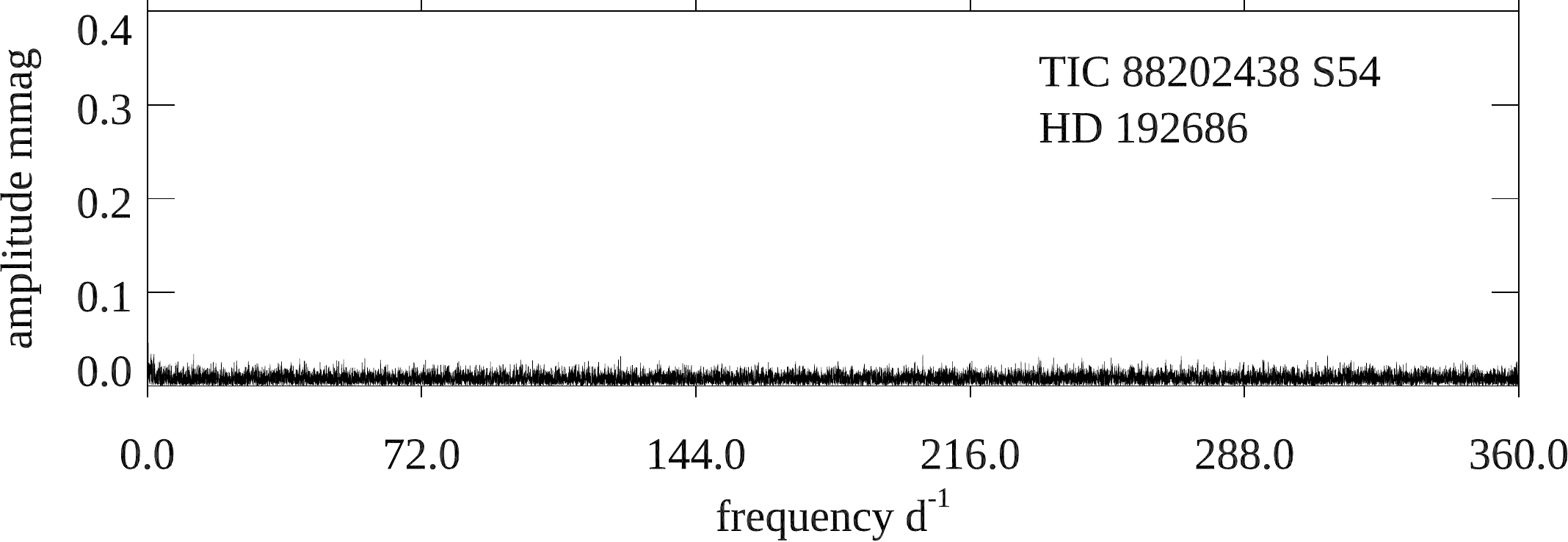}
    \includegraphics[width=0.48\linewidth,angle=0]{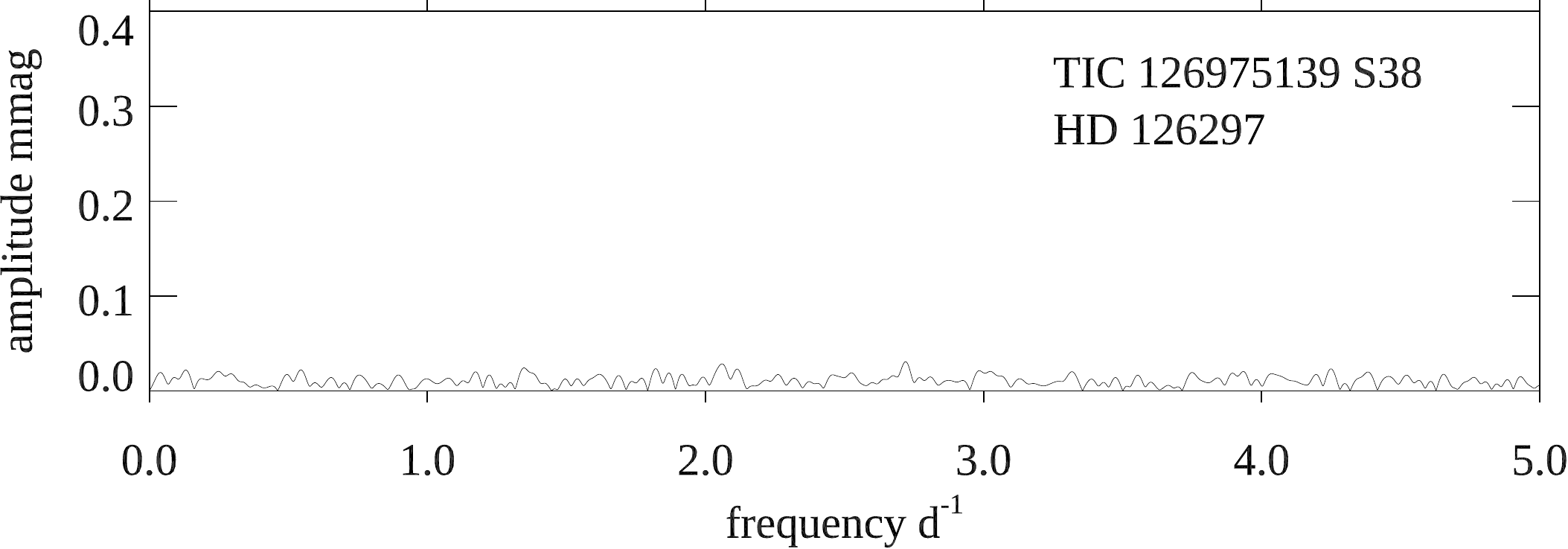}
  \includegraphics[width=0.48\linewidth,angle=0]{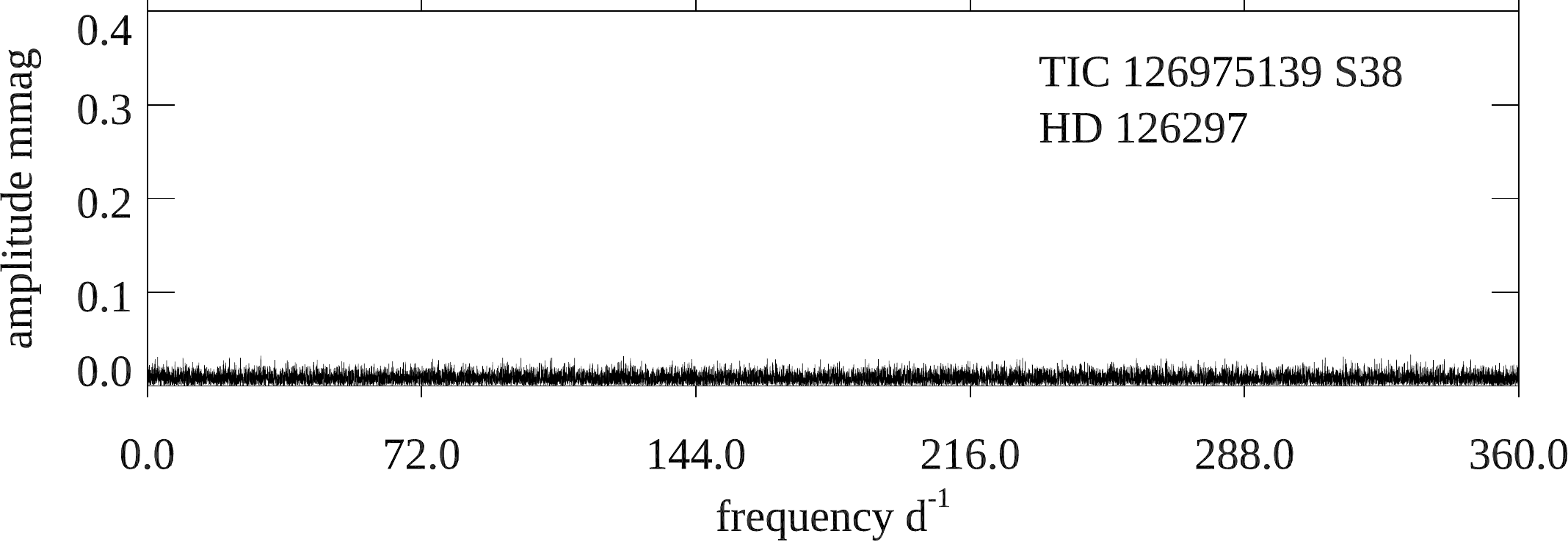}
    \includegraphics[width=0.48\linewidth,angle=0]{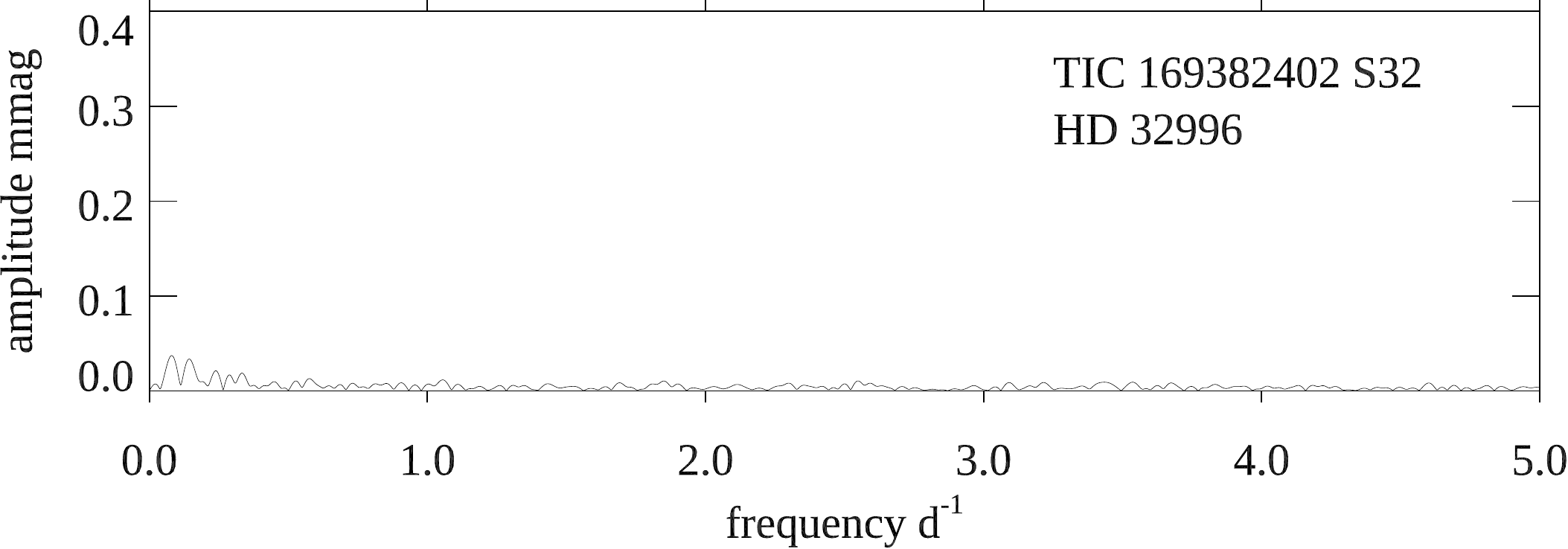}
  \includegraphics[width=0.48\linewidth,angle=0]{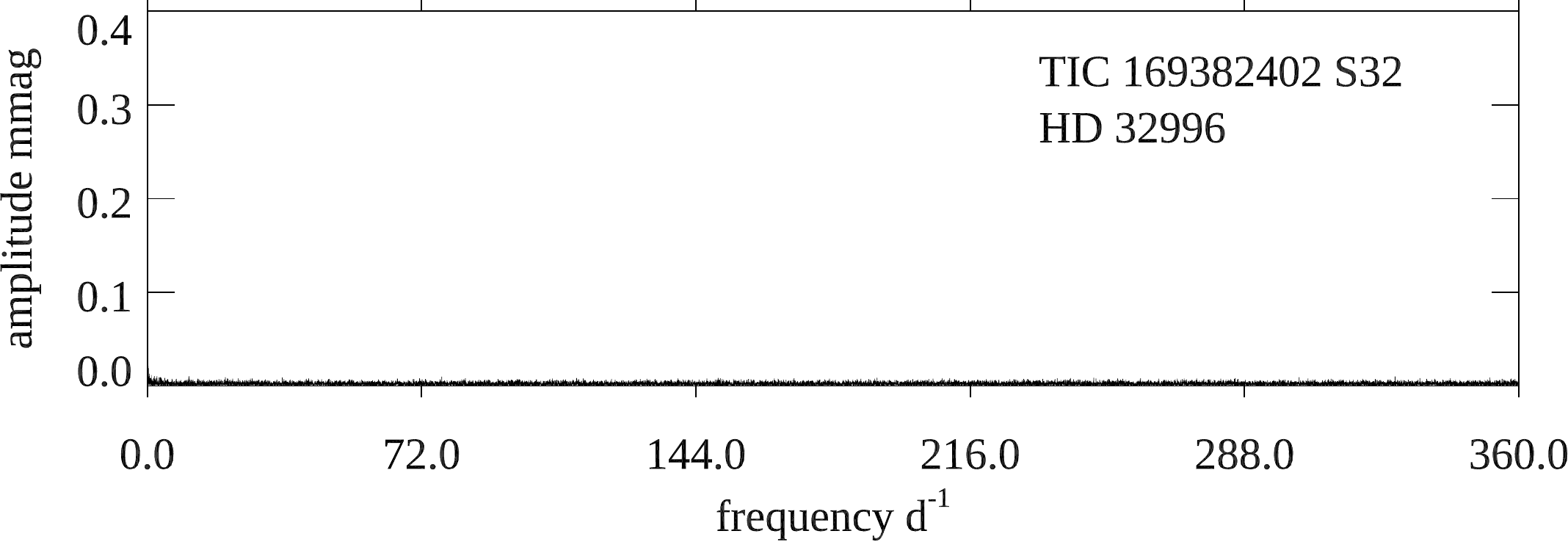}
  \caption{Amplitude spectra for the long-period Ap stars. Each row presents a low-frequency amplitude spectrum showing no rotational variation in the left panel, and a full amplitude spectrum to the Nyquist frequency of 360\,d$^{-1}$ in the right panel, which allows detection of $\delta$~Sct or roAp pulsation. We note the occasional changes of ordinate scale to accommodate pulsation peaks.}
  \label{fig:ssrAp2-1}
\end{figure*}

\begin{figure*}
  \centering
 \includegraphics[width=0.48\linewidth,angle=0]{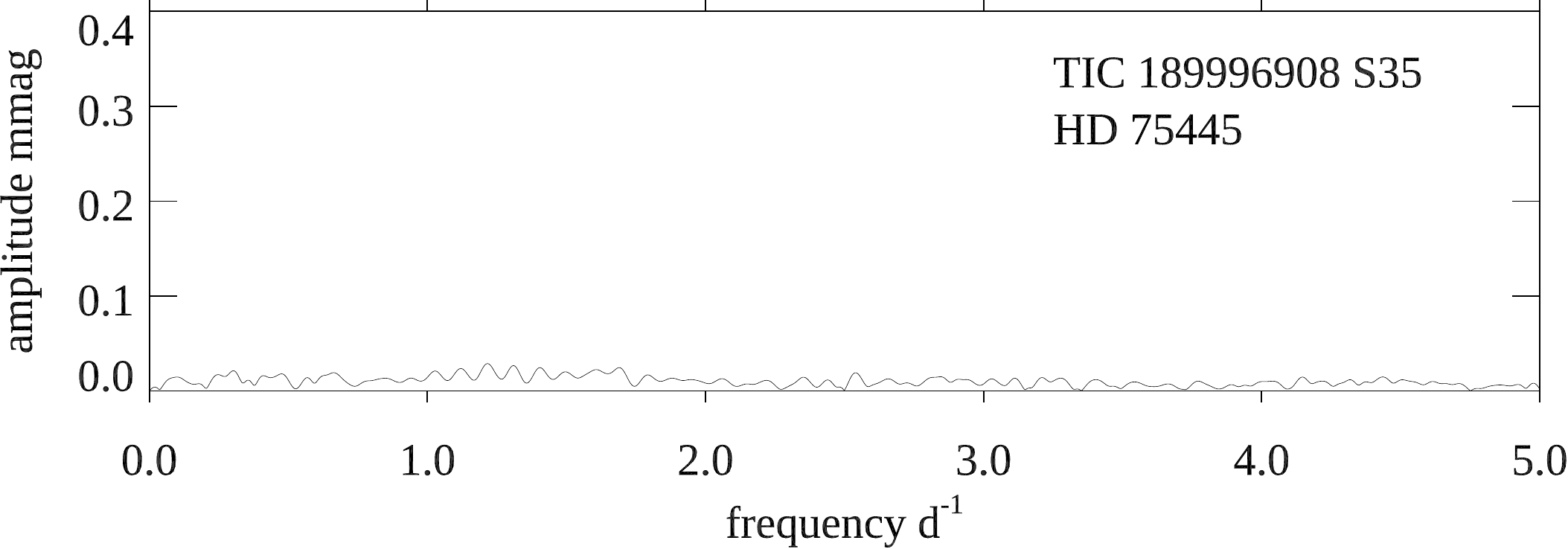}
  \includegraphics[width=0.48\linewidth,angle=0]{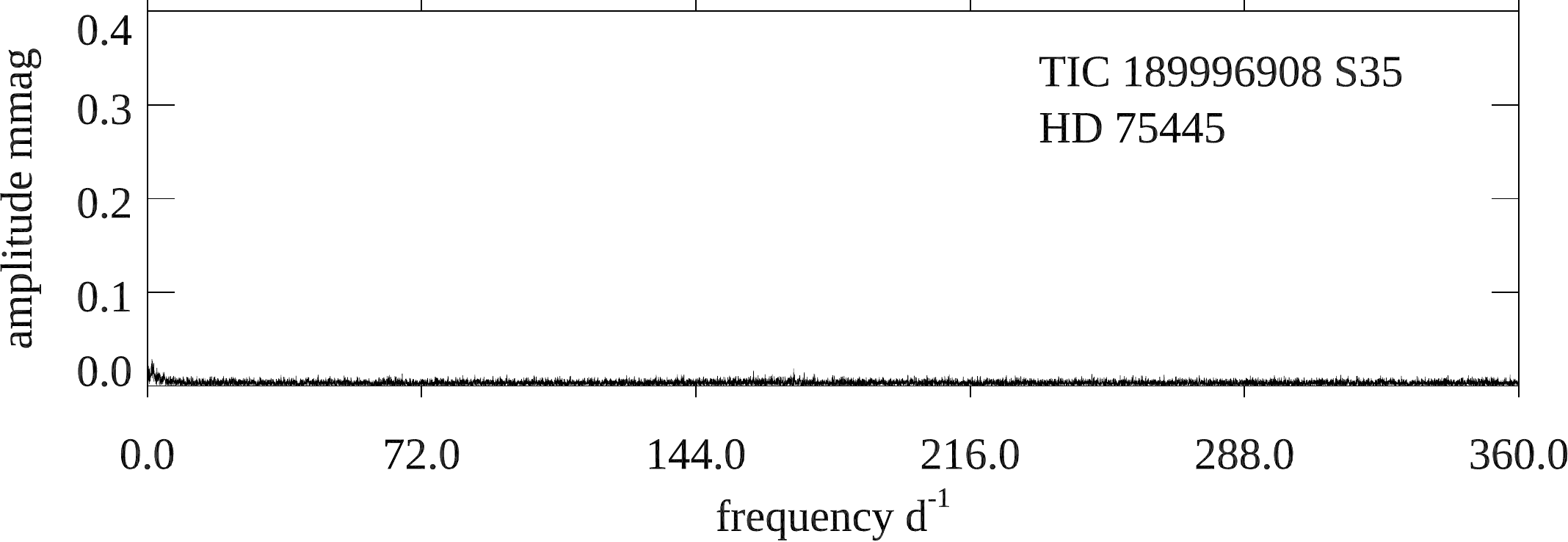}
 \includegraphics[width=0.48\linewidth,angle=0]{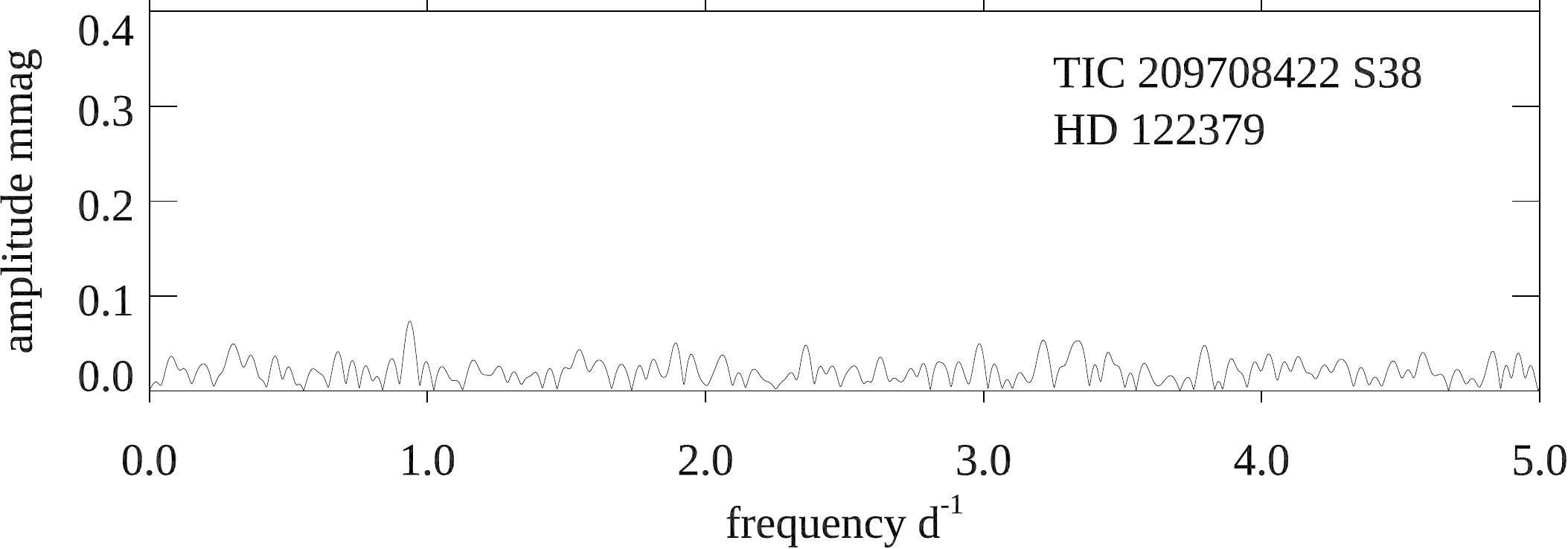}
 \includegraphics[width=0.48\linewidth,angle=0]{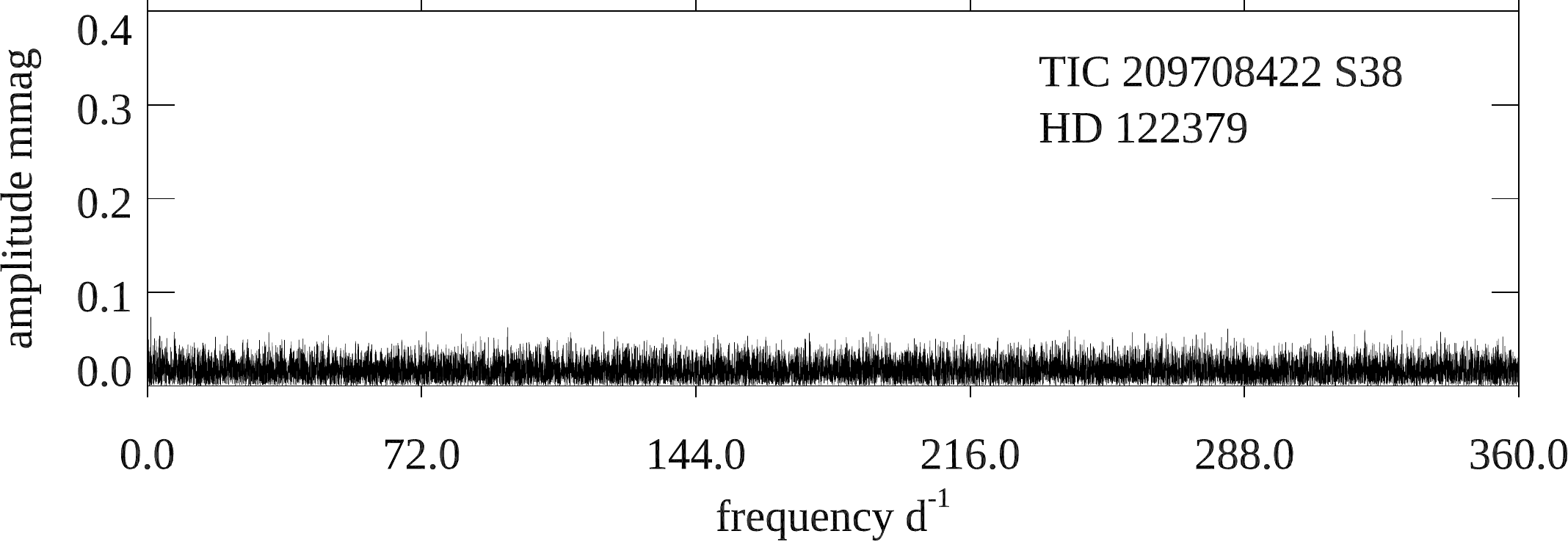}
 \includegraphics[width=0.48\linewidth,angle=0]{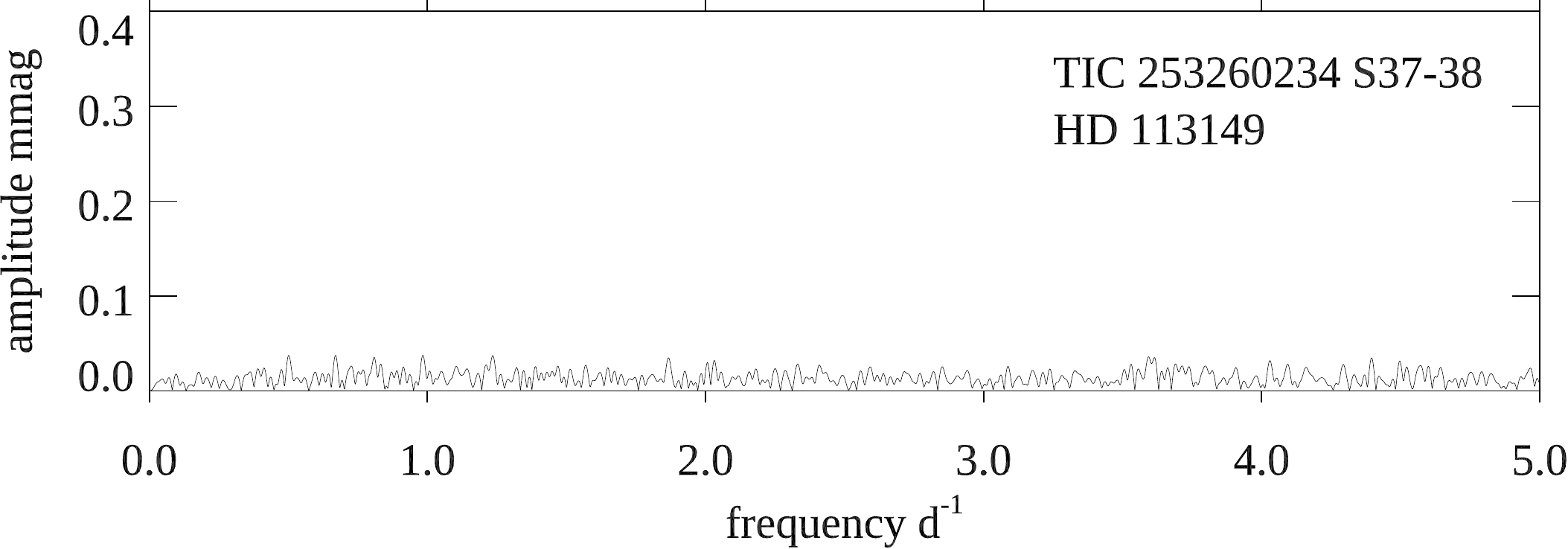}
 \includegraphics[width=0.48\linewidth,angle=0]{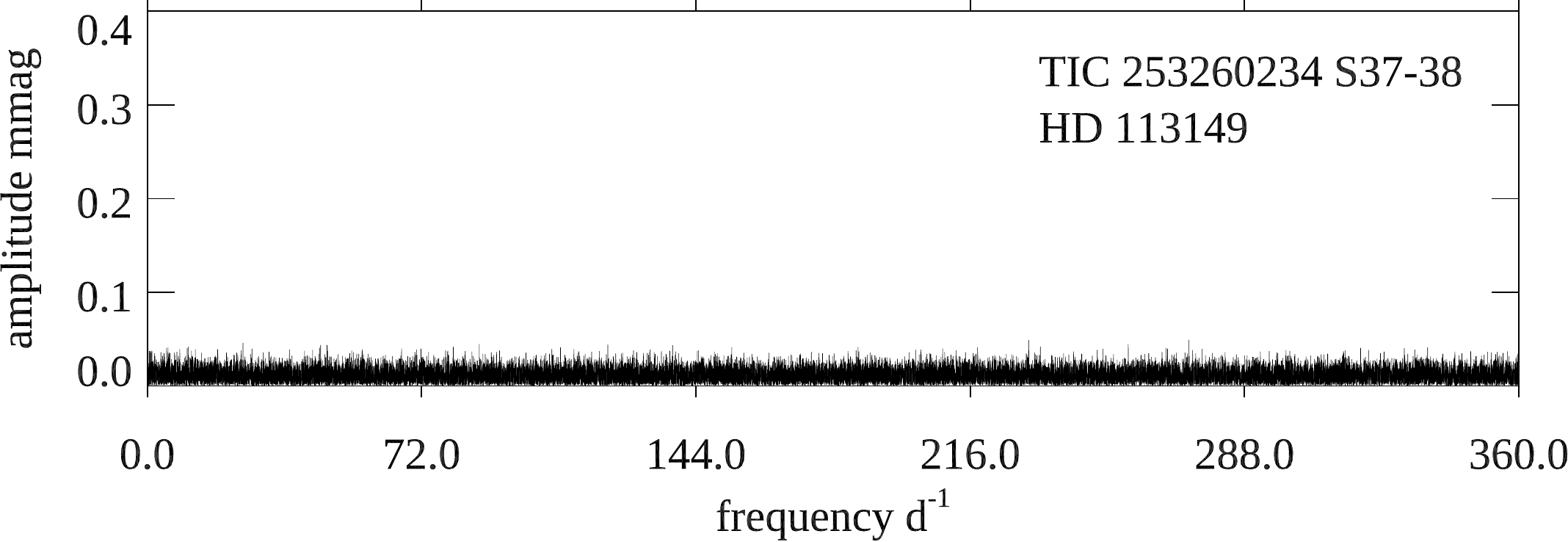}
 \includegraphics[width=0.48\linewidth,angle=0]{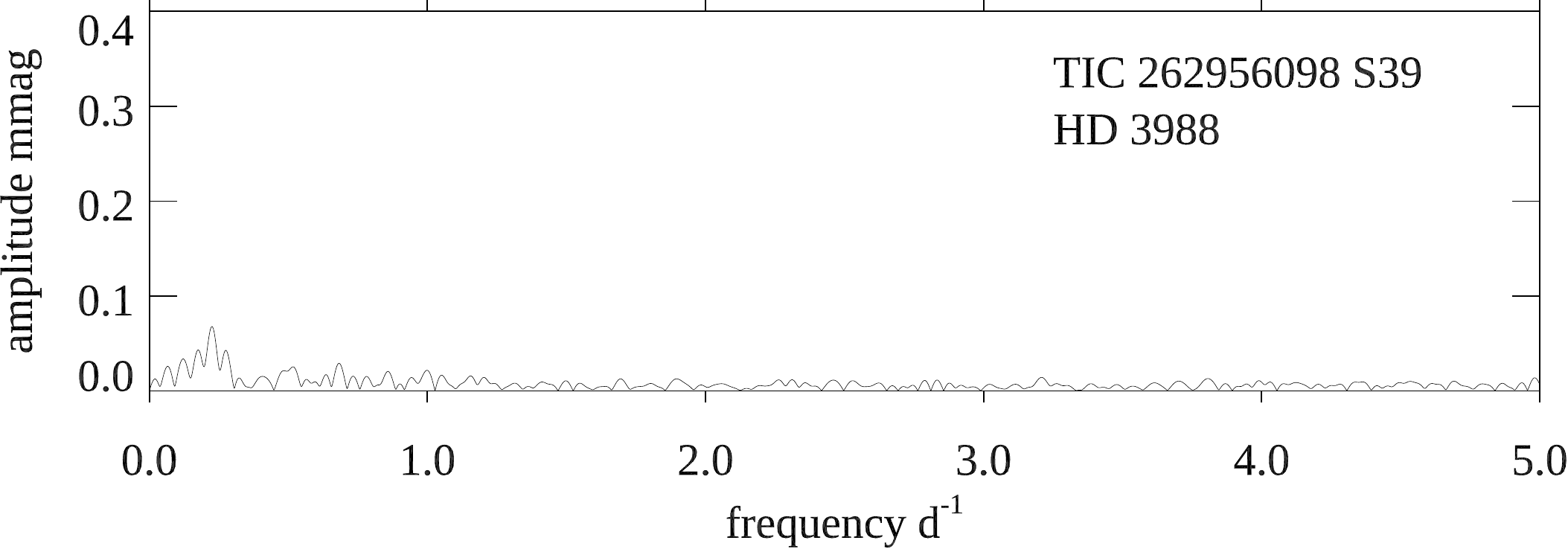}
 \includegraphics[width=0.48\linewidth,angle=0]{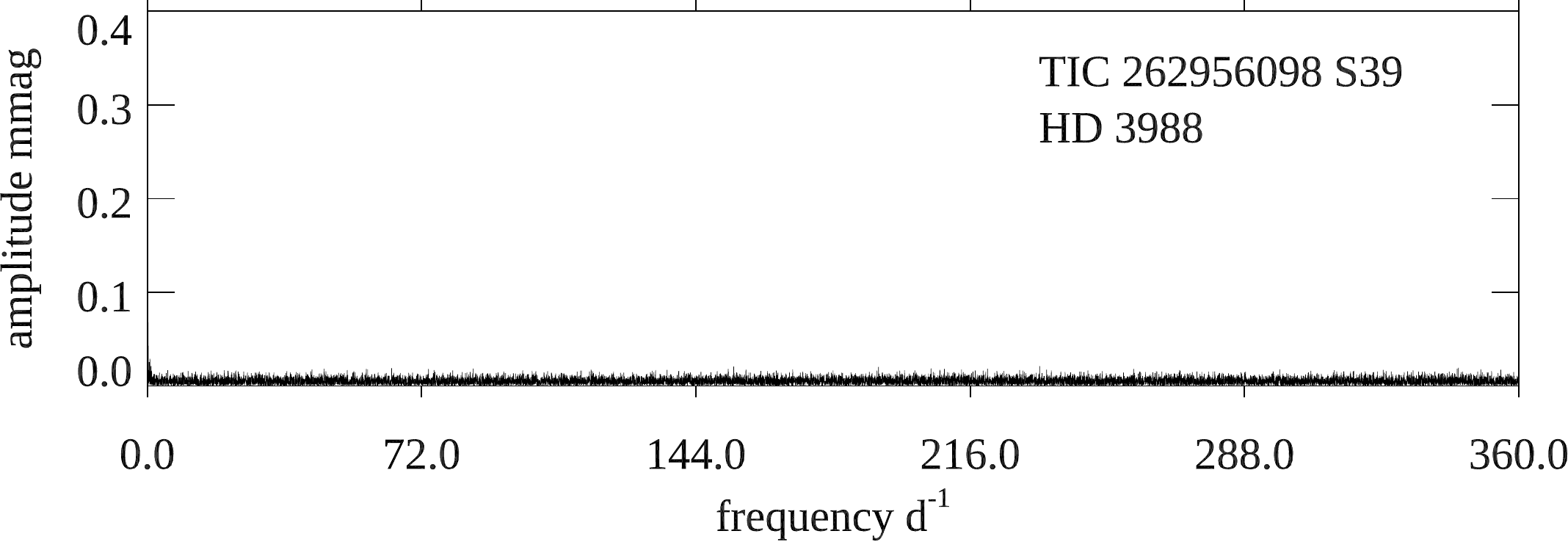}
  \includegraphics[width=0.48\linewidth,angle=0]{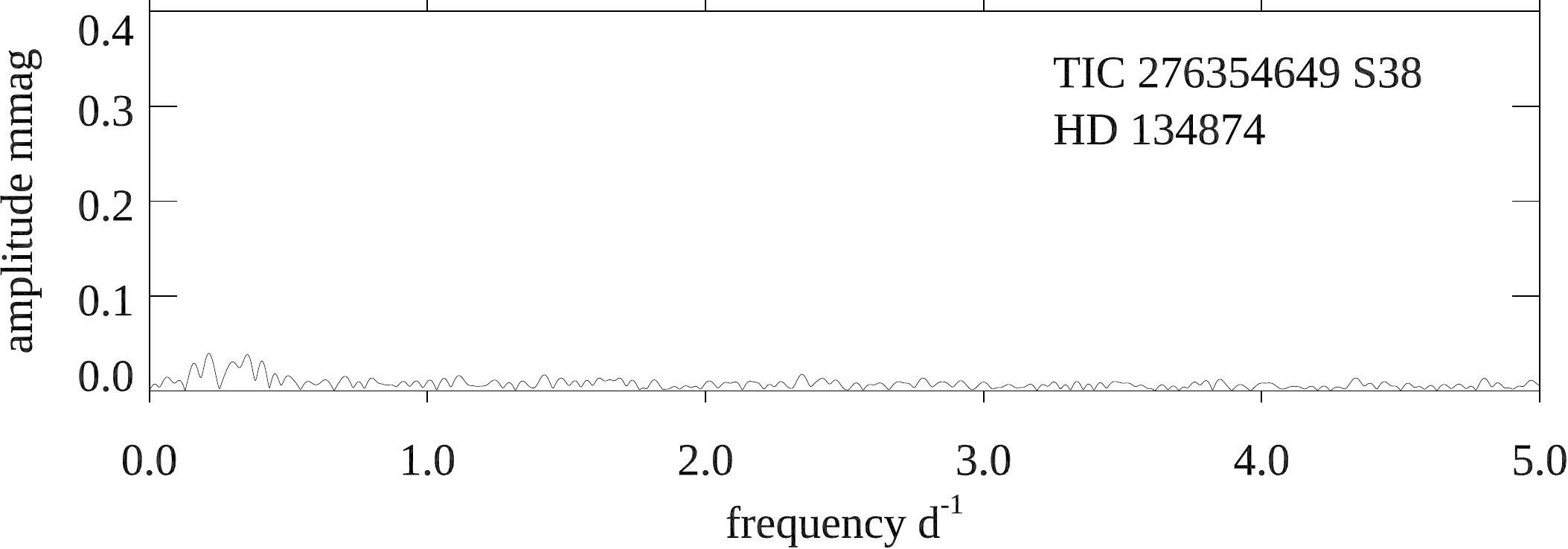}
  \includegraphics[width=0.48\linewidth,angle=0]{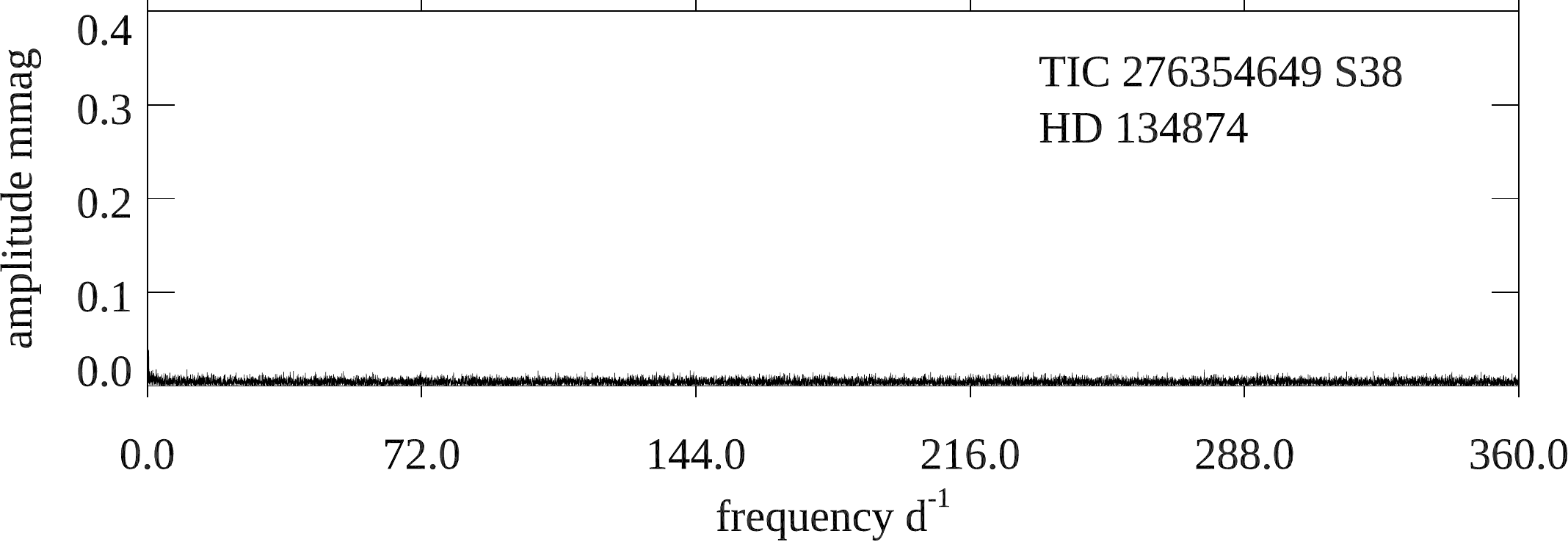}
  \includegraphics[width=0.48\linewidth,angle=0]{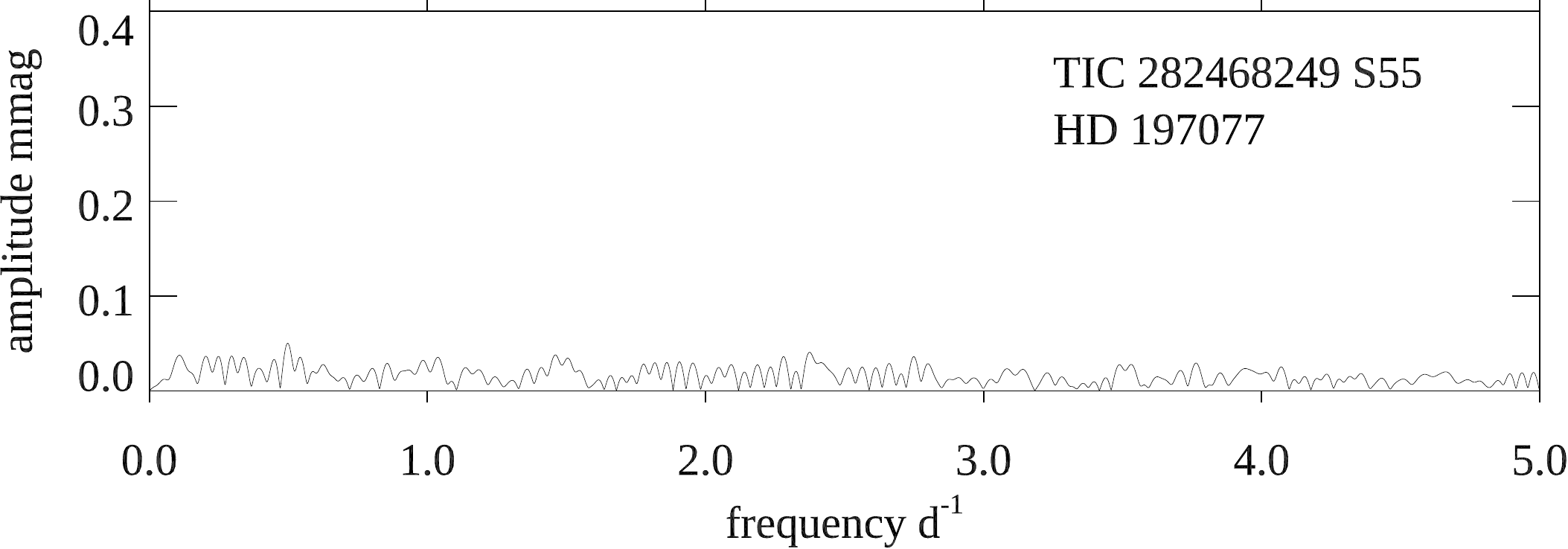}
  \includegraphics[width=0.48\linewidth,angle=0]{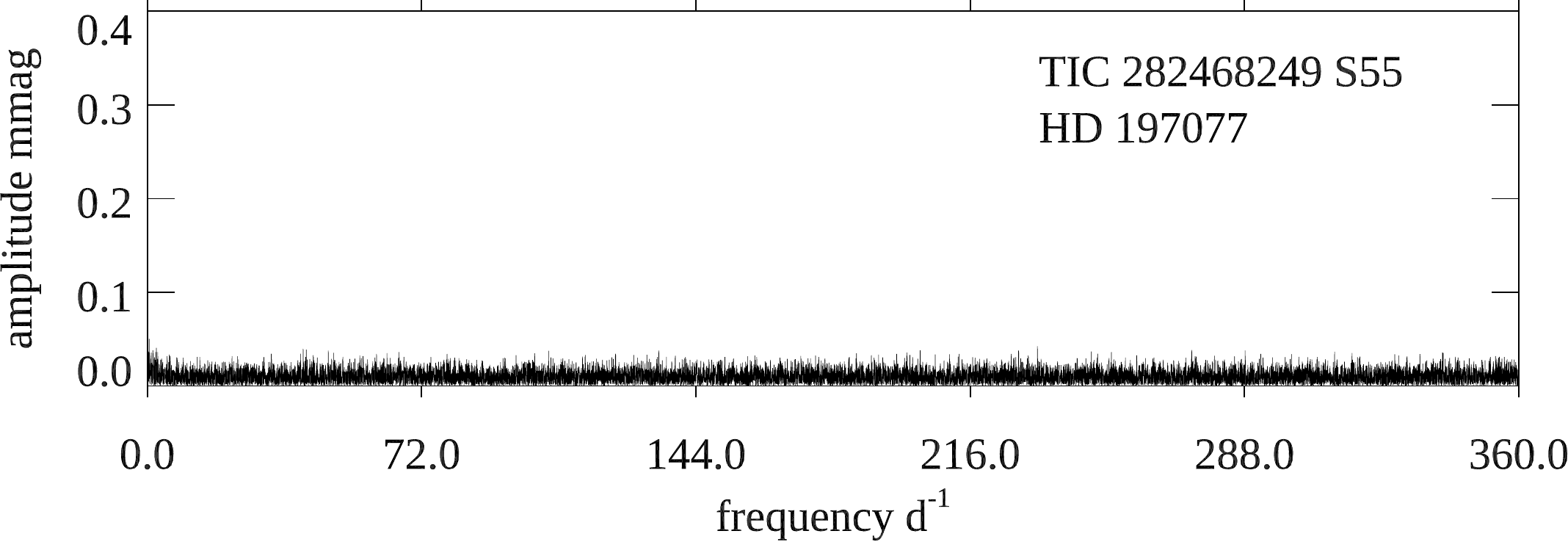}
  \includegraphics[width=0.48\linewidth,angle=0]{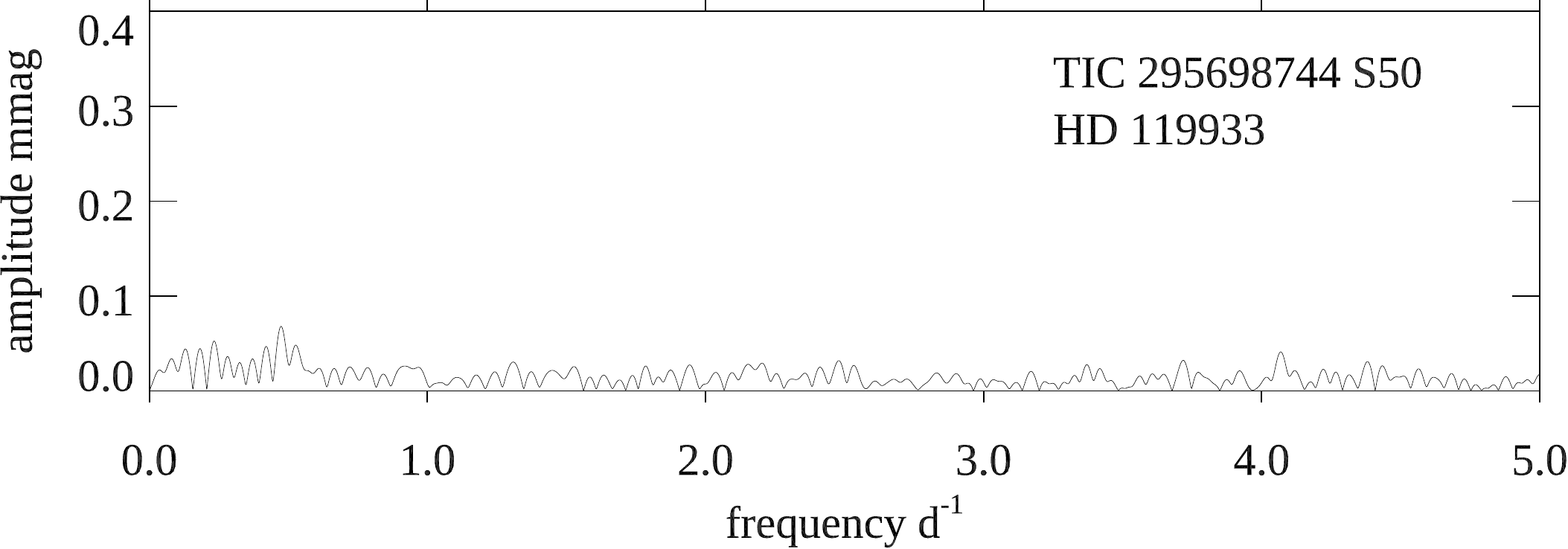}
  \includegraphics[width=0.48\linewidth,angle=0]{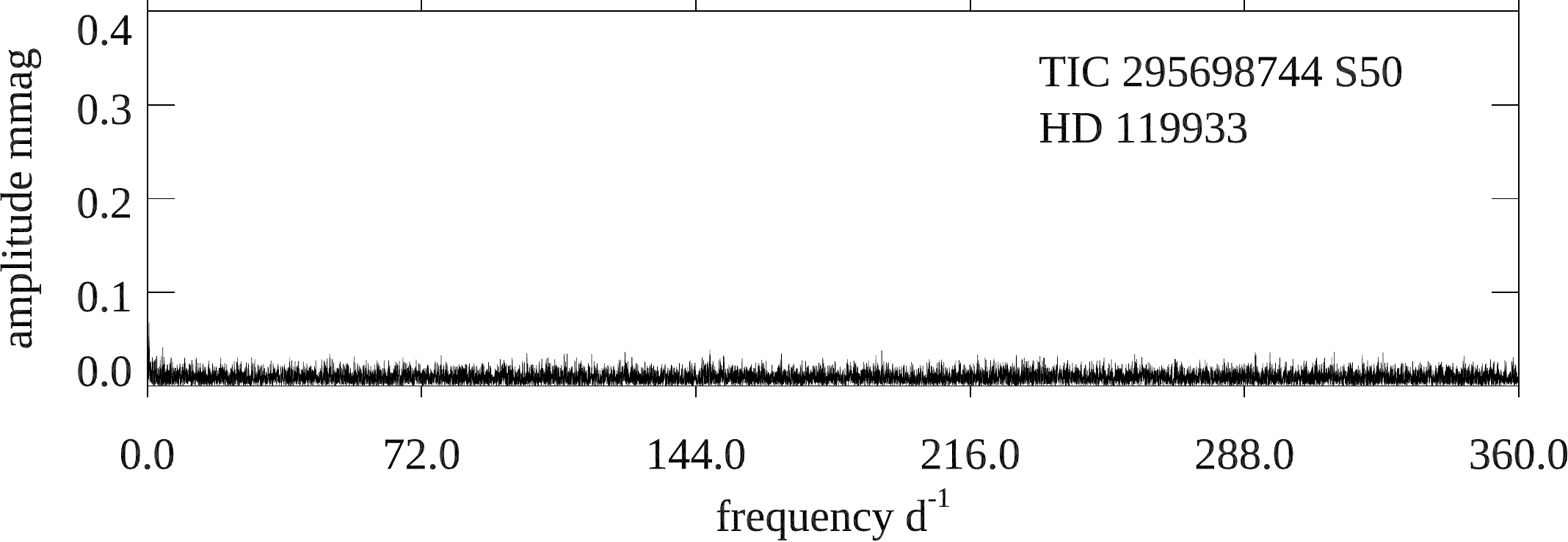}
 \caption{Amplitude spectra for the long-period Ap stars -- continued.
    }
  \label{fig:ssrAp2-2}
\end{figure*}

\begin{figure*}
  \centering
\includegraphics[width=0.48\linewidth,angle=0]{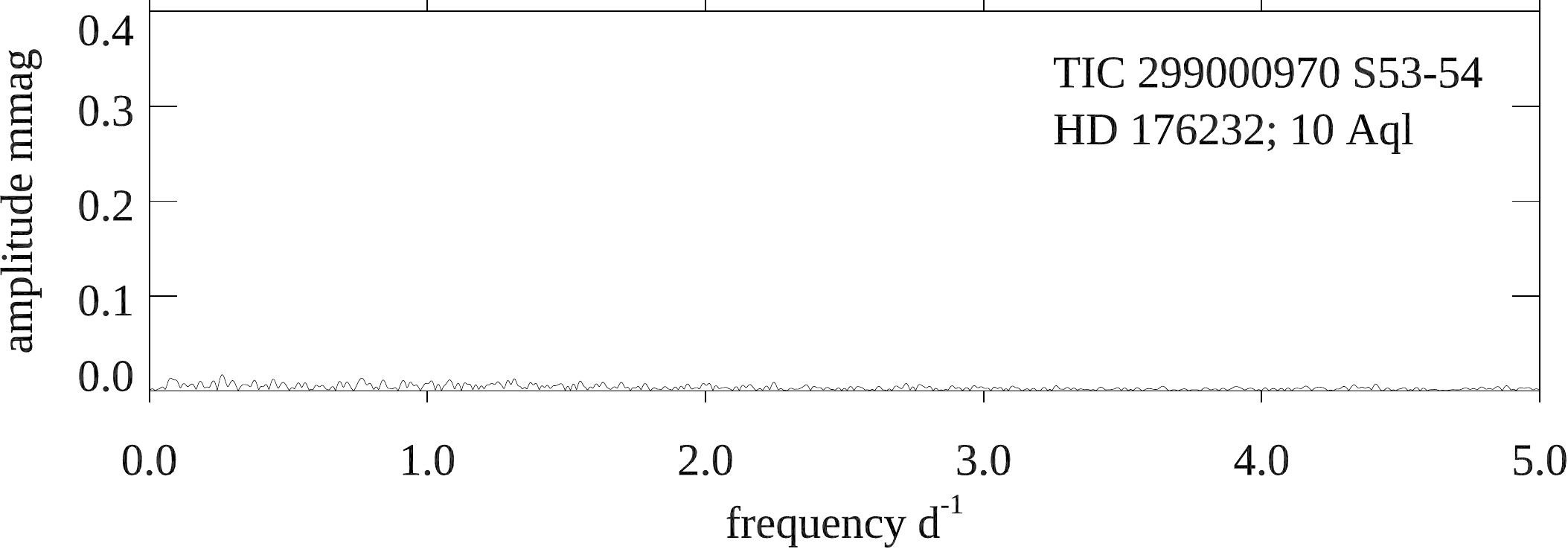}
\includegraphics[width=0.48\linewidth,angle=0]{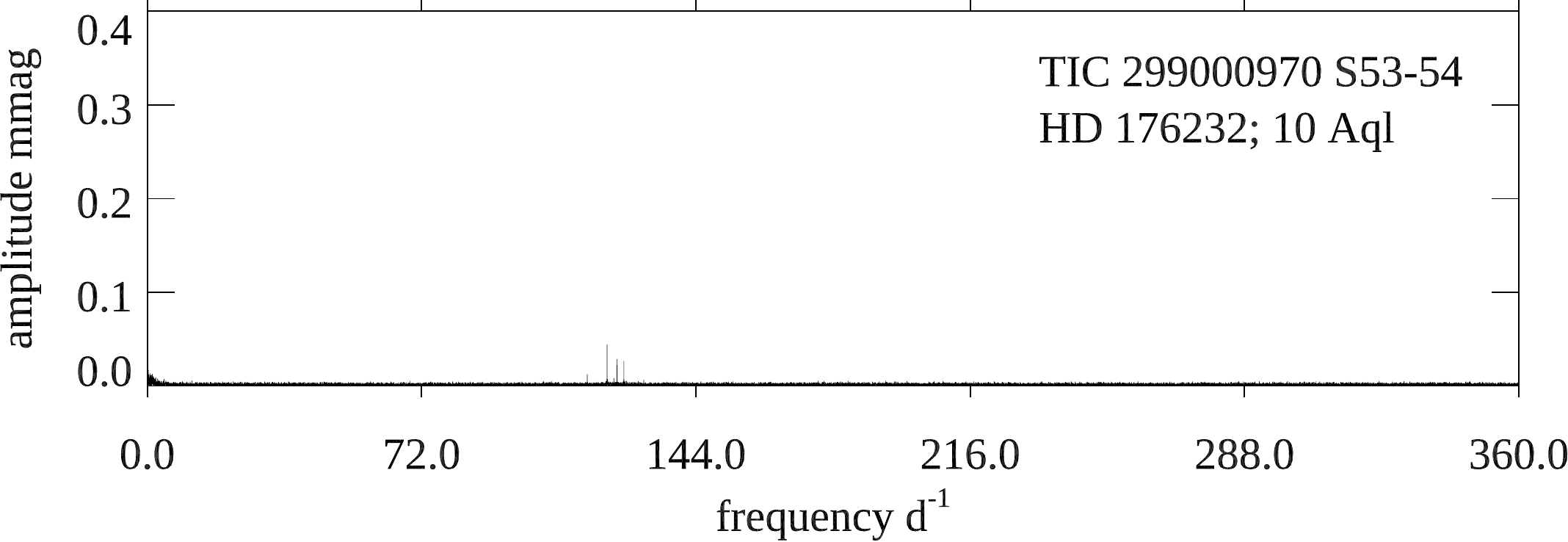}
\includegraphics[width=0.48\linewidth,angle=0]{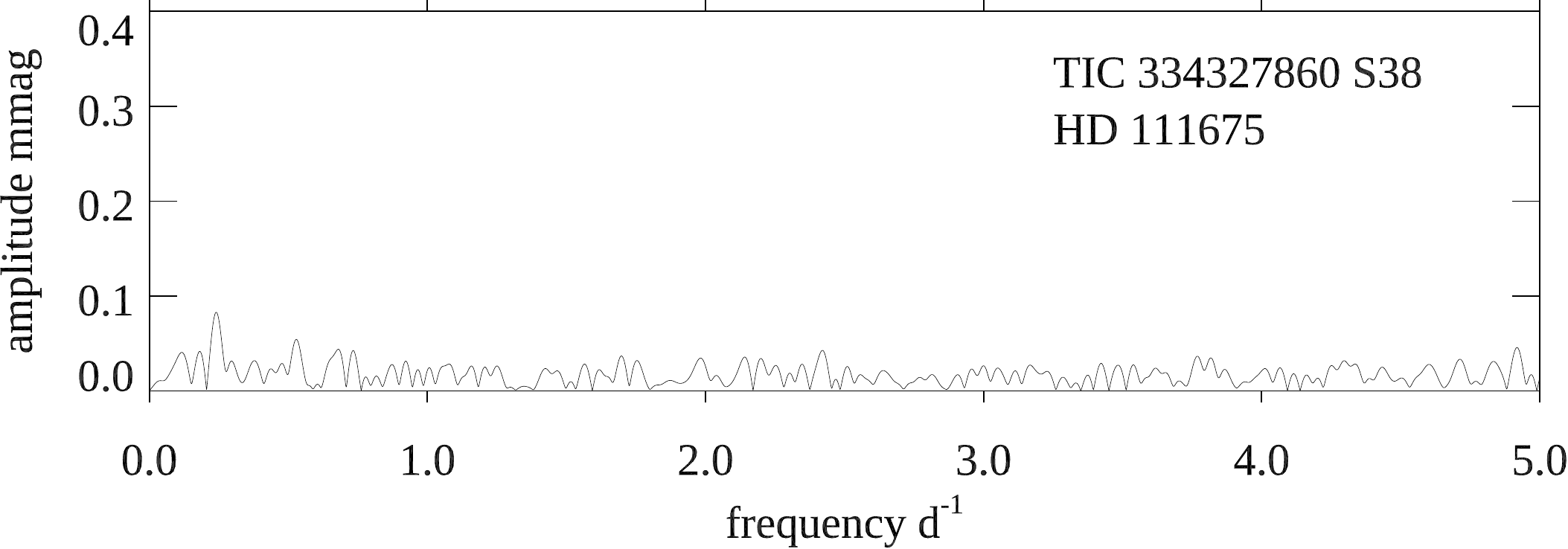}
\includegraphics[width=0.48\linewidth,angle=0]{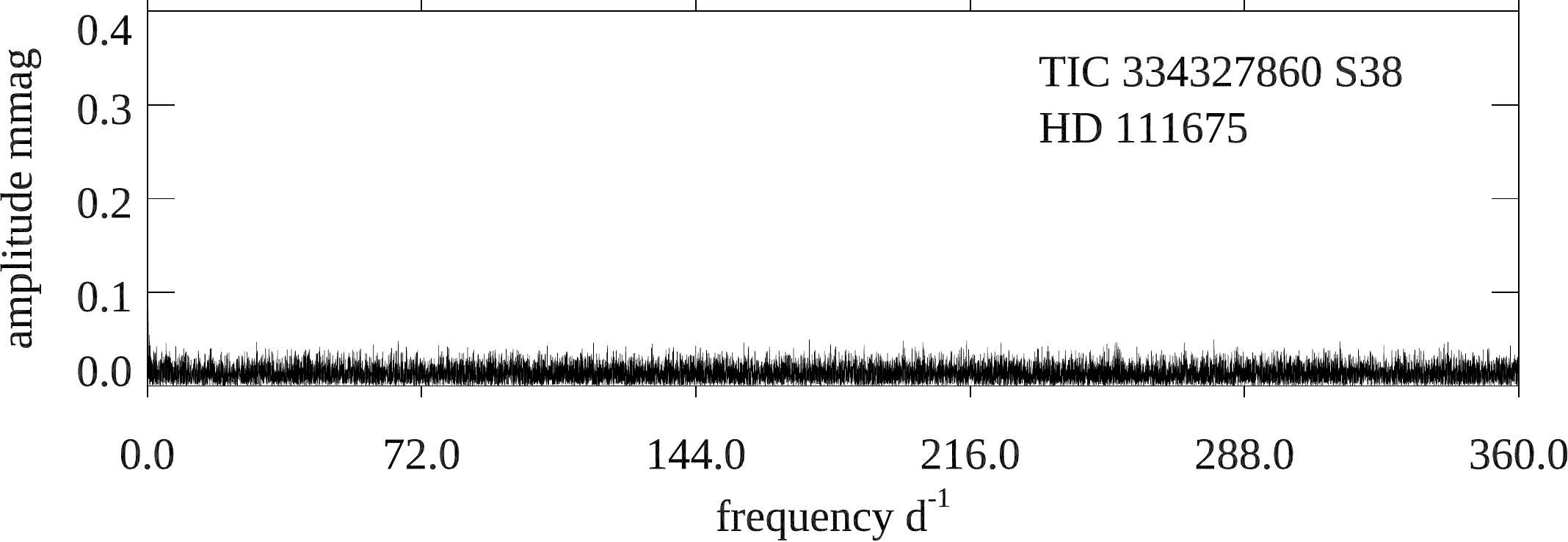}
\includegraphics[width=0.48\linewidth,angle=0]{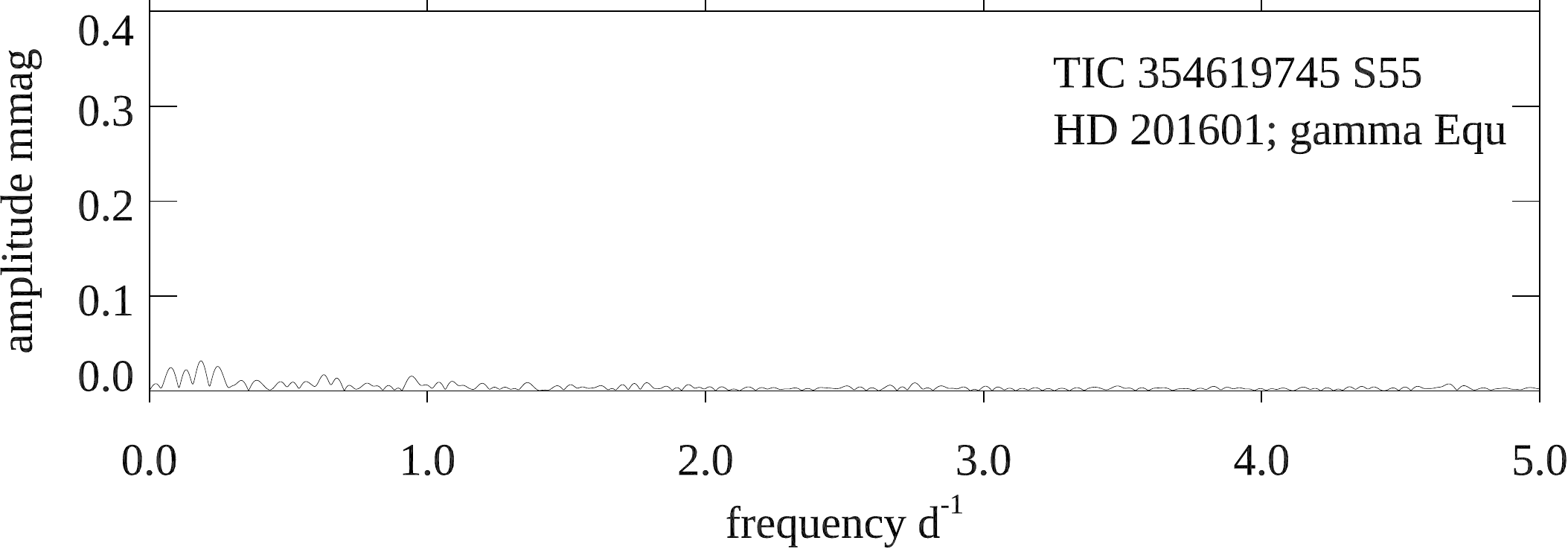}
\includegraphics[width=0.48\linewidth,angle=0]{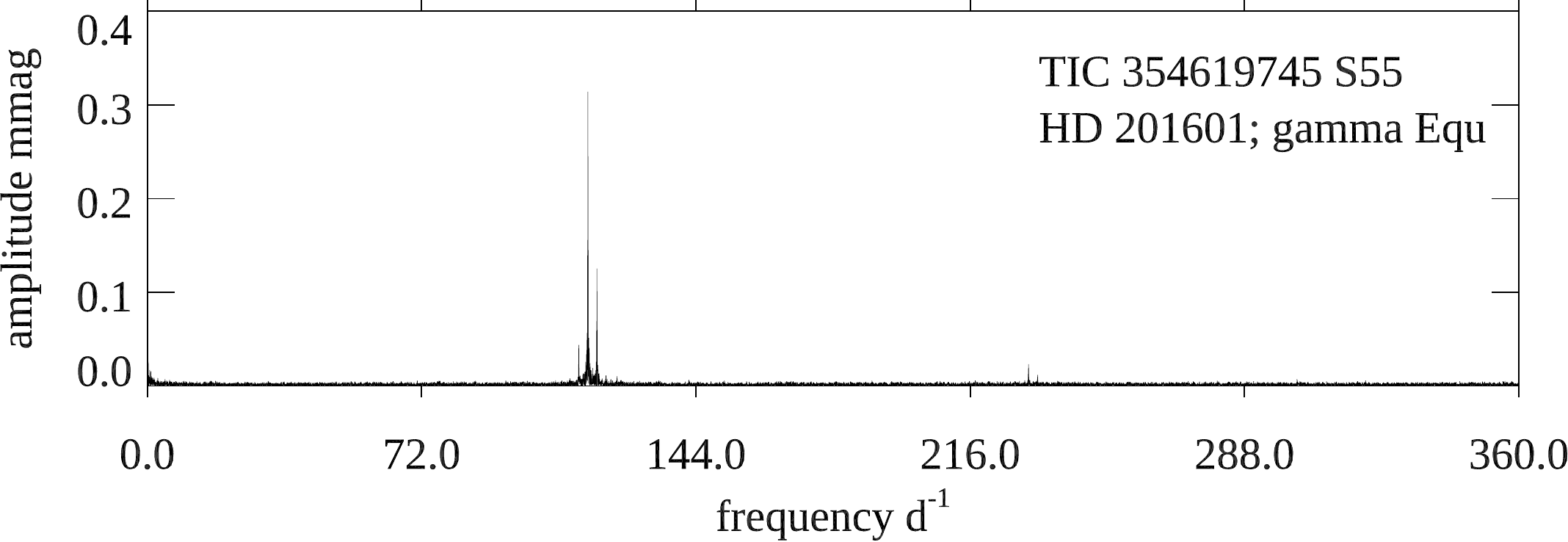}
\includegraphics[width=0.48\linewidth,angle=0]{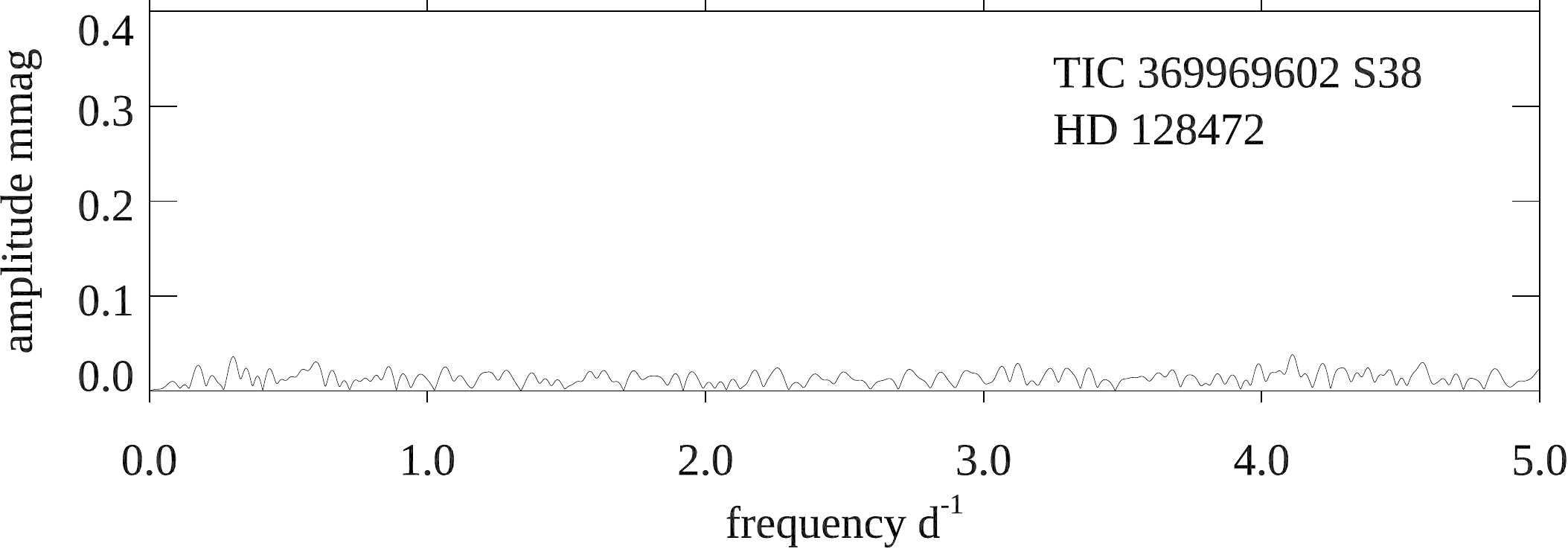}
\includegraphics[width=0.48\linewidth,angle=0]{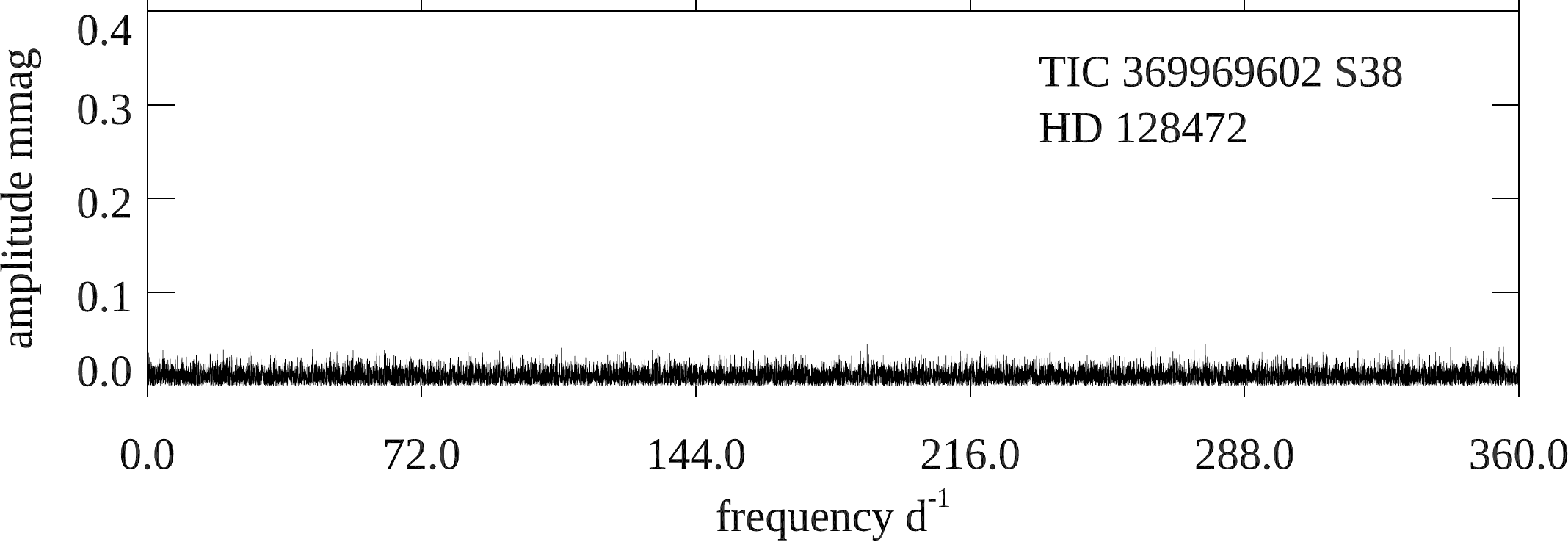}
\includegraphics[width=0.48\linewidth,angle=0]{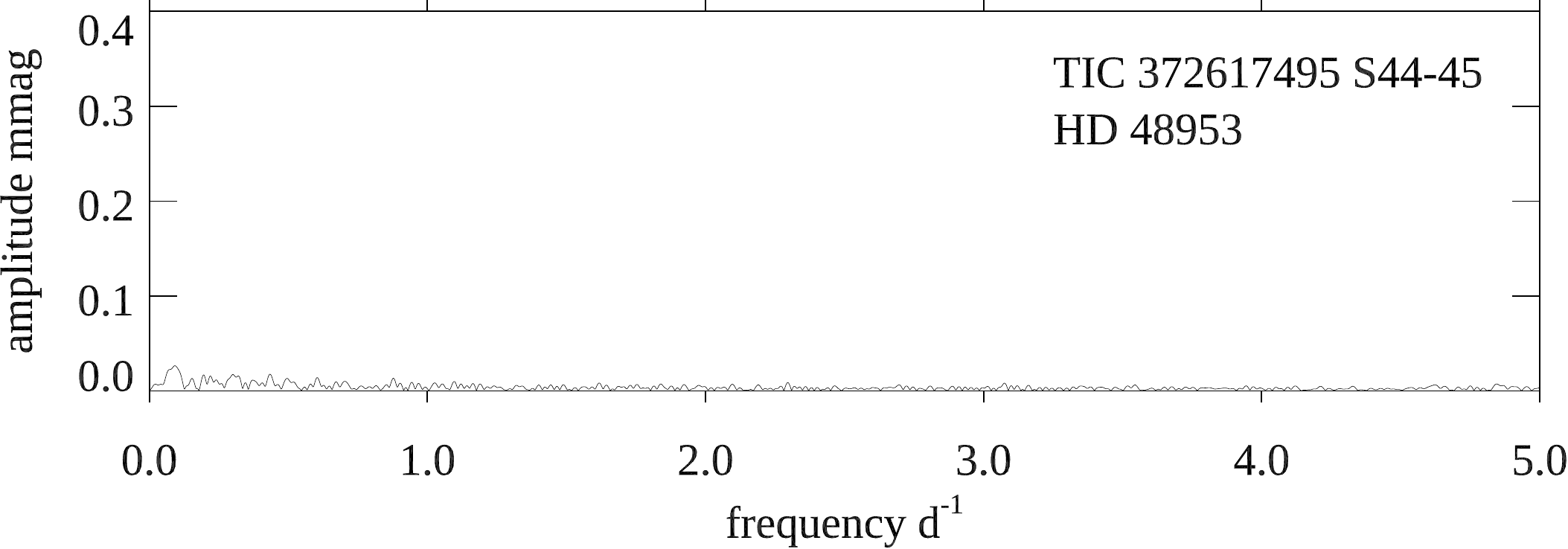}
\includegraphics[width=0.48\linewidth,angle=0]{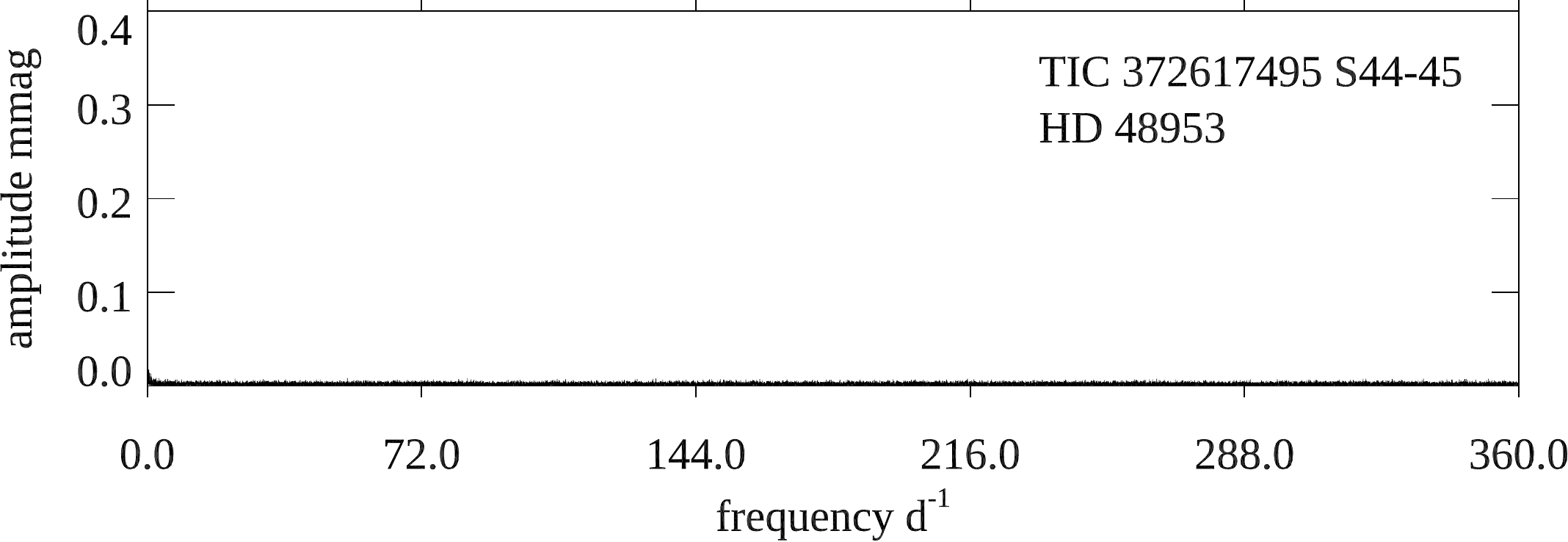}
\includegraphics[width=0.48\linewidth,angle=0]{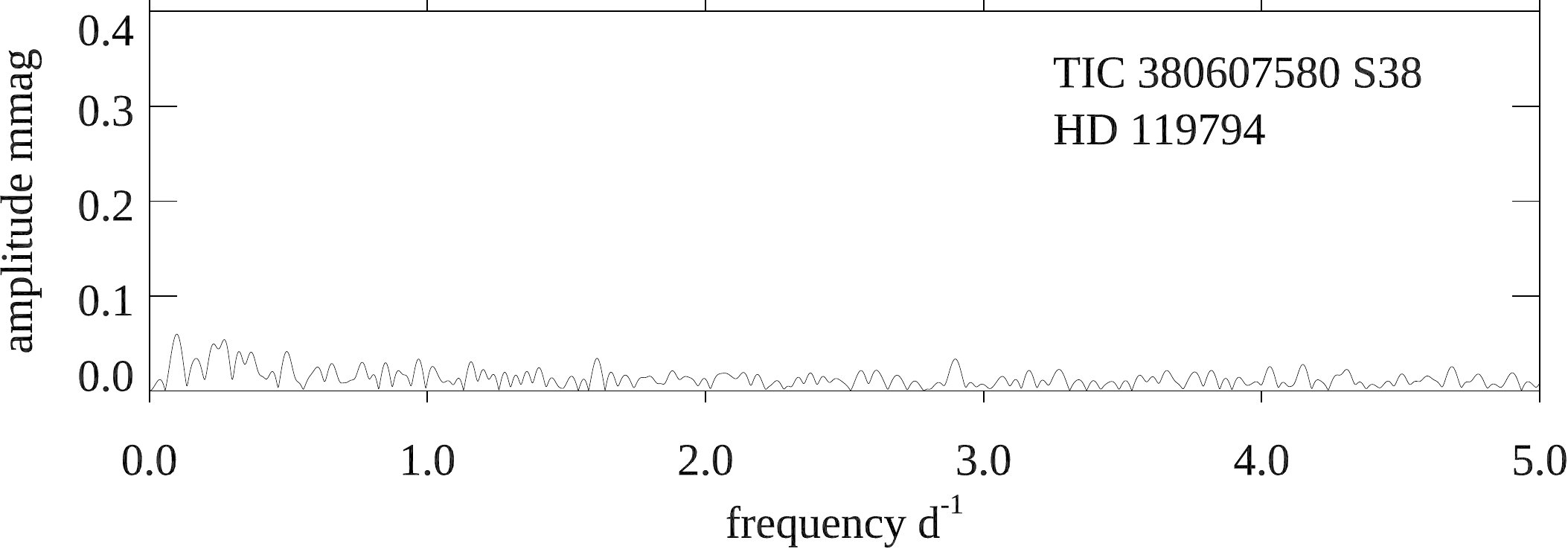}
\includegraphics[width=0.48\linewidth,angle=0]{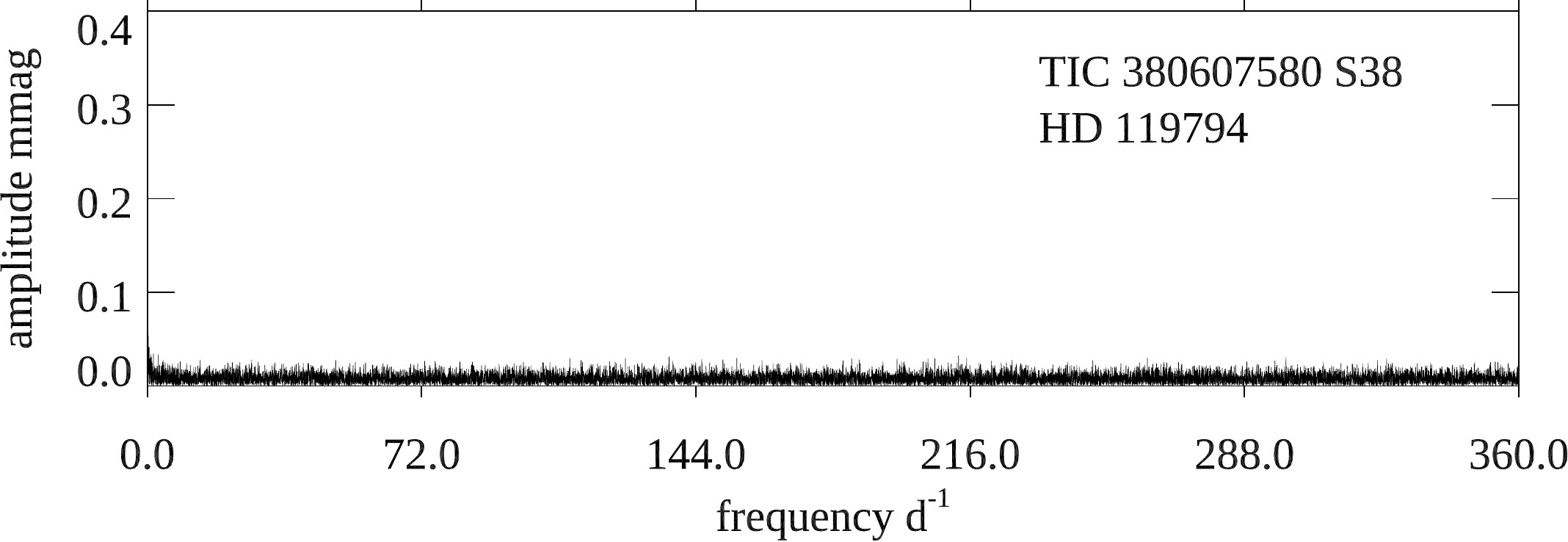}
\includegraphics[width=0.48\linewidth,angle=0]{405516045_S37-38_SAP_adjusted.pdf}
\includegraphics[width=0.48\linewidth,angle=0]{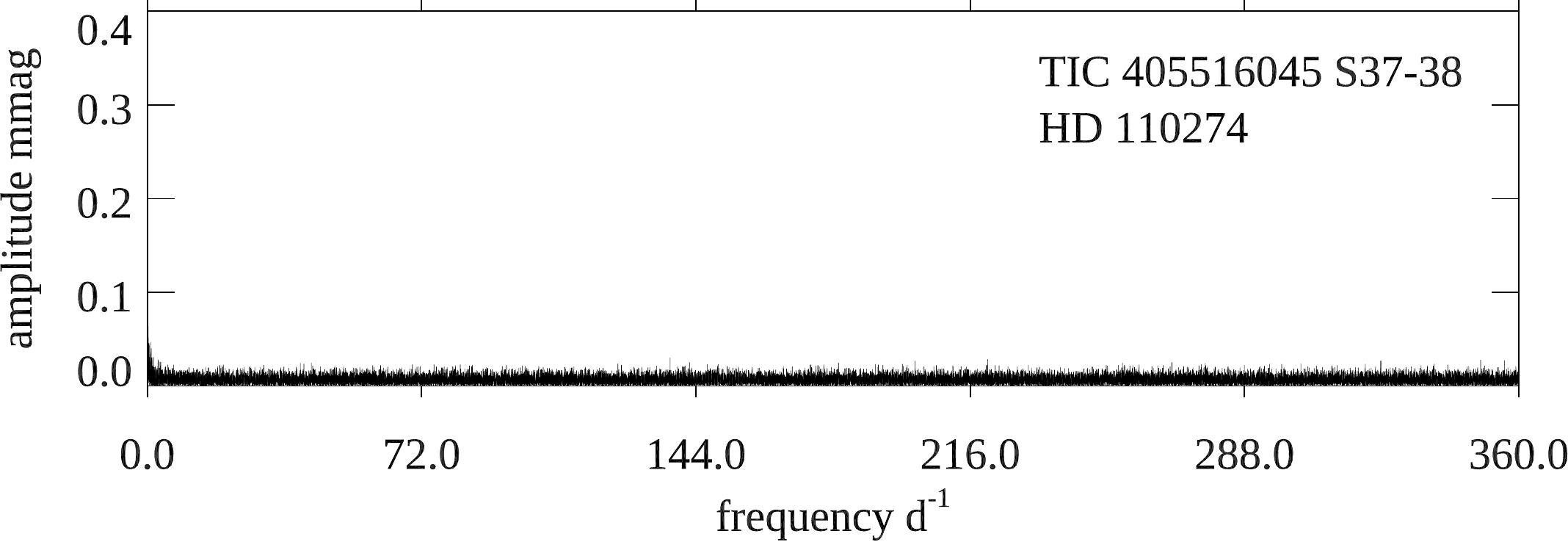}
\caption{Amplitude spectra for the long-period Ap stars -- continued.}
  \label{fig:ssrAp2-3}
\end{figure*}

\begin{figure*}
  \centering
\includegraphics[width=0.48\linewidth,angle=0]{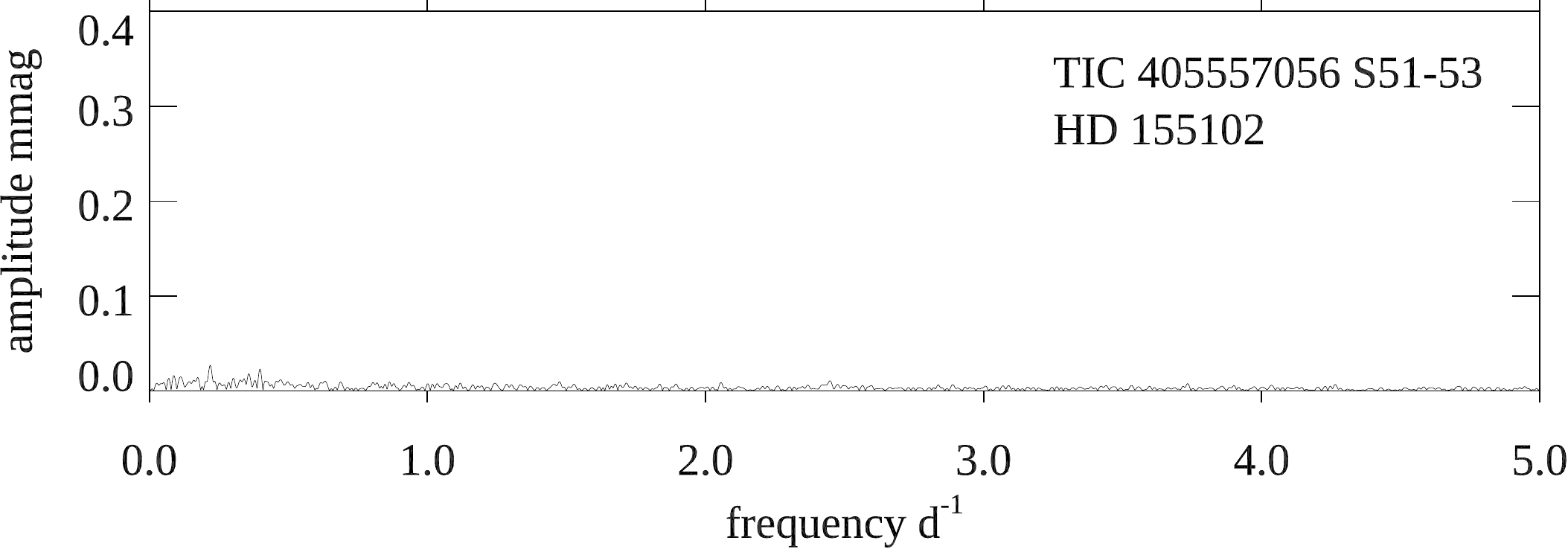}
\includegraphics[width=0.48\linewidth,angle=0]{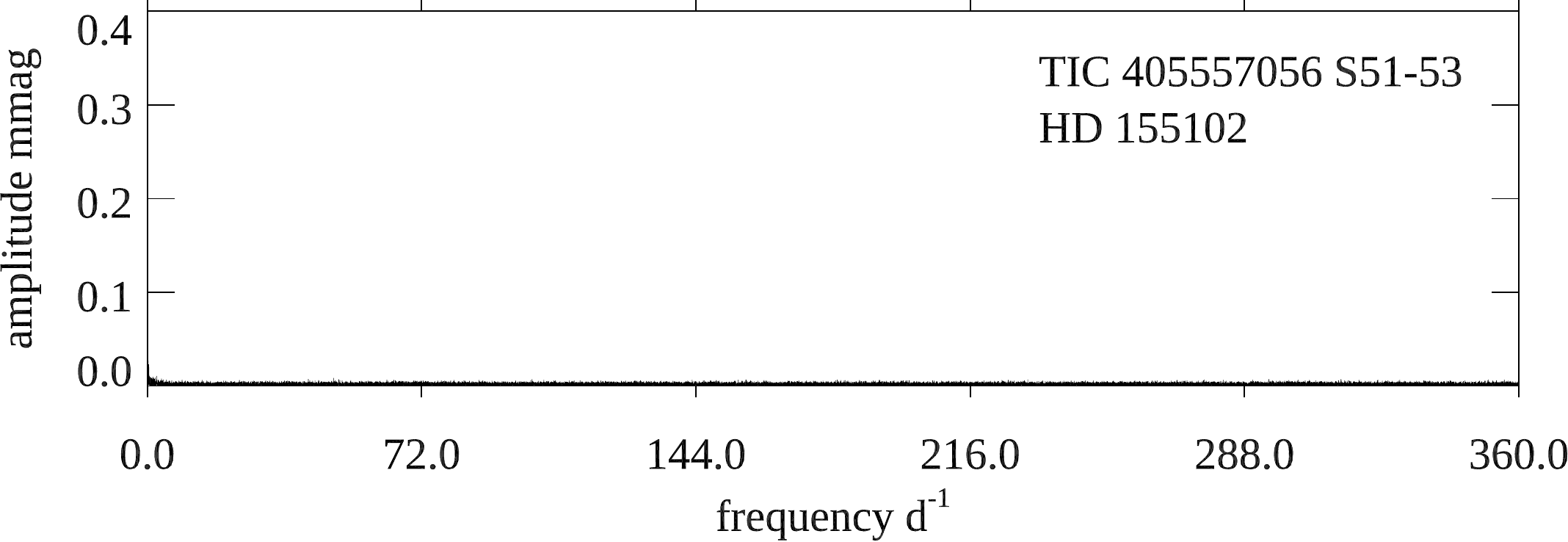}
\includegraphics[width=0.48\linewidth,angle=0]{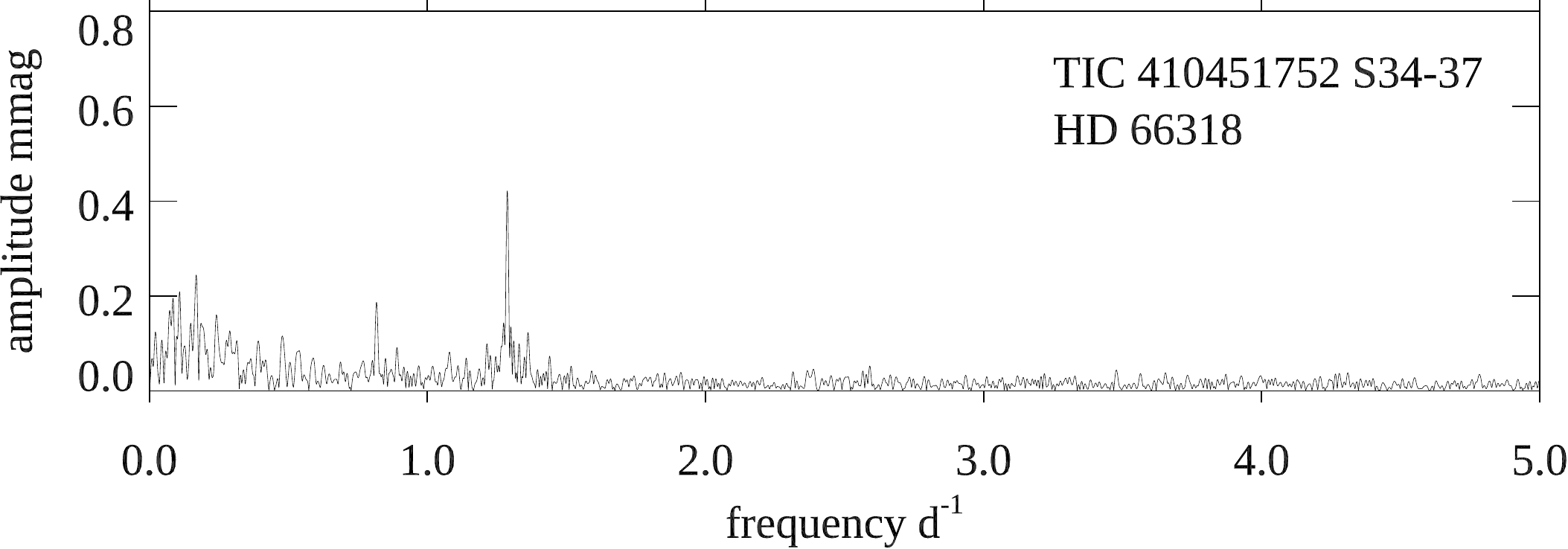}
\includegraphics[width=0.48\linewidth,angle=0]{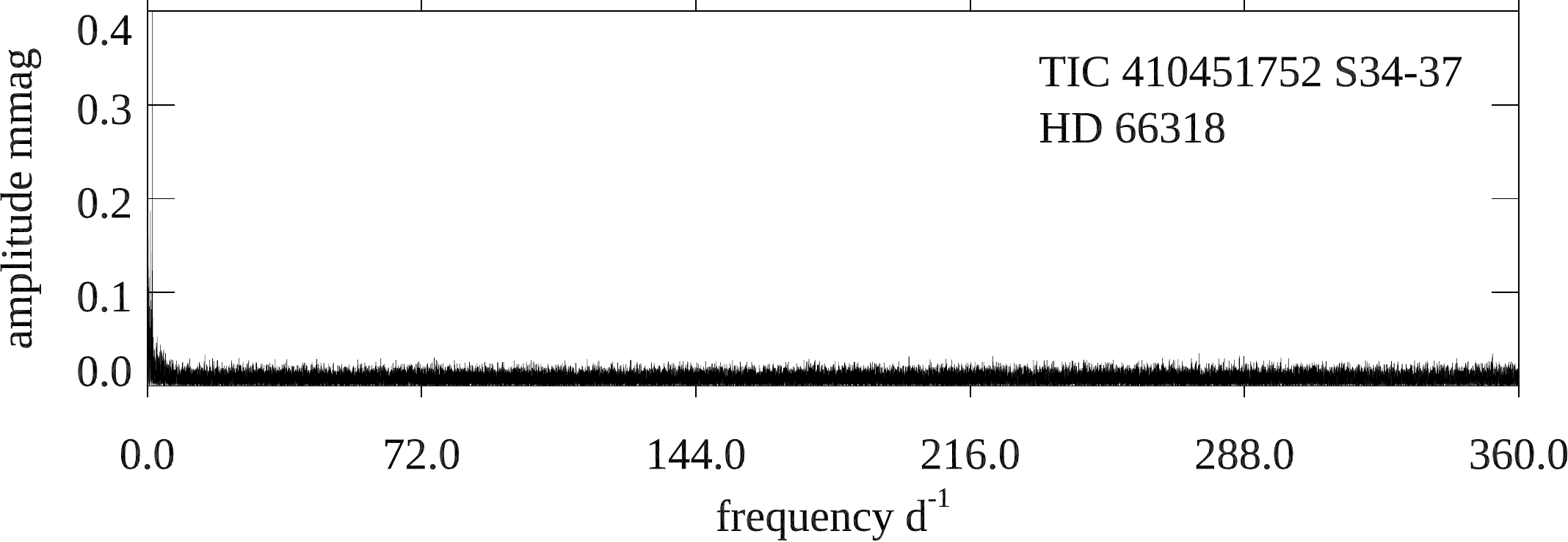}
\includegraphics[width=0.48\linewidth,angle=0]{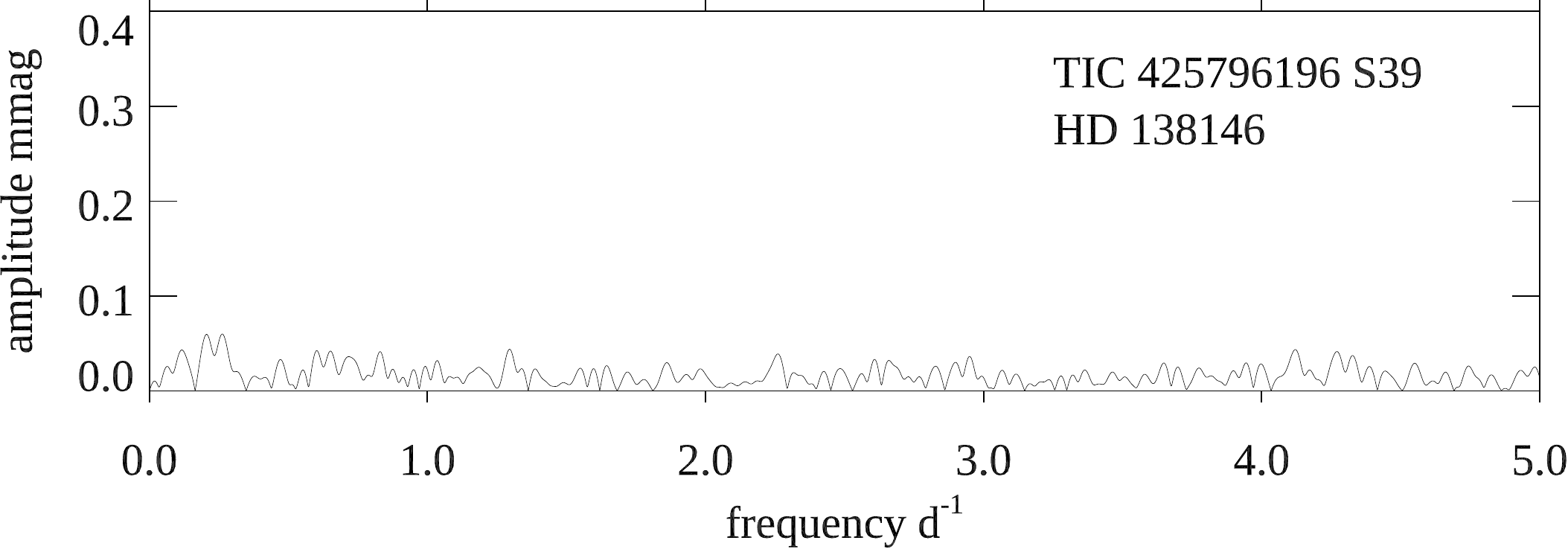}
\includegraphics[width=0.48\linewidth,angle=0]{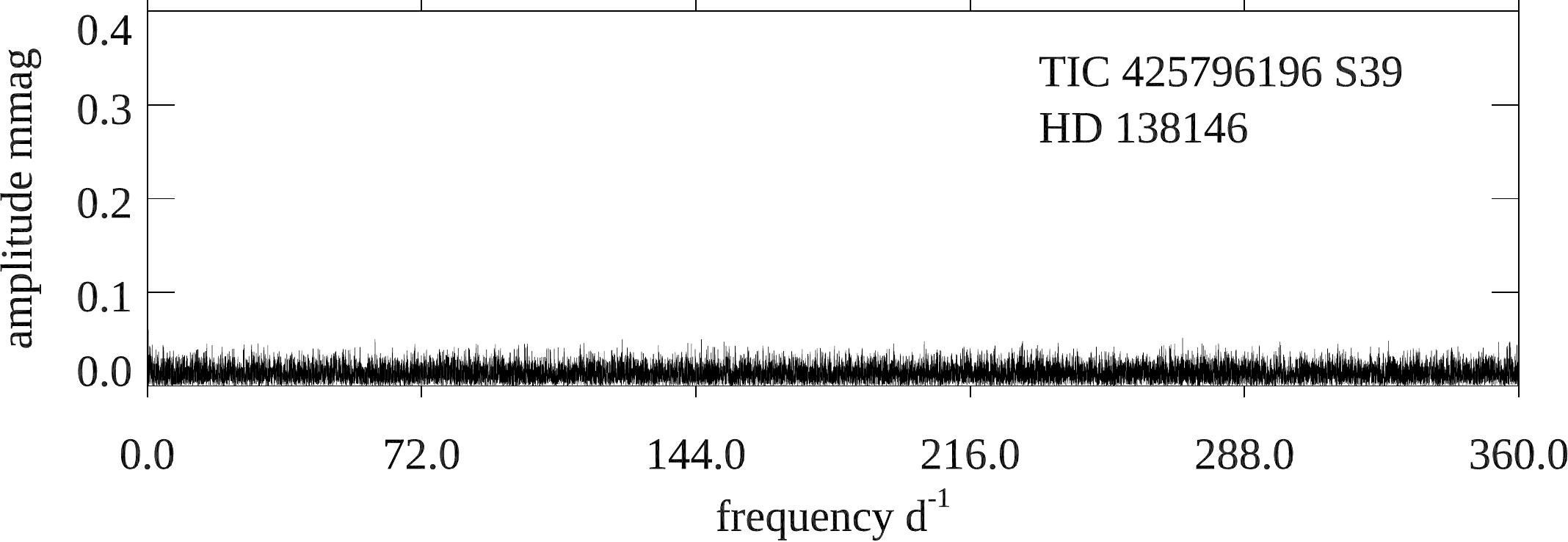}
\includegraphics[width=0.48\linewidth,angle=0]{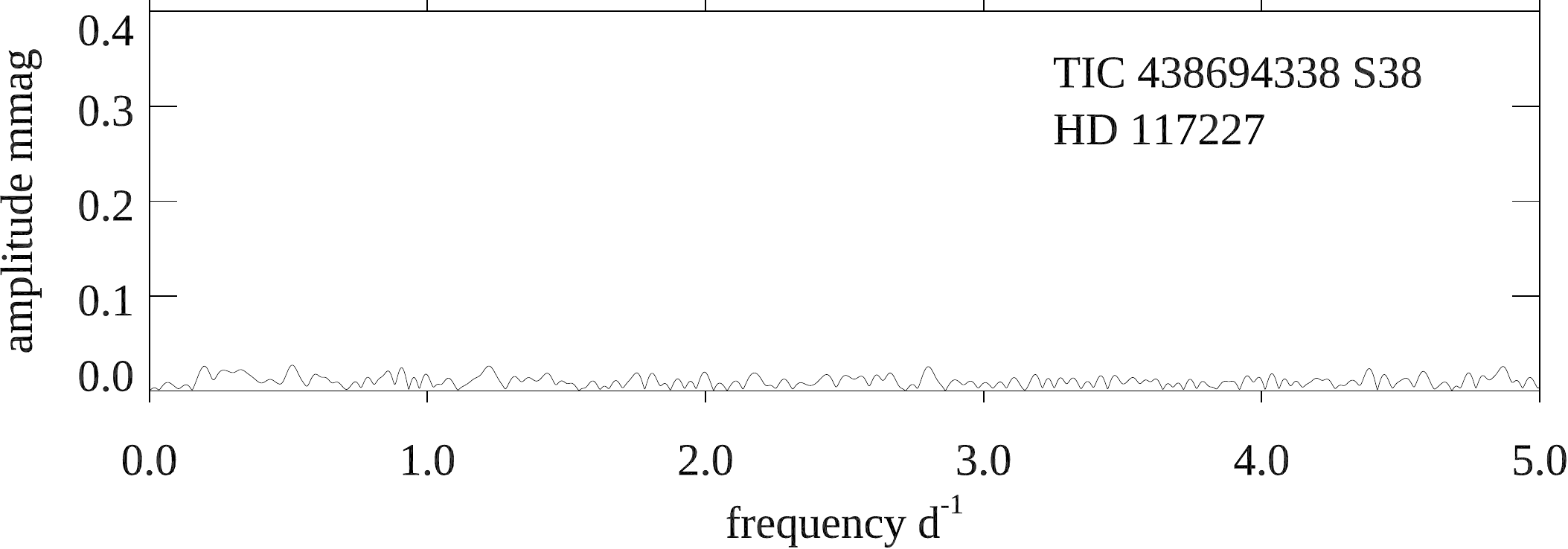}
\includegraphics[width=0.48\linewidth,angle=0]{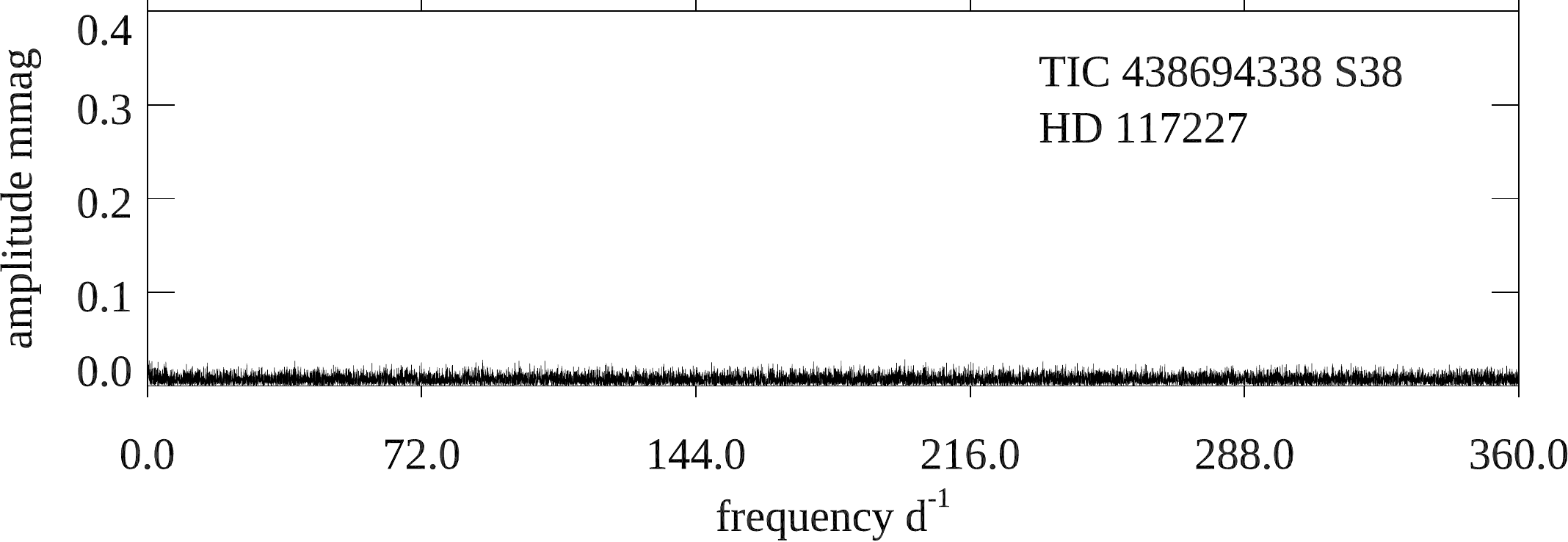}
\includegraphics[width=0.48\linewidth,angle=0]{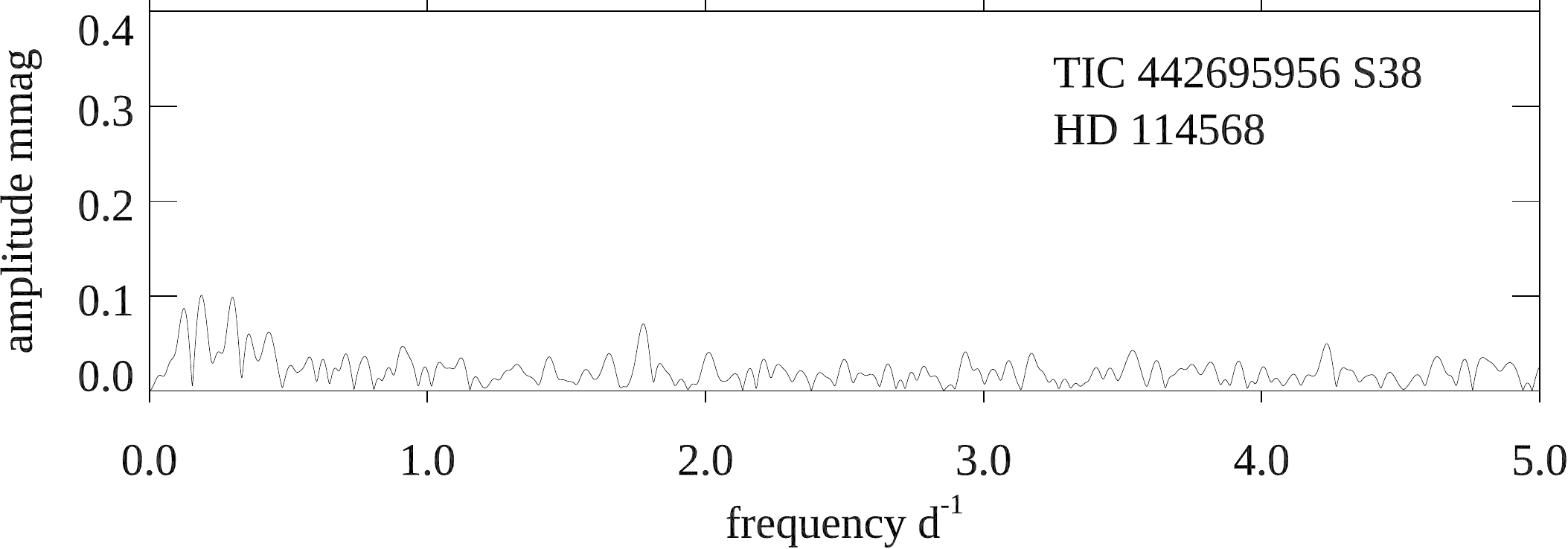}
\includegraphics[width=0.48\linewidth,angle=0]{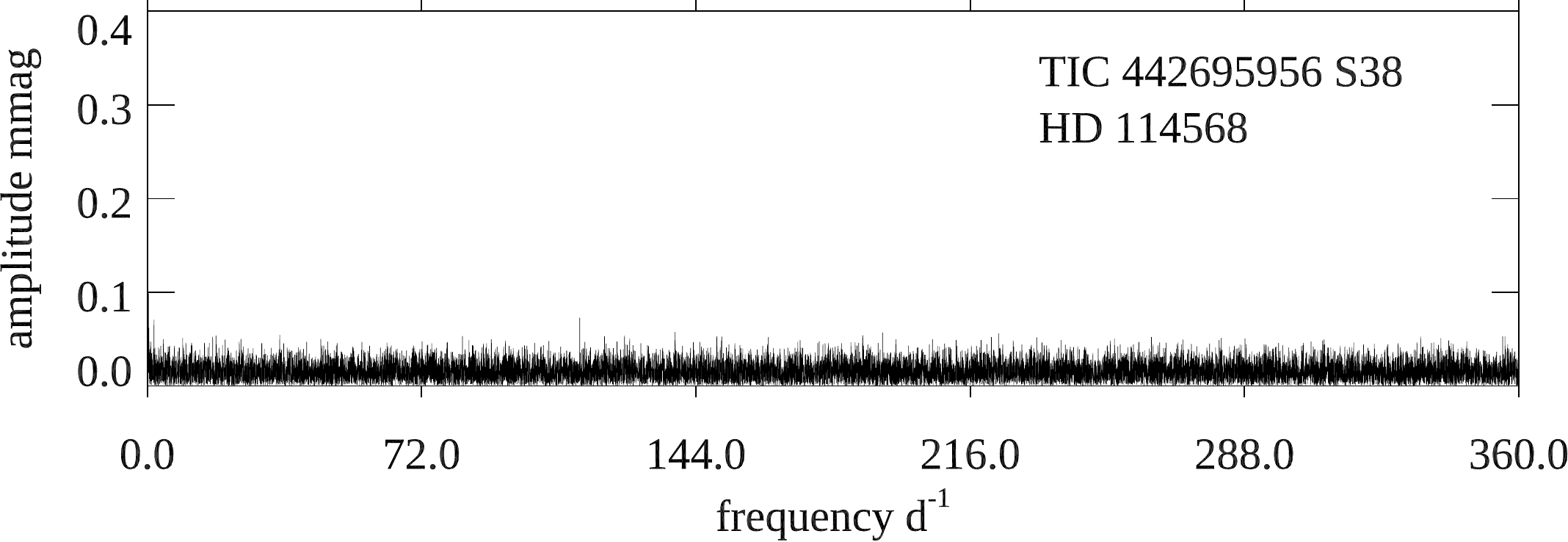}
\caption{Amplitude spectra for the long-period Ap stars -- continued. 
TIC\,410451752 (HD\,66318) has a 55$\sigma$ peak at $1.2874 \pm
0.0001$\,d$^{-1}$ ($P = 0.77675 \pm 0.00006$\,d), and a 24$\sigma$
peak at $0.0174 \pm 0.0002$\,d$^{-1}$. These may be g~modes, which is
unusual for an Ap star.} 
\label{fig:ssrAp2-4}
\end{figure*}

Figs.\,\ref{fig:ssrAp2-1} to \ref{fig:ssrAp2-4} shows amplitude spectra for the long-period Ap stars. Each row presents a low-frequency amplitude spectrum showing no rotational variation in the left panel, and a full amplitude spectrum to the Nyquist frequency of 360\,d$^{-1}$ in the right panel, which allows detection of $\delta$~Sct or roAp pulsation.

\end{document}